\documentclass[longauth]{aa} % for the long lists of 
\usepackage{graphicx}
\usepackage{txfonts}
\usepackage{hyperref}
\setcitestyle{citesep={,}}

\begin{document} 

    % \title{(ASASSN by MUSE) A characterization of ASAS-SN core-collapse supernova environments with VLT+MUSE}

    % \title{Supernovae and revelations: A characterization of ASAS-SN core-collapse supernova environments with VLT+MUSE}

   \title{A characterization of ASAS-SN core-collapse supernova environments with VLT+MUSE\thanks{Tables 1, 2, D.1, D.2, and G.1 are only available in electronic form at the CDS via anonymous ftp to \url{cdsarc.cds.unistra.fr} (130.79.128.5) or via \url{https://cdsarc.cds.unistra.fr/cgi-bin/qcat?J/A+A/}. The parameter maps are also available in \url{https://sites.google.com/view/theamusingasassnsample}.}}

   \subtitle{I. Sample selection, analysis of local environments, and correlations with light curve properties}

   \author{T. Pessi
          \inst{1,2}
          \and
          J. L. Prieto
          \inst{1,3}
          % \fnmsep
          % \thanks{Just to show the usage
          % of the elements in the author field}
          \and
          J. P. Anderson
          \inst{2,3}
          \and
          L. Galbany
          \inst{4,5}
          \and 
          J. D. Lyman
          \inst{6}
          \and
        C. Kochanek
        \inst{7,8} 
        \and
        S. Dong
        \inst{9}
        \and
        F. Forster
        \inst{3,10,11}
        \and
        R. González-Díaz 
        \inst{4,12}
        \and
        S. Gonzalez-Gaitan
        \inst{13}
        \and
        C. P. Gutiérrez 
        \inst{14, 15}
        \and
        T. W.-S. Holoien
        \inst{16}
        \and
        P. A. James
        \inst{17}
        \and
        C. Jiménez-Palau
        \inst{4,5}
        \and
        E. J. Johnston   
        \inst{1}
        \and
        H. Kuncarayakti 
        \inst{14,15}
        % \and
        % E. Perez
        \and
        F. Rosales-Ortega
        \inst{12}
        \and
        S. F. S\'anchez
        \inst{18}
        \and
        S. Schulze
        \inst{19}
        \and 
        B. Shappee
        \inst{20}}

   \institute{Instituto de Estudios Astrof\'isicos, Facultad de Ingenier\'ia y Ciencias, Universidad Diego Portales, Av. Ej\'ercito Libertador 441, Santiago, Chile\\ %1
              \email{thallis.pessi@mail.udp.cl}
         \and % 2
             European Southern Observatory, Alonso de Córdova 3107, Vitacura, Casilla 19001, Santiago, Chile
             % \email{c.ptolemy@hipparch.uheaven.space}
         \and % 3
            Millennium Institute of Astrophysics MAS, Nuncio Monse\~nor Sotero Sanz 100, Off. 104, Providencia, Santiago, Chile
        \and % 4
            Institute of Space Sciences (ICE, CSIC), Campus UAB, Carrer de Can Magrans, s/n, E-08193 Barcelona, Spain
        \and  % 5
            Institut d’Estudis Espacials de Catalunya (IEEC), E-08034 Barcelona, Spain
        \and % 6
            Department of Physics, University of Warwick, Coventry CV4 7AL, UK
        \and % 7
        Department of Astronomy, The Ohio State University, 140 West 18th Avenue, Columbus, OH 43210, USA
        \and  % 8
        Center for Cosmology and Astroparticle Physics, The Ohio State University, 191 W. Woodruff Avenue, Columbus, OH 43210, USA
        \and % 9
        Kavli Institute for Astronomy and Astrophysics, Peking University, Yi He Yuan Road 5, Hai Dian District, Beijing 100871, People's Republic of China
        \and % 10
        Data and Artificial Intelligence Initiative (IDIA), Faculty of Physical and Mathematical Sciences, University of Chile, Santiago, Chile
        \and % 11
        Center for Mathematical Modeling (CMM), Universidad de Chile, Santiago, Chile.
        \and % 15
        Instituto Nacional de Astrof{\'i}sica, {\'O}ptica y Electr{\'o}nica (INAOE CONACyT), Luis E. Erro 1, 72840, Tonantzintla, Puebla, Mexico
        \and
        CENTRA-Centro de Astrofísica e Gravitação and Departamento de Física, Instituto Superior Técnico, Universidade de Lisboa, Avenida Rovisco Pais, P-1049-001 Lisboa, Portugal
        \and % 12
         Finnish Centre for Astronomy with ESO (FINCA), FI-20014 University of Turku, Finland 
         \and % 13
         Tuorla Observatory, Department of Physics and Astronomy, FI-20014 University of Turku, Finland
        \and
        Carnegie Observatories, 813 Santa Barbara Street Pasadena, CA 91101, USA
        \and % 14
        Astrophysics Research Institute, Liverpool John Moores University, Liverpool L3 5RF,  UK
        \and % 16
        Instituto de Astronom\'ia, Universidad Nacional Aut\'onoma de  M\'exico, A.~P. 70-264, C.P. 04510, M\'exico, Ciudad de México, Mexico.
        \and % 17
        The Oskar Klein Centre, Department of Physics, Stockholm University, Albanova University Center, SE 106 91 Stockholm, Sweden
        \and
        Institute for Astronomy, University of Hawai‘i, 2680 Woodlawn Dr., Honolulu, HI 96822, USA}

   \date{Received 28 March 2023; accepted 15 June 2023}

% \abstract{}{}{}{}{} 
% 5 {} token are mandatory
 
  \abstract
  % context heading (optional)
  % {} leave it empty if necessary  
   {The analysis of core-collapse supernova (CCSN) environments can provide important information on the life cycle of massive stars and constrain the progenitor properties of these powerful explosions. The MUSE instrument at the Very Large Telescope (VLT) enables detailed local environment constraints of the progenitors of large samples of {CCSNe}.  
    Using a homogeneous SN sample from the All-Sky Automated Survey for Supernovae (ASAS-SN) survey, an untargeted and spectroscopically complete transient survey, has enabled us to perform a minimally biased statistical analysis of CCSN environments.}
  % aims heading (mandatory)
   {We analyze 111 galaxies observed by MUSE that hosted 112 CCSNe -- 78 II, nine IIn, seven IIb, four Ic, seven Ib, three Ibn, two Ic-BL, one {ambiguous} Ibc, and one superluminous SN -- detected or discovered by the ASAS-SN survey between 2014 and 2018.
    The majority of the galaxies were observed by the the All-weather MUse Supernova Integral field Nearby Galaxies (AMUSING) survey.  
    Here we analyze the immediate environment around the SN locations and compare the properties between the different CCSN types and their light curves.
    }
  % methods heading (mandatory)
   {We used stellar population synthesis and spectral fitting techniques to derive physical parameters for all \ion{H}{ii} regions detected within each galaxy, including the star formation rate (SFR), H$\alpha$ equivalent width (EW), oxygen abundance, {and extinction.}}
  % results heading (mandatory)
   {We found that stripped-envelope supernovae (SESNe) occur in environments with a higher {median} SFR, H$\alpha$ EW, and oxygen abundances than SNe~II and SNe~IIn/Ibn. {Most of the distributions have no statistically significant differences, except} between oxygen abundance distributions of SESNe and SNe~II, and between H$\alpha$ EW distributions of SESNe and SNe~II.
    The distributions of SNe II and IIn are very similar, indicating that these events explode in similar environments. 
    For the SESNe, SNe Ic have higher median SFRs, H$\alpha$ EWs, and oxygen abundances than SNe Ib.
    SNe IIb have environments with similar SFRs and H$\alpha$ EWs to SNe Ib, and similar oxygen abundances to SNe Ic.
    % While there are some indications that the different CCSN types have different progenitor properties based on their environments, the fact that the differences between the distributions are generally not statistically significant suggests that the progenitors differences in masses, age and metallicity is not large.
    % We note that our results suffer from low-number statistics with respect to SESN types.
     We also show that the postmaximum decline rate, $s$, of SNe~II correlates with the H$\alpha$ EW, and that the luminosity and the $\Delta m_{15}$ parameter of SESNe correlate with the oxygen abundance, H$\alpha$ EW, and SFR at their environments. This suggests a connection between the explosion mechanisms of these events to their environment properties.}
  % conclusions heading (optional), leave it empty if necessary 
   {}


   \maketitle
%
%-------------------------------------------------------------------

\section{Introduction}

Core-collapse supernovae (CCSNe) are luminous explosions associated with the deaths of massive ($\geq 8$ M$_\odot$) stars \citep{1979NuPhA.324..487B, 1986ARA&A..24..205W, 1989ARA&A..27..629A}. The study of these events has a deep impact in many fields of astronomy, as these explosions are responsible for dramatic changes in the evolution of a galaxy, triggering or ending the formation of new stars \citep{1986A&A...154..279M, 2010ApJ...714L.275M}. 
CCSNe display a variety of properties in their light curves (LCs) and spectra that are directly connected to the evolution of their progenitor stars and to explosion properties (for example, explosion energy and $^{56}$Ni mass). Type II SNe (hereafter SNe II) are characterized by broad hydrogen features in their spectra, while SNe Ib and SNe Ic are devoid of hydrogen, with the latter showing little or no helium. 
Some SNe~Ic show broad lines in their spectra, being labeled SNe~Ic-BL \citep{2002ApJ...572L..61M}, and they are well-known for being associated with gamma-ray bursts \citep[GRBs,][]{2006ARA&A..44..507W, 2001ApJ...555..900P}.
SNe IIb are transitional events that are similar to SNe II at very early times but the hydrogen lines vanish during their later evolution \citep[][]{1994AJ....108.2220F}. 
{The lack of hydrogen in SNe IIb, Ib, and Ic reflects the loss of the progenitor star's envelope during the last stages before death, and hereafter are called together stripped-envelope SNe (SESNe). For a review of spectroscopic classification of SNe, readers can refer to \citet{1997ARA&A..35..309F} and \citet{2017hsn..book..195G}.
Some SNe occur inside a dense circumstellar medium (CSM), giving rise to narrow emission lines in their spectra due to the interaction of the ejecta with the dense CSM. If the medium is H-rich, these are called SNe IIn \citep[][]{1990MNRAS.244..269S}, while when the medium is He-rich, they are labeled SNe Ibn \citep[][]{2008MNRAS.389..113P}. Recently, SNe Icn have been discovered, and lack both H and He, but with a C- and O-rich CSM \citep[][]{2021TNSAN..76....1G, 2022ApJ...927..180P}.

SNe II are thought to arise from single red supergiants (RSGs) with a mass range between $\sim 8 - 20$~M$_\odot$ \citep{2015PASA...32...16S}.
Stellar models indicate that stars with zero-age main sequence (ZAMS) masses larger than $8 - 9 \ \textrm{M}_\odot$ develop an Fe core at the end of their evolution. Since Fe nuclei have strong binding energies, the process of nuclear fusion can no longer occur in the core of these stars, leading to gravitational collapse \citep[See reviews by][]{1990RvMP...62..801B, 2005ARNPS..55..467M, 2021Natur.589...29B}. 
This scenario is supported by several direct observations of progenitor stars.
\citet{2009Sci...324..486M} found a RSG to be the progenitor of SN 2003gd and \citet{2015MNRAS.447.3207M} showed that the progenitor of SN 2008cn was consistent with a RSG with a mass of $<16$~M$_\odot$.
\citet{2014MNRAS.438..938M} analyzed five SN II progenitors and found them to be consistent with RSGs with masses between $8.4-12.0$~M$_\odot$. \citet{2012AJ....143...19V} found the progenitor of SN 2008bk to be a RSG with $\sim 8$~M$_\odot$, and \citet{2016MNRAS.456L..16F} showed that the disappeared progenitor of SN 2012aw was a RSG with $\sim 12.5$~M$_\odot$ \citep[also see the review by ][]{2015PASA...32...16S}.

{SESNe have only a few progenitors detected through direct imaging.
For SNe~IIb, some of their observed progenitors are consistent with yellow supergiant (YSG) or blue supergiant (BSG) stars with masses between $\sim 13$ to $19$M$_{\odot}$ \citep[e.g., ][]{2011ApJ...739L..37M, 2014AJ....147...37V, 2015ApJ...811..147F}.} \citet{2009Sci...324..486M} found a RSG to be the progenitor of SN 1993J and interacting binary systems have also been proposed for some of the observed progenitors of these events \citep[e.g., ][]{2014ApJ...790...17F}.
{The type of stars that become SNe Ib and Ic is also not clear.} The mechanism needed to explain the shedding of the envelopes of hydrogen and/or helium before the core-collapse can be due to two distinct pathways: they might be lost through strong winds within a Wolf-Rayet (WR) or similar star, or they might lose mass through binary interactions \citep[][]{2010ApJ...725..940Y}. 
While the former scenario requires more massive progenitors -- a 20 to 40~M$_\odot$ star for SNe Ib, and a $\gtrsim 40$~M$_\odot$ star for SNe Ic, at solar metallicity, following the rotating single-star model from \citet{2009A&A...502..611G} -- the latter could arise from relatively lower mass progenitors, similar to normal SNe II. Only a few candidate companion stars have been detected in preexplosion images of SESNe.
% \citet{2014AJ....148...68B} showed that the progenitor of the SN Ib iPTF13bvn was consistent with a $\sim 8$~M$_\odot$ star in an interacting binary system, while
{\citet{2021MNRAS.504.2073K} found the progenitor of the SN Ib 2019yvr to be consistent with a single luminous star, although the nature of the preexplosion counterpart of SN 2019yvr is still under debate \citep[see e.g.,][]{2022MNRAS.510.3701S}, and recently \citet[][]{2022ApJ...929L..15F} have shown the candidate companion of the SN Ibc 2013ge to be a main-sequence B supergiant star.}

Given the lack of a significant sample of direct detections of progenitor stars and binary companions of SESNe, indirect methods such as the analysis of the environment around these events can aid in constraining their properties \cite[for a historical review of SN environment studies, see][]{2015PASA...32...19A}.
Pixel statistics of H$\alpha$ emission in SN host galaxies showed that SNe Ic traced H$\alpha$ emission, while SNe II and Ib do not show a strong correlation with these regions \citep[][]{2012MNRAS.424.1372A}, implying that SNe~Ic arise from more massive stars.
By analyzing the host galaxies of SESNe, \citet[][]{2010ApJ...721..777A} and \citet[][]{2011ApJ...731L...4M} found that SNe Ic happen in environments with higher metallicities than SNe Ib and IIb. 
\citet[][]{2010MNRAS.407.2660A}, \citet[][]{2011A&A...530A..95L}, and \citet[][]{2012ApJ...758..132S}, however, did not find any significant difference in the environment metallicity between SNe Ic, Ib, and IIb \citep[also see][for recent analyses]{2021ApJ...923...86C,2023ApJ...944..110M}.

SNe IIn are expected to arise from progenitor stars with a considerably larger mass than the progenitors of normal SNe II, as the observed amount of CSM observed in these events require extreme episodes of mass-loss in the pre-SN stages \citep[e.g.,][]{2014ARA&A..52..487S}.
A possible class of progenitor stars for SNe IIn are luminous blue variables (LBVs); {massive stars that are characterized by very strong and episodic mass-loss}. This scenario is consistent with the progenitor detections of SNe IIn and the LBV-like variability observed in some preexplosion observations \citep[][]{1993ApJ...416..167R, 2009Natur.458..865G, 2013MNRAS.430.1801M, 2013ApJ...767....1P, 2016MNRAS.463.3894E}. However, \citet[][]{2012MNRAS.424.1372A}, \citet{2014MNRAS.441.2230H}, and \citet[][]{2022MNRAS.513.3564R} showed that SNe~IIn have very similar degree of association to host galaxy H$\alpha$ emission to SNe II, and \citet[][]{2013A&A...555A..10T}  demonstrated that they also happen in environments with very similar metallicity to SNe II. By analyzing the locations of LBV stars in the Large Magellanic Cloud (LMC) and M33 galaxies, \citet[][]{2017A&A...597A..92K} found no correlation between the spatial distribution of these massive stars and SNe IIn, favoring less massive progenitors. In fact, a RSG star is thought to be the progenitor of the SN IIn 1998S, indicating that these stars could generate at least a fraction of these transients \citep[][]{2012MNRAS.424.2659M, 2015A&A...580A.131T}. However, \citet[][]{2019MNRAS.489.4378S} also argue that LBV stars are usually isolated and are not found in the most star-forming regions of galaxies, clouding the interpretation of SNe~IIn environment studies. {A similar scenario of distinct progenitor stars is also possible for SNe~Ibn. Although massive H-poor WR stars were the first proposed progenitor for this class of transients \citep{2007Natur.447..829P}, recent environment studies found a connection to older stellar populations and to a diversity of potential progenitors \citep{2019ApJ...871L...9H, 2023ApJ...946...30B}.}

Recently, many SN environment analyses have used the capabilities of Integral Field Units (IFUs) to perform spatially resolved spectroscopy over a large field-of-view, allowing for the study of individual \ion{H}{ii}  regions within host galaxies and for a more precise study of the host galaxy as a whole. 
Using the CALIFA survey, \citet{2014A&A...572A..38G} showed that there is no significant difference in the metallicities of the environments of SNe II, Ib, and Ic. In addition, the star formation rate (SFR) and the H$\alpha$ equivalent width (EW) also showed no significant difference between the CCSN types.
The remarkable capabilities of the Multi-unit Spectroscopic Explorer \citep[MUSE, ][]{2014Msngr.157...13B} instrument at the Very Large Telescope (VLT) in the study of SN environments were demonstrated by \citet{2016MNRAS.455.4087G}. They analyzed six nearby galaxies that hosted 11 SNe, and presented a new method for comparing \ion{H}{ii}  region properties within SN host galaxies, which was made possible by the high angular resolution of the instrument. 
A larger analysis combining MUSE and VIMOS data \citep[][]{2003SPIE.4841.1670L} was made by \citet[][]{2018A&A...613A..35K}, who analyzed 83 nearby SN explosion sites. They found no significant difference between the metallicity of the different CCSN environments, but showed that SNe Ic are associated with the youngest stellar populations, followed by SNe Ib, IIb, and II.

Most of the SN environment studies used SNe compiled from various surveys, many of which were of a galaxy-targeted nature, including pointed and panoramic surveys, as well as amateur discoveries, with little or no control for biases and systematics. Such surveys are biased toward SNe discovered in bigger and therefore more metal rich galaxies, a factor that might affect the final result. 
Heterogeneous samples might also have classification and publication biases \citep[see e.g.,][for a discussion on the selection biases on their analysis]{2016ApJ...830...13P}.
Sample selection has a large effect in the analysis of SN environments, as demonstrated by, for example,
\citet{2012ApJ...758..132S}, \citet{2016A&A...591A..48G}, and \citet{2018ApJ...855..107G}. 
\citet[][]{2012ApJ...758..132S} showed that, when combining different spectroscopic measurements from the literature, SNe Ib, Ic, and IIb appear to have very similar values of metallicity, but significant differences are seen if SNe from untargeted samples are used. 
\citet[][]{2018ApJ...855..107G} used the PISCO SN host compilation of 232 galaxies and found similar results to \citet{2014A&A...572A..38G}, but also demonstrated that significant differences in the environments of the different CCSN types were found when applying an untargeted selection of host galaxies (also see \citeauthor{2021MNRAS.503.3931T} \citeyear{2021MNRAS.503.3931T} and \citeauthor{2021ApJS..255...29S} \citeyear{2021ApJS..255...29S} for discussions on integrated properties of SN host galaxies drawn from homogeneous samples). 

In our new sample, we use a homogeneous SN sample from the All-Sky Automated Survey for Supernovae (ASAS-SN) survey, an untargeted and spectroscopically complete survey of transients, which allows us to perform a minimally biased statistical analysis of their environments. The importance of untargeted surveys can already be seen in the ASAS-SN study of SNe~Ia rates as a function of host galaxy properties \citep{2019MNRAS.484.3785B}, which found no dependence of SN~Ia occurrence to star formation activity in the galaxies. 

In the current work (Paper I), we analyze the local environment of 112 ASAS-SN CCSNe, exploring the physical properties of {\ion{H}{ii} regions centered at the SN positions.} We also look for correlations between their environments and LC properties, when photometry is available. In Paper II we will fully exploit the IFU nature of our data, and compare the local environment of each SN to all the \ion{H}{ii}  regions in their host galaxies. 
We present the sample selection and data reduction in Section \ref{sec:sample}. In Section \ref{sec:ana}, we discuss the different analysis methods used to study the physical parameters of our sample, and in Section \ref{sec:res} we present our results. A comparison of our results to previous studies and the implications for our understanding of CCSN progenitors is given in Section \ref{sec:dis}. Our conclusions and final remarks are presented in Section \ref{sec:conc}.

\section{Sample selection and data reduction} \label{sec:sample}

The All-weather MUse Supernova Integral field Nearby Galaxies (AMUSING) survey is a long-term ``filler'' programme that uses MUSE in suboptimal atmospheric conditions to obtain integral field observations of {galaxies that hosted SNe} \citep[][Galbany et al. in prep.]{2016MNRAS.455.4087G}.
With a spatial sampling of $0.2 \times 0.2 \ \textrm{arcsec}$, a field of view of $1\ \textrm{arcmin}^{2}$ (in Wide Field Mode), and a mean spectroscopic resolution of $R\sim 3000$ with wide wavelength coverage ($480-930$~nm), MUSE is one of the most advanced Integral Field Units (IFUs) in operation. 

Each semester AMUSING has focused on a different specific science case. In semesters P96 {(2015 October 1 to 2016 March 31)} and P103 {(2019 April 1 to 2019 September 30)}, AMUSING observed galaxies that hosted SNe discovered or recovered by the ASAS-SN survey, with the aim of constraining differences in SN progenitor properties through studying their parent stellar populations, together with investigating the rates of SNe with respect to environmental properties.
{ASAS-SN started operation in 2011} (with transient alerts beginning in 2013), with the goal of monitoring the entire sky with rapid cadence in search of nearby and bright transients \citep[see e.g.,][for technical details on the survey]{2014ApJ...788...48S, 2017PASP..129j4502K}. Given the relatively shallow depth of the survey, the SNe are all relatively bright, and spectral classifications are possible in nearly all cases. 
This makes the survey almost spectroscopically complete, in the sense that almost all the discovered transients have an associated spectrum and classification: as of 2019, 97 per cent of ASAS-SN discoveries were spectroscopic classified \citep[][]{2019MNRAS.484.1899H}. 
ASAS-SN is also an all-sky survey that does not target particular galaxies. As evidenced in \citet[][]{2019MNRAS.484.1899H}, one quarter of the SN host galaxies had no previous spectroscopic redshift estimates, and some SN hosts were not described in any existing galaxy catalog. This allows the ASAS-SN survey to serve as an ideal tool for the study of nearby SN populations, as it is not biased toward brighter and more extended galaxies. In addition, ASAS-SN discovered more SNe in the central parts of galaxies compared with previous professional and amateur surveys, again lowering observational biases \citep[][]{2017MNRAS.464.2672H, 2017MNRAS.467.1098H, 2017MNRAS.471.4966H, 2019MNRAS.484.1899H, 2022arXiv221006492N}.

Between 2013 and 2018, ASAS-SN detected a total of 449 transients that were spectroscopically confirmed as CCSNe \citep[][]{2017MNRAS.464.2672H, 2017MNRAS.467.1098H, 2017MNRAS.471.4966H, 2019MNRAS.484.1899H, 2022arXiv221006492N}.
{The events selected to be observed by AMUSING had a heliocentric redshift of $\leq 0.02$ and an observability constraint from the Paranal observatory (DEC $\leq 25^{\circ}$).
For the 2014 to 2017 events, only SNe with an apparent magnitude at peak luminosity of $V \leq 17.0$~mag were selected \citep[following ][]{2017MNRAS.464.2672H, 2017MNRAS.467.1098H, 2017MNRAS.471.4966H, 2019MNRAS.484.1899H}, while for the 2018 events a cut on $V \leq 18.0$~mag was applied \citep[following ][]{2022arXiv221006492N}}.
{For the period P96, AMUSING proposed observations of 13 SNe~II, four SNe IIn, and one SN~Ibn. From those, eight SNe were observed (six II, one Ibn, and one IIn). 
For the period P103, AMUSING proposed observations of 52 SNe~II, 18 SESNe, and nine interacting SNe. From those, 71 SNe were observed (47 II, seven IIb, seven IIn, five Ib, four Ic, and one Ibn).}
{AMUSING obtains observations with different observing conditions, with variable  sky transparency and seeing. The data presented here were obtained with a median seeing of $\sim 0.9\arcsec$.}
  An additional 39 galaxies that hosted ASAS-SN SNe (29 II, five Ic, two IIb, one IIn, one Ibn, and one SLSN) between 2014 and 2018 were obtained from the ESO Archive Science Portal\footnote{\url{http://archive.eso.org/scienceportal/home}}.
  {While using archival data could lead to biases in the final sample selection, we recalculated our results with a reduced sample only containing those SN hosts observed through AMUSING. No difference in the results were seen and thus we choose to keep the archival data in our sample.}

The MUSE datacubes were reduced using the ESO reduction pipeline\footnote{\url{http://www.eso.org/sci/software/pipelines/}} \citep[version 1.2.1, ][]{2014ASPC..485..451W}. The pipeline applies corrections for the bias, flat-field, geometric distortions and illumination, and performs wavelength calibration and sky subtraction. All of the data obtained in P96 were reduced by the AMUSING collaboration, while data from P103 were directly obtained from the ESO Archive Science Portal {(for P103, observations were processed by ESO and uploaded in semi-real time at the Science Portal)}. 
Further astrometric corrections were applied in the reduced data: the cubes were registered to the Gaia DR2 \citep[][]{2018A&A...616A...1G} source catalog via VizieR Queries\footnote{\url{https://astroquery.readthedocs.io/en/latest/vizier/vizier.html}} \citep{2019AJ....157...98G} and compressed in the wavelength direction to create an image and Gaussians are fit at a 10-pixel box centered at the position of all cataloged sources within the datacube field of view (FoV). Finally, the average astrometric shift is calculated and the WCS is updated.
Additional flux calibration correction was achieved by producing synthetic \textit{r-} and \textit{i-} band images from the MUSE datacube, and comparing aperture photometry performed on the synthetic images to photometry using the same aperture from PanSTARRS \citep[][]{2002SPIE.4836..154K, 2016arXiv161205560C, 2020ApJS..251....7F}, SDSS \citep[][]{2000AJ....120.1579Y}, or DES \citep[][]{2005IJMPA..20.3121F, 2005astro.ph.10346T} \textit{r} and \textit{i} images. The MUSE datacubes were then scaled to the fluxes to the survey values (Galbany et al. in prep.).

From our initial sample of 117 CCSN hosted by 116 galaxies (with the galaxy NGC 5962 hosting two SNe: SN 2016afa and SN 2017ivu), in three SN host galaxies (NGC 1566, NGC 6907, and ESO 560-G013) the region where the SN was located is affected by contamination due to an ionizing source other than star formation, and is therefore not useful for our analysis (see section \ref{sec:IFUanal} for details of our emission-line extraction and analysis techniques). 
This is determined by a Baldwin, Phillips, and Terlevich (BPT)-diagram \citep[][]{1981PASP...93....5B} test (see Figure \ref{fig:BPT} and a description of the BPT analysis of our SN host galaxy \ion{H}{ii}  regions is given in Appendix \ref{app:BPT}), where these regions fall above the \citet[][]{2003MNRAS.346.1055K} line.
The galaxy NGC 3256 was also excluded from the final sample, as it contains galactic-scale outflows that could affect the interpretation of the physical properties of the environment close to the SN position \citep[][]{2020AJ....159..167L}.
Finally, the host galaxy of ASASSN-15lv (GALEXASC J015900.57-322225.2) was excluded, as the signal-to-noise ratio (SNR) was low for any meaningful analysis (see Section \ref{sec:IFUanal} for a discussion on the SNR threshold on our analysis).
After these cuts we are left with 112 SNe in 111 galaxies.
We did not exclude any galaxy based on its inclination and thus the sample might include highly inclined galaxies, which could include extra uncertainties to the SN environment analyses due to extinction and line of sight confusion.

Table \ref{tab:sn_prop} shows the general properties of the 112 CCSNe and their host galaxies. 
The SN names, types, coordinates and host galaxy names were obtained from \citet[][]{2017MNRAS.464.2672H, 2017MNRAS.467.1098H, 2017MNRAS.471.4966H, 2019MNRAS.484.1899H}, and \citet[][]{2022arXiv221006492N}.
The galaxy coordinates and Galactic extinctions were taken from the NASA-IPAC Extragalactic Database (NED)\footnote{\url{https://ned.ipac.caltech.edu}}, while the host $B$ band absolute magnitudes were taken from HyperLeda\footnote{\url{http://leda.univ-lyon1.fr/}} \citep[][]{2014A&A...570A..13M}. The host redshifts were estimated using a spectrum from a $1\arcsec$ diameter aperture centered at the galaxy core, extracted from each datacube, using the peaks of the main emission lines (Balmer H$\alpha$ and H$\beta$, and [OIII]$\lambda5007$) or the minimum of the H$\alpha$ and H$\beta$ absorption features, {and are reported in Table \ref{tab:sn_prop}}. The initial redshifts were converted to heliocentric redshifts by using Astropy SkyCoord\footnote{\url{https://docs.astropy.org/en/stable/api/astropy.coordinates.SkyCoord.html}} and to CMB frame redshifts using the dipole coordinates from \citet[][]{1997mba..conf...69L}.
Table \ref{tab:HII_dist} reports the redshift estimated in the cosmic microwave background frame ($z_{cmb}$) and the derived luminosity distance (D$_L$). 
Since our sample comprises nearby galaxies, the D$_L$ values were estimated using a cosmic flow model for the Local Universe \citep[][]{2015MNRAS.450..317C}\footnote{\url{https://cosmicflows.iap.fr/}}, assuming a flat curvature, H$_0 = 70 \ \textrm{km} \ \textrm{s}^{-1} \ \textrm{Mpc}^{-1}$ and $\Omega_M = 0.3$.

Figure \ref{fig:pie_chart} shows the distribution of the different CCSN types in our sample. There are {78 II}, nine IIn, seven IIb, seven Ib, four Ic, three Ibn, two Ic-BL, {one {ambiguous} Ibc \citep[SN 2017gax has an ambiguous classification, as reported in][]{2017TNSCR.870....1J}}, and one SLSN. SNe II dominate the sample, comprising $69 \%$ of all the SNe. 
{In the magnitude-limited sample of \citet[][]{2011MNRAS.412.1441L}, the ratio between SNe II and SESNe (including SNe~IIb) was estimated to be $\sim 2.0$ . The same ratio in our sample is of $3.4 \pm 1.9$, including the seven SNe~IIb, seven Ib, four Ic, one {ambiguous} Ibc, and two Ic-BL. This ratio is estimated using the median of a binomial distribution, with the uncertainty given by the 90\% confidence range. The ratio of $0.4 \pm 0.2$ between SNe Ic and Ib is lower than the estimated by \citet[][]{2011MNRAS.412.1441L} ($\sim 1.6$). 
On the other hand, the ratios of SNe II to SNe IIn ($8.8 \pm 1.7$) and IIb ($8.7 \pm 1.4$) are higher than \citet[][]{2011MNRAS.412.1441L} ($\sim 1.9$ and $\sim 3.2$, respectively), indicating an underrepresentation of the latter types. We also note that low number statistics may dominate the relative numbers of all SNe other than SNe~II. }

In Figure \ref{fig:sample} we show a comparison of 77 SNe II and their host galaxy properties in our sample to 116
SNe II from \citet[A14, ][]{2014ApJ...786...67A} and \citet[G17, ][]{2017ApJ...850...89G}. The latter samples are heterogeneous and comprise SNe detected and followed by a number of different surveys between 1986 and 2009.
We also show a comparison with 19 SNe II to be within low-luminosity galaxies from \citet[G18, ][]{2018MNRAS.479.3232G}.
The top and middle panel of Figure \ref{fig:sample} show, respectively, the distribution of peak apparent and $V-$band absolute magnitudes of the SNe.
One can see that the ASAS-SN CCSN sample is shifted to brighter magnitudes, with a median $V-$band apparent magnitude of $16.2$~mag, as compared to the $17.3$~mag for A14/G17 and $16.5$~mag for G18.
This could be due to ASAS-SN being a magnitude-limited survey, with a limit of detection of $V \sim 17.5$, thus discovering transients with brighter apparent magnitudes (Desai et al. in prep). Another possibility is the intrinsic differences in the SNe due to the observation selection effect.
The middle panel shows that the ASAS-SN events are shifted to intrinsically brighter values, with the median absolute magnitude of the ASAS-SN being $\sim -17.6$~mag, while the A14/G17 and G18 samples show a median $\sim -16.7$~mag. 
Because ASAS-SN detects brighter events, making a redshift cut on the sample decreases the bias toward brightness and makes it almost complete in a given volume. 
The bottom panel of Figure \ref{fig:sample} shows the distribution of 60 SNe II host absolute magnitudes in $B$ band (M$_B$) and CMB redshifts ($z_{cmb}$) in our sample. 
{The redshifts span between $\sim 0.002$ and $\sim 0.025$, with a median at $\sim 0.012$.}
The ASAS-SN sample is concentrated at relatively nearby host galaxies, with a median redshift slightly smaller than the A14/G17 galaxies (median $\sim 0.015$), but slightly larger than the G18 galaxies (median $\sim 0.011$). 
The most important difference of the ASAS-SN CCSNe to heterogeneous and targeted samples is evident also in the bottom panel of Figure \ref{fig:sample}: as a heterogeneous sample, the A14/G17 SNe are biased toward brighter galaxies and do not cover the lower end of the host M$_B$ values (with a median absolute magnitude at $\sim -20.6$~mag). As a targeted sample, on the other hand, the G18 sample is biased toward fainter galaxies (with a median absolute magnitude at $\sim -17.7$~mag). The ASAS-SN host galaxies show a more spread distribution of M$_B$, with values between $\sim -16.5$ and $\sim -22.0$~mag, and a median at $\sim -19.6$~mag. 
This shows the completeness of our sample in terms of galaxy brightness and size, which highlights the unbiased nature of our analysis of the CCSN environments.

\begin{table*}
\caption{General properties of the SNe and their host galaxies. \label{tab:sn_prop}}     
\centering          
\begin{tabular}{lccclccc}  
\hline       
Name & Type$^a$ & RA$^b$& DEC & Host Galaxy Name & $z_{host}$ & $A_{v,MW} \ ^c$ & $M_{B,host} \ ^d$\\ 
\hline                    
SN2018eog & II & 20 28 11.970 & -03 08 13.47 & 2MASS J20281135-0308096 & 0.01886 & - & - \\
SN2018ant & II & 08 36 31.450 & -11 49 40.73 & MCG-2-22-22 & 0.019784 & 0.192 & -20.67 \\
ASASSN-18oa & II & 01 30 27.120 & -26 47 05.98 & ESO476-G16 & 0.019744 & 0.055 & -20.94 \\
SN2018bl & II & 08 24 11.590 & -77 47 16.55 & ESO18-G9 & 0.017773 & 0.357 & -20.94 \\
SN2018evy & IIn & 18 22 38.170 & +15 41 47.66 & NGC6627 & 0.017647 & 0.611 & -20.94 \\
SN2018cho & IIn & 00 13 26.515 & +17 29 19.57 & IC4 & 0.016558 & 0.173 & -20.72 \\
SN2018ie & II & 10 54 01.060 & -16 01 21.40 & NGC3456 & 0.013954 & 0.211 & -20.76 \\
... \\
\hline                  
\end{tabular}
\tablefoot{The entire table is available online. $^a$ All the supernova types, coordinates, and host galaxy names are taken from \citet[][]{2017MNRAS.464.2672H, 2017MNRAS.467.1098H, 2017MNRAS.471.4966H, 2019MNRAS.484.1899H}, and \citet{2022arXiv221006492N}; $^b$ RA and DEC are given in the J2000 epoch; $^c$ The Galactic extinction, $A_{v,MW}$, was taken from
the NASA-IPAC Extragalactic Database (NED); $^d$ The host galaxy absolute magnitudes in $B$ band, $M_B$, were taken from HyperLeda.}
\end{table*}

\begin{table*}
\caption{Distance properties of the \ion{H}{ii} regions related to each SN. \label{tab:HII_dist}}
\centering
\begin{tabular}{lccccc}
\hline
Name & z$_{cmb}$ & D$_{L}$ (Mpc) \ $^a$ & d$_{proj}$ (pc) \ $^b$ & r$_{\ion{H}{ii} }$ (pc) \ $^c$ & Area$_{\ion{H}{ii} }$ (kpc$^2$)\\ 
\hline
SN2018eog & 0.0180 & 79.1 & 0.0 & 460.2 & 5.99 \\
SN2018ant & 0.0207 & 91.9 & 0.0 & 534.6 & 3.29 \\
ASASSN-18oa & 0.0190 & 86.8 & 0.0 & 504.8 & 2.94 \\
SN2018bl & 0.0181 & 80.2 & 0.0 & 466.3 & 0.68 \\
SN2018evy & 0.0173 & 77.1 & 0.0 & 448.6 & 1.90 \\
SN2018cho & 0.0155 & 68.1 & 0.0 & 396.3 & 0.49 \\
SN2018ie & 0.0151 & 67.7 & 0.0 & 393.6 & 0.49 \\
... \\
\hline
\end{tabular}
\tablefoot{The entire table is available online. $^a$ The luminosity distance, D$_L$ was estimated using the CMB redshift, $z_{cmb}$, and a cosmic flow model for the Local Universe. 
 $^b$ d$_{proj}$ is the projected distance between the SN position and their related \ion{H}{ii} region. d$_{proj} = 0$ means that the \ion{H}{ii} region is centered at the SN position.
 $^c$ The \ion{H}{ii}  region radius, r$_{\ion{H}{ii} }$, and area, Area$_{\ion{H}{ii} }$, are estimated from the projected size in pixels of the identified regions.}
\end{table*}

\begin{figure}[t]
\centering
\includegraphics[scale=0.61]{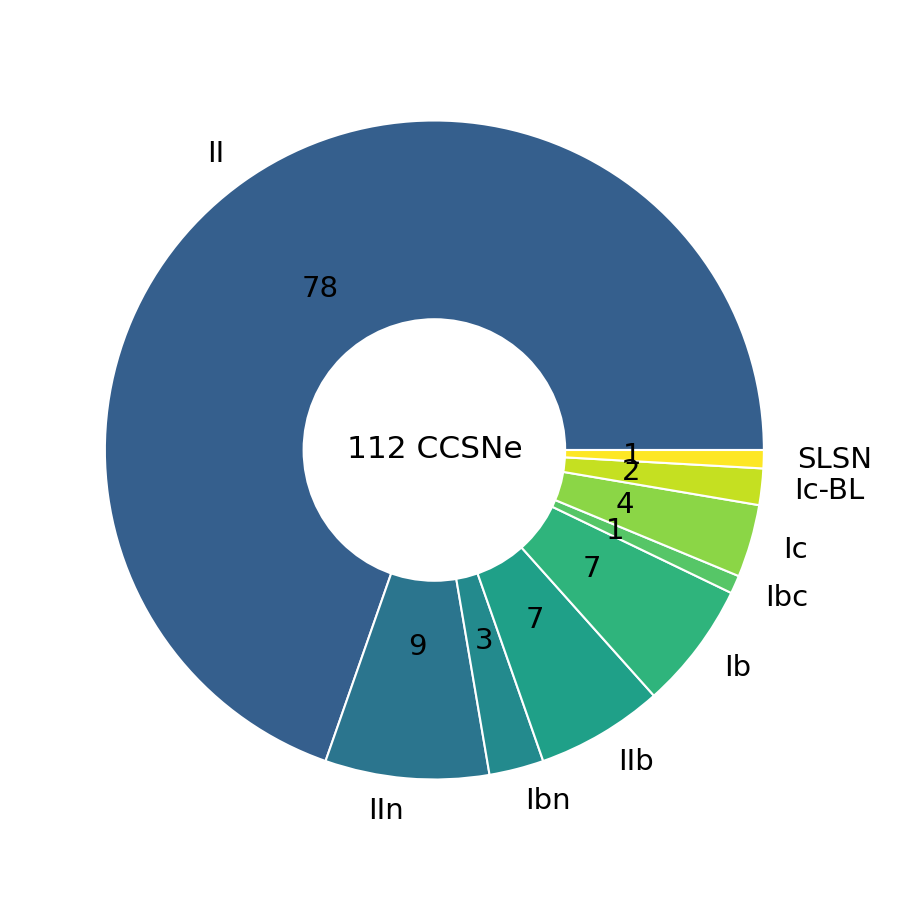}
\caption{Distribution of the different SN types for the 112 events in our sample.  \label{fig:pie_chart}} 
\end{figure}

\begin{figure}[t!]
\centering
\includegraphics[width=0.41\textwidth]{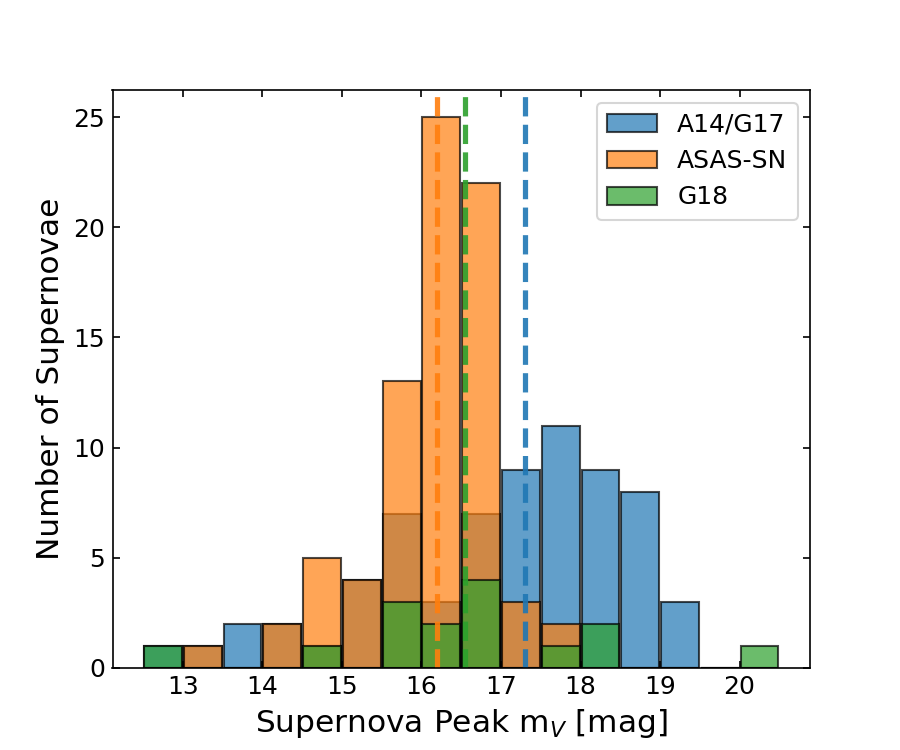}
\includegraphics[width=0.41\textwidth]{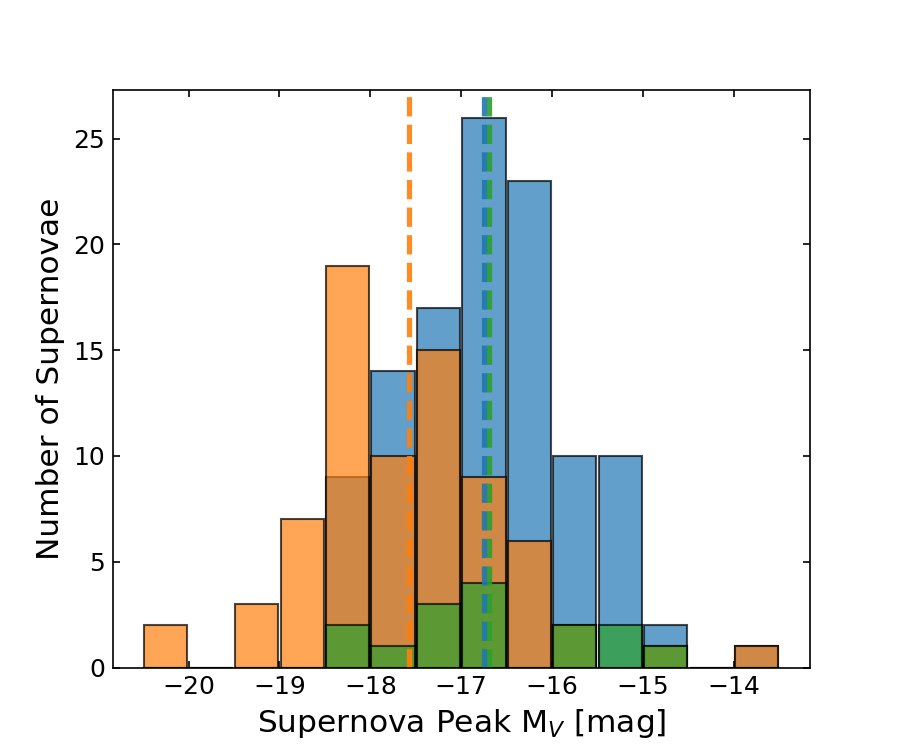}
\includegraphics[width=0.41\textwidth]{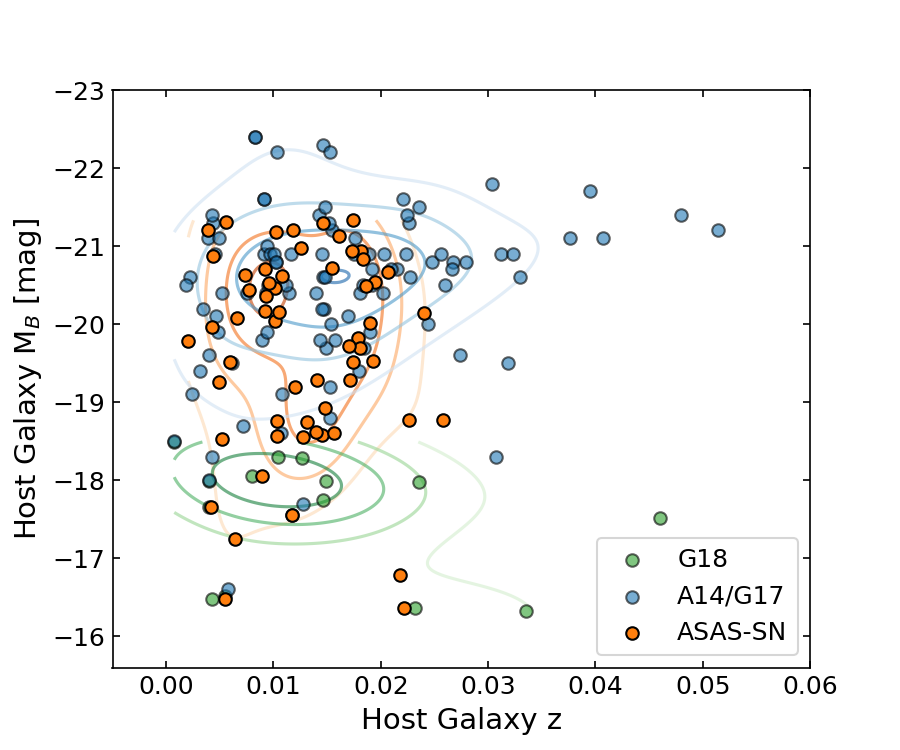}
\caption{Comparison of the properties of the SNe II and their hosts in our sample to the SNe II presented in A14/G17 and G18. 
The top and middle panels show, respectively, the distribution of the apparent (m$_V$) and absolute (M$_V$) magnitudes in $V$ band at peak brightness for the transients. 
Apparent magnitudes for the A14/G17 sample are estimated using values of M$_V$, distance modulus, host and Milky Way extinction provided by the authors, while for the G18 sample apparent magnitudes are estimated using the M$_V$ and Milky Way extinction values provided in the paper. The bottom panel shows the distribution of the SN host galaxy redshift as a function of the host absolute magnitude in $B$ band, M$_B$. For A14/G17 and G18, the host redshift is estimated using the recession velocity, while for the ASAS-SN sample we use the host z$_{CMB}$ reported in Table \ref{tab:HII_dist}.}
\label{fig:sample}
\end{figure}

\section{Analysis} \label{sec:ana}

\subsection{\ion{H}{ii} region segmentation} \label{sec:IFUanal}

CCSNe are known to be generated by relatively young and massive stars, making the delay time between the progenitor formation and its explosion short enough for the star to be close to its birth site \citep[e.g., ][]{2017A&A...601A..29Z}. 
This assumption makes the closest \ion{H}{ii} region a useful probe of the physical properties of their progenitors. 
We therefore analyze the {\ion{H}{ii} region at the SN location} and use it as a proxy for the study of the direct environment around the explosion. 
In order to identify the distinct \ion{H}{ii} regions in each galaxy and to improve their SNR, we use the segmentation method of these regions \citep[as used in, e.g.,][]{2016MNRAS.455.4087G, 2018MNRAS.473.1359L}, which consists in combining the observed signal of adjacent spaxels into one region.
We use \textsc{IFUanal}\footnote{\url{https://ifuanal.readthedocs.io/en/latest/index.html}} \citep{2018MNRAS.473.1359L} through the ``nearest bin'' method to perform \ion{H}{ii}  region segmentation in the datacubes and to obtain emission line and stellar continuum fittings for the spectra in each region. 
\textsc{IFUanal} is optimized to perform \ion{H}{ii}  region segmentation on MUSE datacubes. We run the code on the cubes corrected by the host redshift and Galactic extinction. The Galactic reddening correction was made using the values from \citet[][]{2011ApJ...737..103S} for the Milky Way toward the host galaxy. 

Figure \ref{fig:NGC3456} shows the MUSE RGB image composition of the galaxy NGC 3456, located at $\sim 62.9$~Mpc. The galaxy hosted SN Ic 2018ie, whose position is marked by the yellow star.
The top panel in the right column of Figure \ref{fig:NGC3456} shows the section of the galaxy covered by the MUSE observation. 
The middle panel in the right column of Figure \ref{fig:NGC3456} shows the H$\alpha$ map of the galaxy, and the bottom panel shows the resulting oxygen abundance in the D16 index for the segmented  \ion{H}{ii} regions in the galaxy.

\textsc{IFUanal} segments the \ion{H}{ii} regions in the following steps: First, {an H$\alpha$ map is produced by simulating a $30 \AA$ filter centered on the H$\alpha$ emission line, subtracting the continuum level and applying a Gaussian filter to avoid peaks in the noise.} 
The code uses three main parameters to define the \ion{H}{ii} regions based on the H$\alpha$ emission map: a minimum flux threshold to define seed peaks from which the region is allowed to expand from; a maximum radius in pixels for the region size; and a maximum flux threshold. {All the pixels around the brightest peak will therefore be included in the \ion{H}{ii} region up to the maximum flux or the maximum predefined distance radius. 
We use a minimum threshold of three times the standard deviation of the H$\alpha$ flux in a  background region of the datacube, and a maximum flux of 10 times the standard deviation of the H$\alpha$ flux.} We find that these parameters generate the best region segmentation and, 
 after a series of tests with different parameters,  varying the parameters around these values do not result in any significant difference in the final results (see Appendix \ref{app:param_tests}).

Since the galaxies in our sample have a large range of distances ($\sim 10 \-- 167$ Mpc), we separate them into two groups to define the maximum radius of the segmented  bins: for galaxies closer to $30$~Mpc we use a maximum bin radius of $10$~px, and for more distant galaxies we use a maximum bin radius of $6$~px. 
This leads to a median bin radius size $\sim 336$~pc, which is similar in size to observed \ion{H}{ii} regions \citep[see e.g.,][]{2013MNRAS.428.1927C}. One caveat is that the regions are circular of very similar sizes, while real \ion{H}{ii} regions have different shapes and sizes.

The spectra are averaged using an arithmetic mean, producing one spectrum for each region. \textsc{STARLIGHT} is used to perform stellar continuum fitting  \citep[][]{2005MNRAS.358..363C, 2006MNRAS.370..721M}, {with simple stellar population synthesis models from \citet{2003MNRAS.344.1000B} for the MILES spectral library \citep{2006MNRAS.371..703S} and using an initial mass function (IMF) from \citet{2003PASP..115..763C} with a mass range $0.1 - 100$~M$_\odot$. The components of the base models have ages from 1~Myr to 13~Gyr for four metallicities, $Z = 0.0004, 0.008, 0.02, 0.05$ \citep{2018MNRAS.473.1359L}.}
The emission lines of the continuum-subtracted spectra were fit using multiple Gaussians. Only the emission lines with a SNR\footnote{The SNR is given by the ratio of the line flux estimated from a Gaussian fit and the total flux uncertainty, which is given by a combination in quadrature of the statistical and photon noise uncertainty.} of $> 3$ are fit, so their fluxes, EWs and uncertainties can be properly estimated. The uncertainty in each parameter is estimated through the propagation of uncertainties in the fittings of the individual emission lines.

For the analysis presented in this work, {we selected \ion{H}{ii} regions centered at the SN positions, with most of the regions selected (90 regions) being exactly at the SN coordinates. Although an \ion{H}{ii} region is spatially coincident with a SN, there is a possibility of chance alignment due to the lack of three-dimensional information on the line of sight \citep[see, e.g.,][]{2021MNRAS.504.2253S,2023MNRAS.521.2860S}.
However, it is important to note that our analysis (and in general these types of environment analyses) are statistical in nature. While the environment properties of any given SN may not be directly related to the event's progenitor properties, for a large sample of events the overall distribution of environment properties can be used to constrain the progenitors of that SN type.} 

For 27 of the SN locations, the signal was still dominated by the transient, with broad lines distinct from the narrow emission features expected for the underlying \ion{H}{ii} region. Table \ref{tab:phase} reports the SN names and their phase at the observation date relative to peak luminosity, given in days. 
We masked the region around the SN for these cases, until only narrow emission lines of the underlying \ion{H}{ii} region can be detected. The spectra for these cases are extracted from the annulus resulting from this masking.  

If the fit of the extracted spectrum at the SN position has a SNR of $> 3$, we extract the H$\alpha$ flux and EW, oxygen abundances, and extinction (see Section \ref{sec:phys_param} for a description of the physical parameters used in this work). 
If the SNR is $< 3$, we fit the H$\alpha$ emission profile at the SN position and use the spectrum of the nearest \ion{H}{ii} region up to a distance of 1 kpc (this was necessary for 11 regions) to derive the {oxygen abundance}. A total of 13 SNe have a \ion{H}{ii} region at a distance larger than 1 kpc. In these cases, we extract the spectrum at the SN position and fit the H$\alpha$ emission profile. If the SN is hosted by a dwarf galaxy, we use the oxygen abundance of the nearest \ion{H}{ii} region, as dwarf galaxies usually have homogeneous metallicity distributions \citep[][]{2022A&A...665A..92T}. If the SN is hosted by a spiral galaxy, we use the observed oxygen abundance gradient to extract the value at the SN position, as described in Appendix \ref{app:met_gradients}. One exception is ASASSN-14jb. Although it has a projected distance of 1.4 kpc from the nearest \ion{H}{ii} region, its explosion site was analyzed in detail by \citet[][]{2019A&A...629A..57M}, who showed that the SN was consistent with a low metallicity environment similar to the outer edges of its host galaxy. Based on the arguments from \citet[][]{2019A&A...629A..57M}, we use the physical properties derived from the nearest \ion{H}{ii} region for ASASSN-14jb.

\begin{figure*}[t]
\centerline{
\includegraphics[scale=0.3]{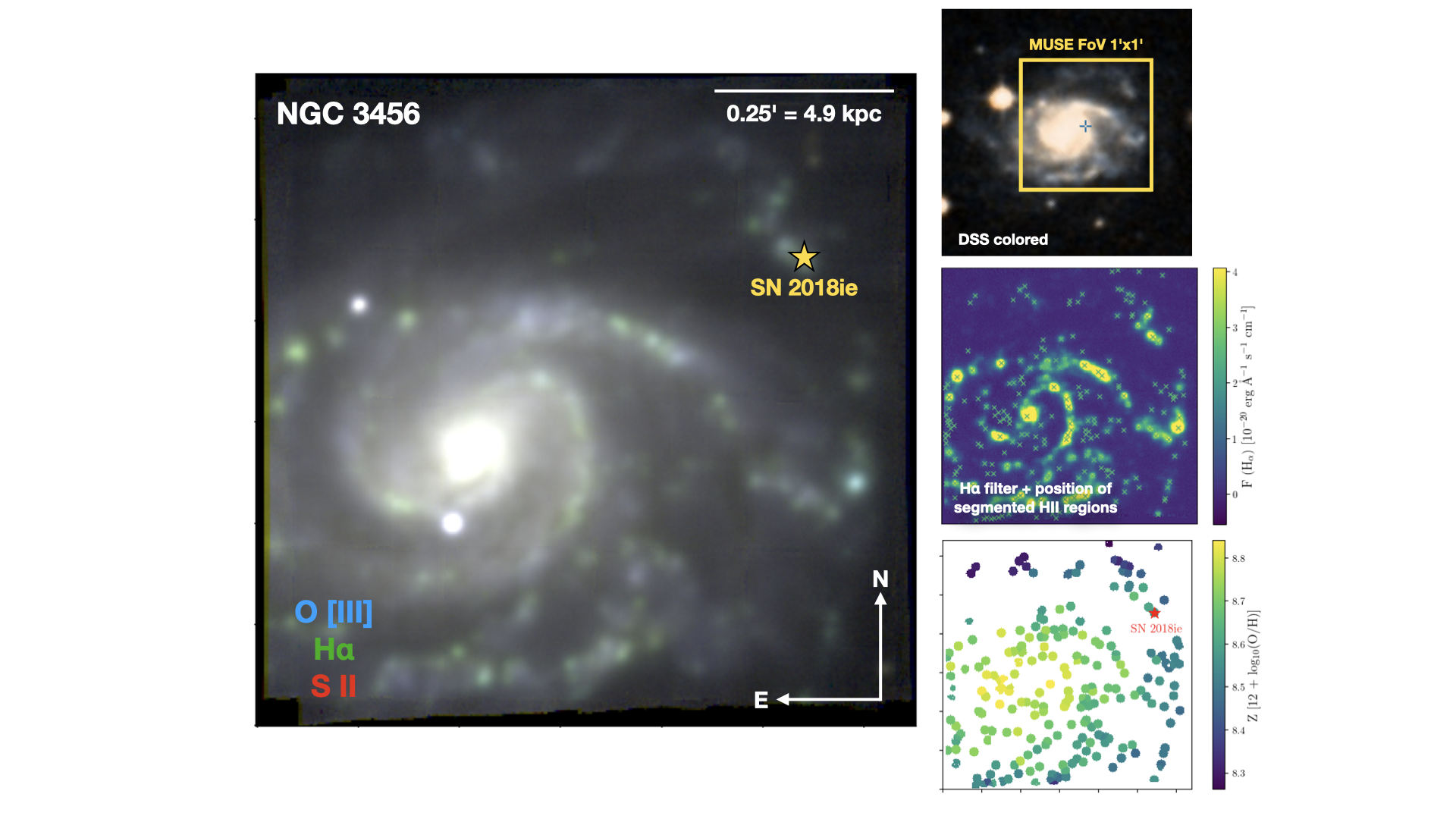}}
\caption{MUSE RGB image composition of the galaxy NGC 3456 where the O [III] emission is blue, H$\alpha$ is green, and \ion{S}{II} is red. The top panel in the right column shows the MUSE pointing from where the observation of the galaxy was taken, obtained from the ESO Archive Science Portal and based on a Digitized Sky Survey (DSS) colored image of the galaxy. The galaxy is located at $z \approx 0.0142$ and hosted the SN Ic 2018ie, where its position is marked by the yellow star. {The middle panel in the right column shows the H$\alpha$ luminosity, with the markers indicating all \ion{H}{ii} regions identified by \textsc{IFUanal} and used to obtain the other parameters from the datacube.} The bottom panel reports the physical parameter map for the oxygen abundance of the segmented \ion{H}{ii}  regions obtained from \textsc{IFUanal}. \label{fig:NGC3456}} 
\end{figure*}

\begin{table}
\caption{Phase of the observations relative to the SN peak luminosity, for the events that presented broad spectral features at their position. \label{tab:phase}}
\centering
\begin{tabular}{ll}
\hline
SN & Phase (d)\\
\hline
ASASSN-14dp & +540 \\
ASASSN-14jb & +391 \\
ASASSN-14kp & +399 \\
ASASSN-15fz & +1576 \\
ASASSN-15lh & +424 \\
SN2015bf & +1277 \\
ASASSN-16at & +1208 \\
ASASSN-16bw & +778 \\
ASASSN-16in & +1076 \\
ASASSN-16jt & +1064 \\
ASASSN-16oy & +1031 \\
ASASSN-16po & +341 \\
SN2017ivu & +527 \\
SN2016bmi & +538 \\
SN2016gkg & +436 \\
ASASSN-17is & +402 \\
ASASSN-17je & +269 \\
ASASSN-17jp & +394 \\
ASASSN-17kx & +320 \\
ASASSN-17ny & +722 \\
ASASSN-17oj & +554 \\
ASASSN-17qp & +667 \\
SN2017ahn & +472 \\
SN2017bif & +447 \\
SN2017bzb & +780 \\
ATLAS17hpc & +673 \\
SN2017gmr & +338 \\
\hline
\end{tabular}
\end{table}

\subsection{Analysis of physical parameters}\label{sec:phys_param}

\subsubsection{Oxygen abundance indicators}

The nebular emission lines of HII regions are tracers of ionization produced by young stellar populations. 
Since oxygen has very strong nebular emission lines that dominate the spectra of \ion{H}{ii} regions together with H lines, its abundance can be used as a proxy to indicate the amount of metals present in the interstellar medium.
One of the most commonly used methods to estimate \ion{H}{ii} region oxygen abundances is the O3N2 index, introduced by \citet[][]{1979A&A....78..200A}, which uses the H$\alpha$, H$\beta$, [\ion{O}{III}]$\lambda 5007$, and [\ion{N}{II}]$\lambda 6584$ emission lines. The index was also used by \citet[][]{2004MNRAS.348L..59P}, who used a calibration based on photoionization models. Here, we use the O3N2 index calibration given by \citet[][M13]{2013A&A...559A.114M}, who used the observations of multiple \ion{H}{ii} regions to estimate the electron temperature. The O3N2 index has the advantage of being insensitive to extinction and flux calibration, because of the similar wavelengths of the H$\alpha$/[\ion{N}{II}] and H$\beta$/[\ion{O}{III}] line pairs.

We also use the N2 index calibration given by M13, which was extensively studied by many authors \citep[e.g.,][]{1994ApJ...429..572S, 1998ApJ...497L...1V, 2000MNRAS.316..559R}. The oxygen abundance given by the N2 index is defined by the line ratio between [\ion{N}{II}]$\lambda 6584$ and H$\alpha$, again minimizing the sensitivity to reddening corrections and flux calibration. 
Finally, we use the oxygen abundance indicator described in \citet[][D16]{2016Ap&SS.361...61D}. The D16 index is calibrated using photoionization models and is also insensitive to reddening. It uses the ratio between [\ion{N}{II}]$\lambda 6584$ to [\ion{S}{II}]$\lambda  \lambda  6717,31$ and [\ion{N}{II}]$\lambda 6584$ to H$\alpha$ as an estimate of chemical abundance. This method provides the highest values of oxygen abundance in our analysis (see discussion in Section \ref{sec:met}) and, as described in \citet[][]{2016Ap&SS.361...61D}, the index model provides a good linear fit up to $12 \ +$ log(O/H) $\sim9.05$~dex. 

{The uncertainties in oxygen abundances are derived from the propagation of line fitting errors from the spectra. We note, however, that each calibration index has intrinsic uncertainties of $\sim 0.1-0.2$~dex, as shown by \citet{2013A&A...559A.114M} and \citet{2016Ap&SS.361...61D}. Such uncertainties might affect the comparison between the different distributions and indicators, and are taken into account in our conclusions.}

\subsubsection{H$\alpha$ emission}

Tracing of H$\alpha$ emission is a well established method to look for correlations between SN locations and the star forming regions in a galaxy \citep[e.g., ][]{2011A&A...530A..95L, 2012MNRAS.424.1372A, 2013AJ....146...30K, 2013AJ....146...31K, 2017A&A...597A..92K}. As demonstrated by \citet[][]{1998ApJ...498..541K}, the H$\alpha$ luminosity alone can be used as an estimate of the recent star formation rate (SFR). 
Here, we use the conversion given by \citet{1998ApJ...498..541K} to obtain the SFR in M$_\odot \ \textrm{yr}^{-1}$ from the {extinction corrected} H$\alpha$ luminosity: $\rm SFR (M_{\odot}\,yr^{-1}) \simeq 7.9\times 10^{-42}\, L(H\alpha) $. Because the galaxies in our sample have different sizes, we use the star formation rate surface density ($\Sigma$SFR) in order to estimate the SF intensity normalized by the physical size of each \ion{H}{ii} region, dividing the SFR at each location by the area, as reported in Table \ref{tab:HII_dist}. 
{We report the uncertainty in SFR propagated from the H$\alpha$ line fitting. We do not correct the H$\alpha$ luminosities by the host galaxy inclination. We also do not take into account the dependence of the conversion factor on the properties of the stellar populations (for example, age, metallicity, and single vs binary stars; see \citealt{2017PASA...34...58E}). }

The H$\alpha$ EW gives a measurement of the strength of the line relative to the continuum. Assuming a stellar population born at the same epoch, the contribution to the stellar population light of young and massive stars will decrease with time, decreasing the ionization of the \ion{H}{ii} region and therefore the line strength relative to the continuum emission. 
Consequently, the EW can be used as a tracer of the age and indicator of the level of the current SFR in an \ion{H}{ii}  region \citep[][]{1999ApJS..123....3L, 2019MNRAS.482..384X}. 
{It is important to note that H$\alpha$ EW is subject to the photon-leakage effect and contamination from older stellar populations, which might input uncertainties in our conclusions \citep{2018MNRAS.477..904X, 2019MNRAS.490.4515S, 2021MNRAS.504.2253S}.}
We use the H$\alpha$ EW measurements from the Gaussian fits to the line in Section \ref{sec:res}.

\subsubsection{Extinction}

We use the Balmer decrement as an indicator of the dust extinction at the line of sight, $E(B-V)_{host}$. We estimate it using Equation 3 from \citet[][]{2013ApJ...763..145D}, assuming $(\textrm{H}\alpha/\textrm{H}\beta) = 2.86$ for Case B recombination, and $k(\lambda)$ parameters obtained from a fitting to the \citet[][]{1989ApJ...345..245C} extinction law. 
{The extinction is extracted only from \ion{H}{ii} regions coincident with the SN position.}

The values of $E(B-V)_{host}$ for the different SN host \ion{H}{ii} regions also are related to the extinction suffered by the SNe themselves. We therefore compare the values of extinction for the different CCSN types and analyze which subtype environment is more reddened, and therefore more affected by the presence of preexisting dust.

\section{Results} \label{sec:res}

In this section, we present the normalized cumulative distributions for the physical parameters of the {\ion{H}{ii} regions at the SN positions.} 
To estimate the measured errors associated to each normalized cumulative distribution, we have performed a resampling of $3 \times 10^4$ trials of each distribution, using the uncertainty associated to each parameter as one sigma of a Gaussian distribution. 
We also show the Kolmogorov-Smirnov (KS) statistic matrix and report the results from the Anderson-Darling (AD) tests for the combination of the different SN types at each distribution. The $p$-value of these tests express the statistical significance of two distributions being drawn from the same parent population. While a KS test is more sensitive to the center of the distributions, the AD test is a modification more sensitive to the tail ends of the distributions.
We consider a $p \leq 0.05$ as indicating a statistically significant result. 
We show the measured fluxes for the different emission lines and the resulting physical parameter values in Appendix \ref{sec:fluxes_params}, and the statistics, KS and AD results for these parameters in Appendix \ref{app:stats_KS_AD}.

{We first separate the CCSNe into types II, IIb, Ib, Ic (including SNe Ic-BL), IIn (including ASASSN-16jt, classified as a SN 2009ip-like event), and Ibn, based on their spectroscopic classification from Table \ref{tab:sn_prop}.
In order to make a more statistically significant analysis, we also group the CCSNe into three different groups: SNe II, consisting of the 78 II; SESNe, including the four Ic, two Ic-BL, seven IIb, seven Ib, plus SN 2017gax, classified as an {ambiguous} SN Ibc; and IIn/Ibn SNe, including the nine IIn and three Ibn}

Because SN 2015bn is classified as a SLSN, it does not belong to any of the other subtypes. As it is one single event, it cannot be included in the statistical analyses used for the other groups. {Therefore, we present an analysis of its environment in Appendix \ref{ref:app_15bn}, and make a comparison with the results for the other subtypes of CCSNe presented in this section.}

\subsection{Oxygen abundance} \label{sec:met}

Figure \ref{fig:cum_dist_metallicity_1} shows the cumulative distributions of the oxygen abundances for the \ion{H}{ii} regions. In Appendix \ref{app:stats_KS_AD}, Tables \ref{tab:Z_stats}, \ref{tab:Z_KS}, and \ref{tab:Z_AD} present the median and average values, the KS test results, and the AD test results, respectively.
SNe Ic, IIb, and {Ibn} have the highest median metallicity values for all three indicators, while SNe II have the lowest median value for the D16 index, SNe IIn for the N2 index, and SNe Ib for the O3N2 index. 
The D16 indicator leads to the largest range of oxygen abundance values (going from $12 \ +$ log(O/H) $\sim 7.6$~dex to $\sim 9.2$~dex). 
This index also provides a KS $p$-value $\sim 0.05$ between SNe Ic and II, suggesting a statistically significant difference.
The other comparisons between the CCSN distributions do not show statistically significant differences. 
No comparisons between the different types show  statistically significant difference for the N2 and O3N2 indexes.

In Figure \ref{fig:cum_dist_metallicity_2}, we show the cumulative distributions the CCSNe grouped into SNe II, SESNe, and IIn/Ibn. The SESN events have the higher median metallicities for all three indexes. 
None of the indicators show any statistically significant differences.

\begin{figure*}[t!]
\centering
\includegraphics[width=0.40\textwidth]{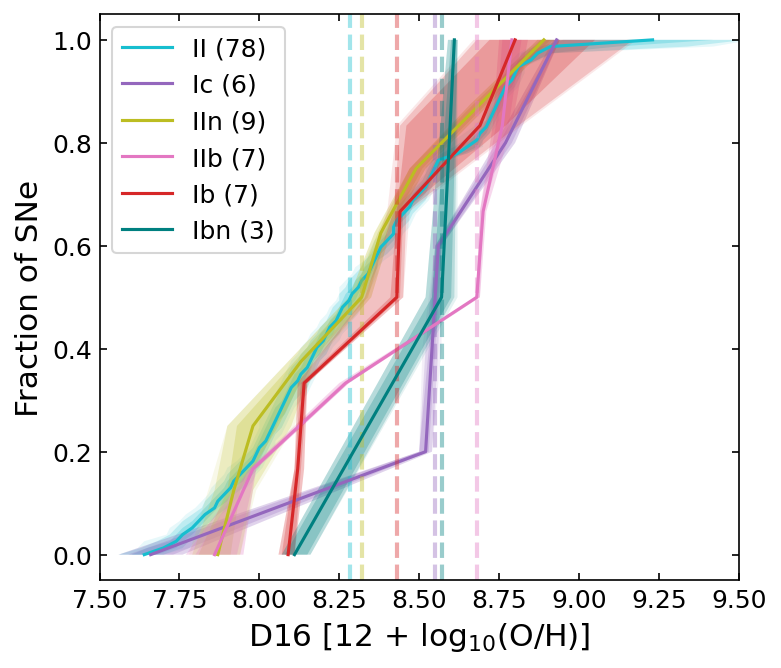}
\includegraphics[width=0.42\textwidth]{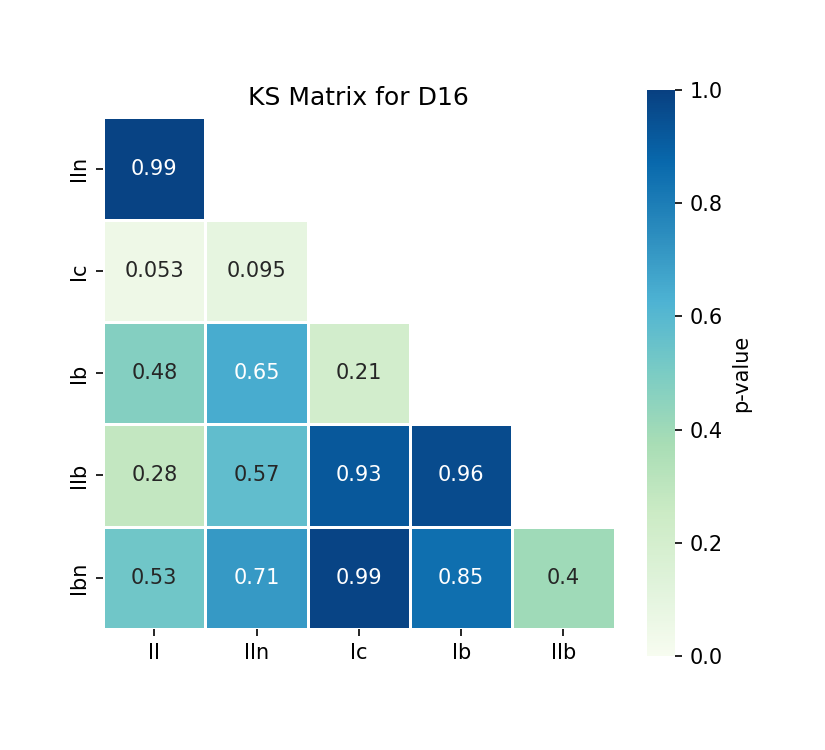}
\includegraphics[width=0.40\textwidth]{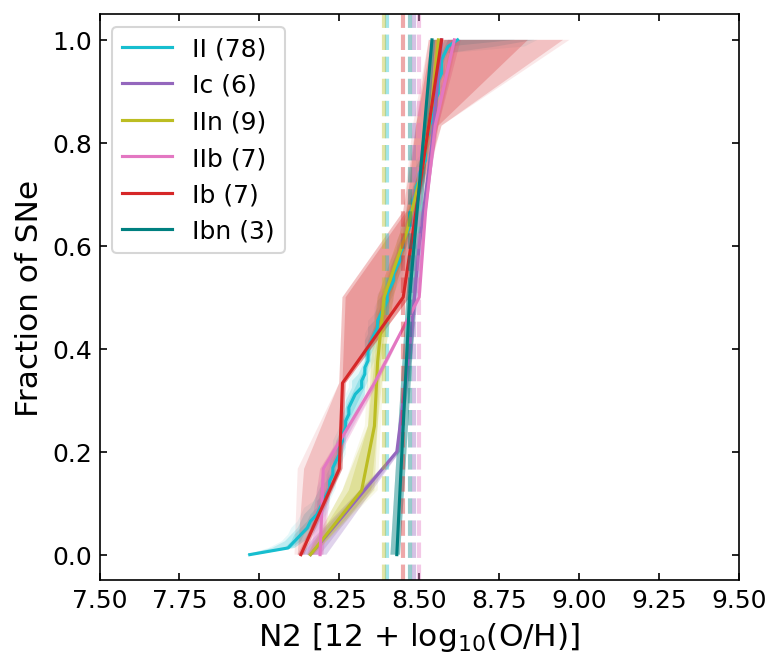}
\includegraphics[width=0.42\textwidth]{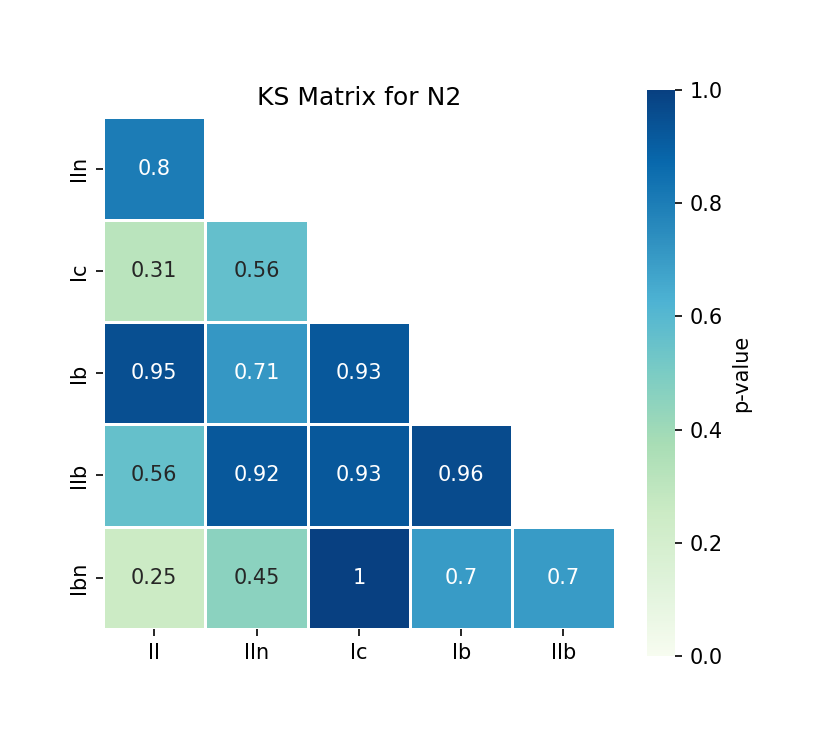}
\includegraphics[width=0.40\textwidth]{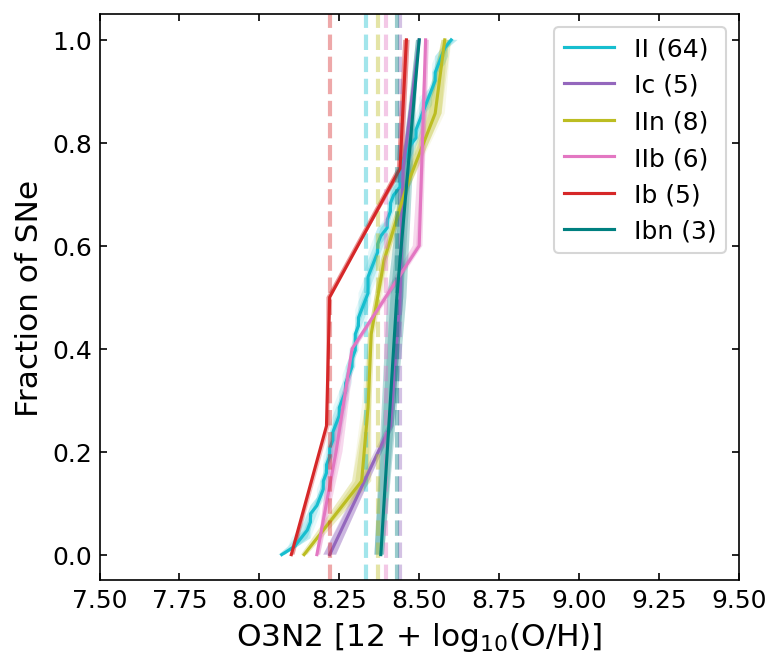}
\includegraphics[width=0.42\textwidth]{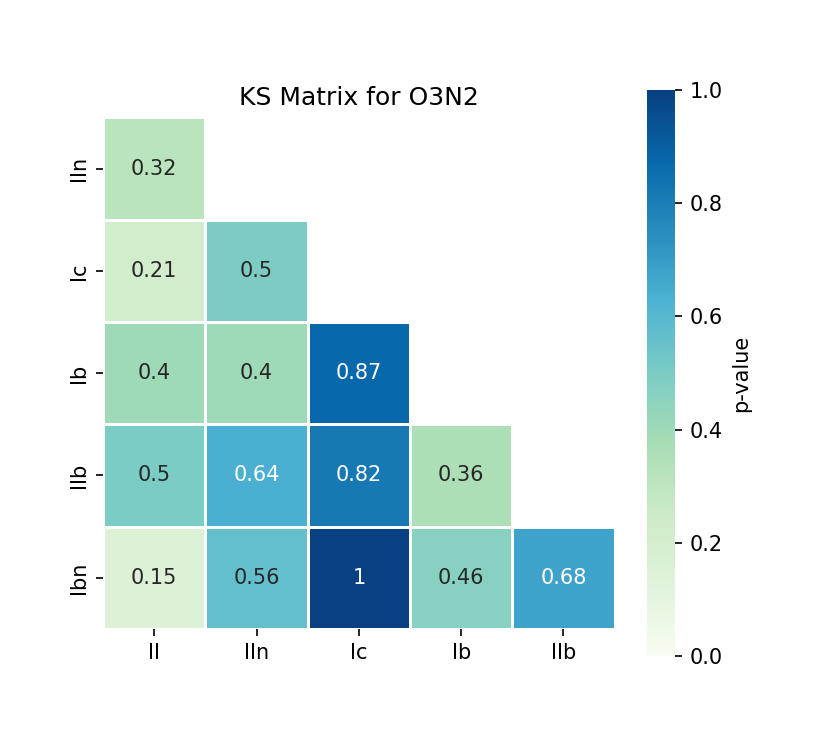}
\caption{Cumulative distributions for the D16, N2, and O3N2 oxygen abundance indicators of the different types of CCSNe. The right column shows the Kolmogorov-Smirnov (KS) statistic matrix for the combination of the different SN types. The legend gives the number of SNe used in each distribution and the median oxygen abundance is marked as the dashed linse. The color scale and the matrix values report the KS p-values. \label{fig:cum_dist_metallicity_1}} 
\end{figure*}

\begin{figure*}[t!]
\centering
\includegraphics[width=0.40\textwidth]{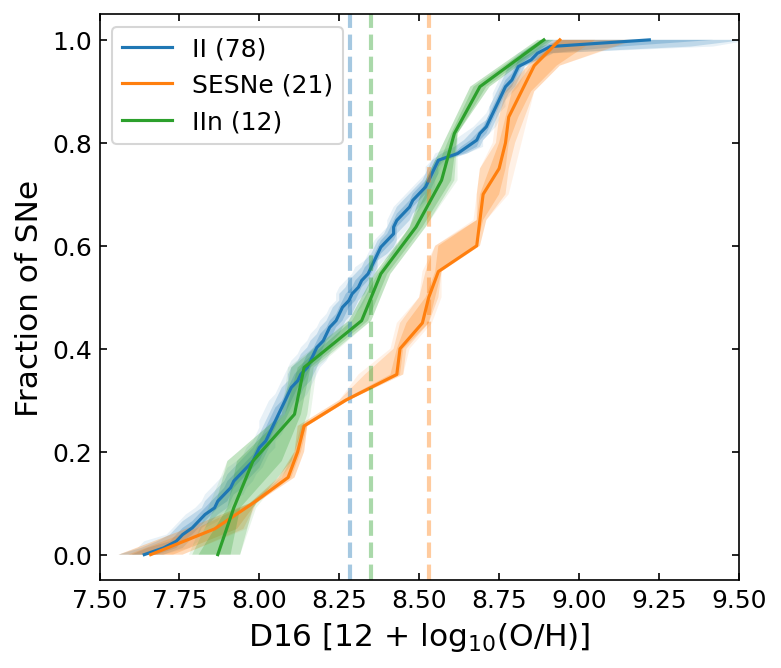}
\includegraphics[width=0.42\textwidth]{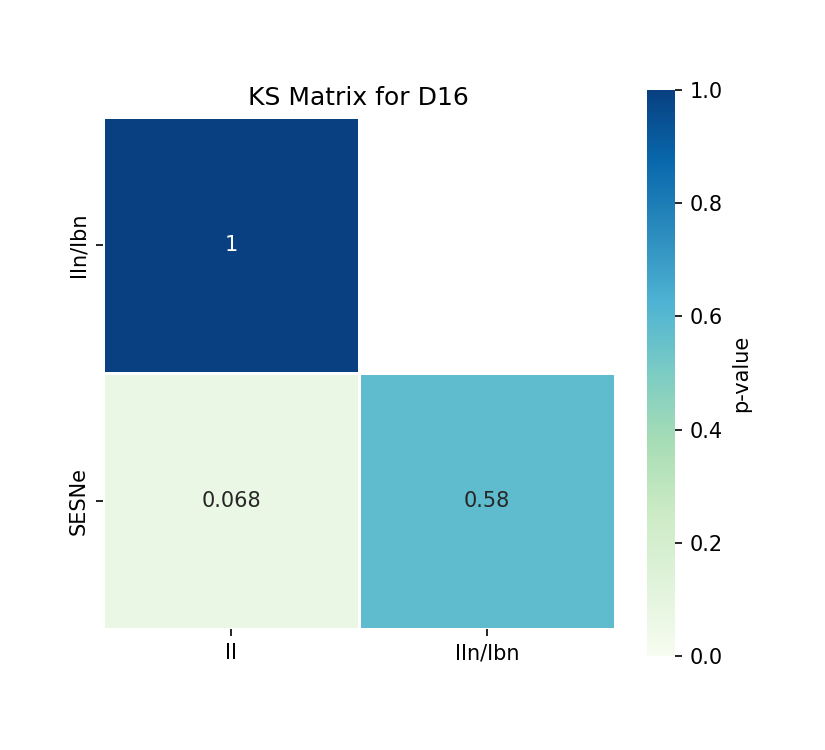}
\includegraphics[width=0.40\textwidth]{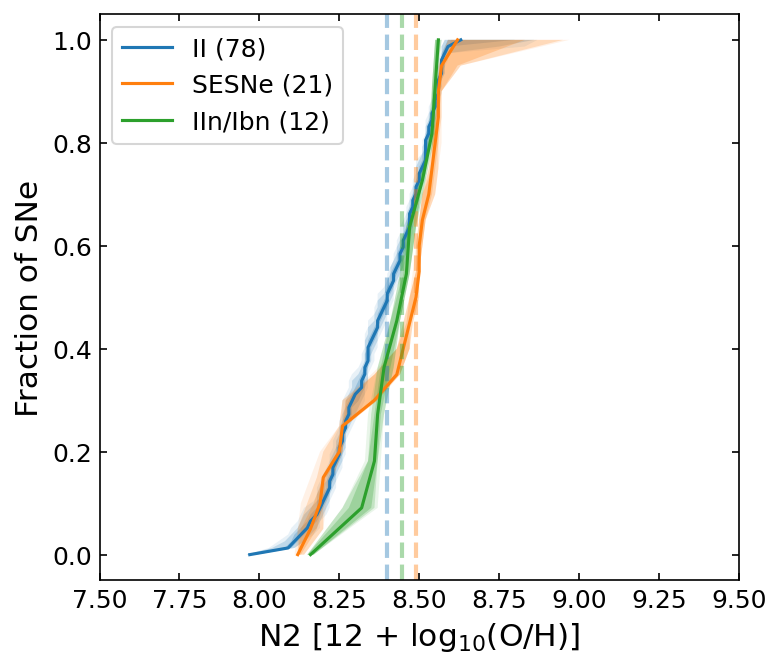}
\includegraphics[width=0.42\textwidth]{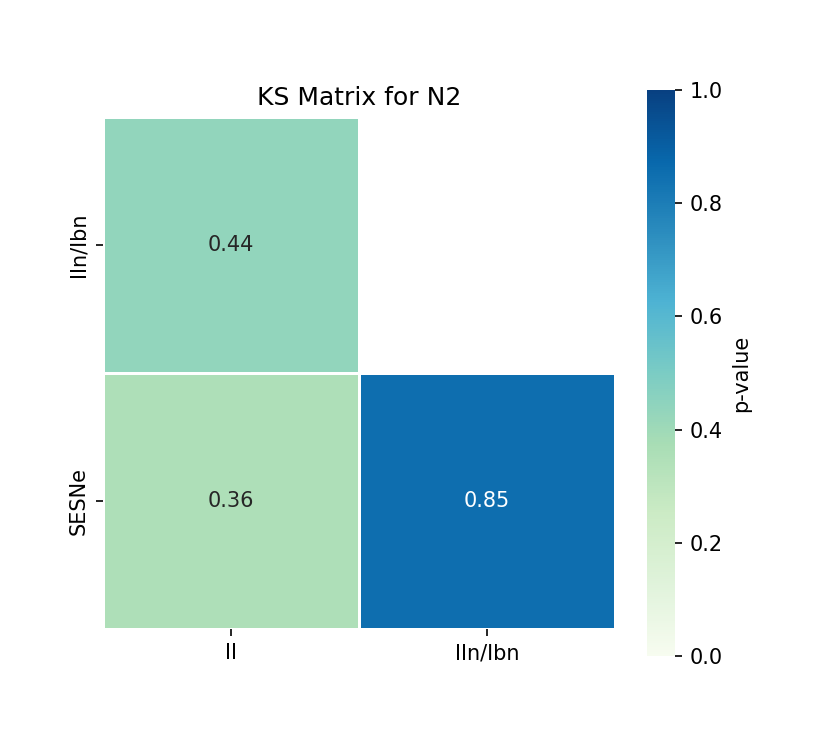}
\includegraphics[width=0.40\textwidth]{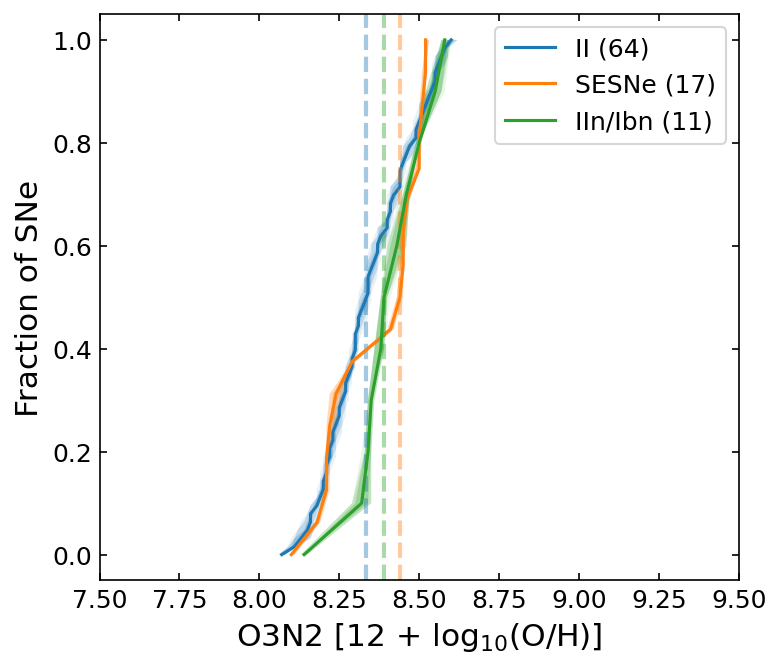}
\includegraphics[width=0.42\textwidth]{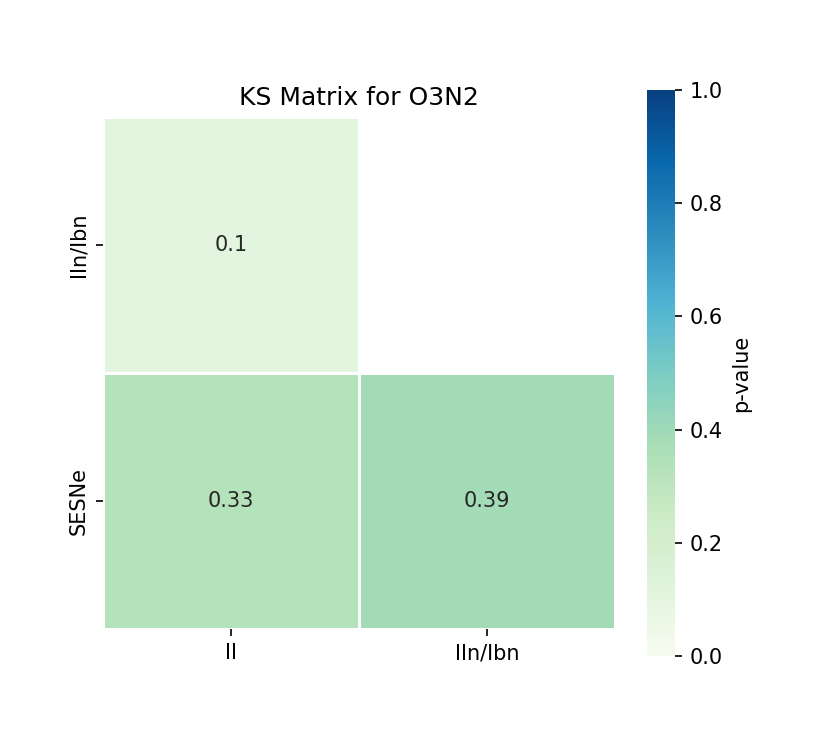}
\caption{Cumulative distributions for the D16, N2, and O3N2 oxygen abundance estimates of the CCSNe grouped as SNe II, SESNe, and IIn/Ibn. The right column shows the Kolmogorov-Smirnov (KS) statistic matrix for the combination of the different SN types.   \label{fig:cum_dist_metallicity_2}} 
\end{figure*}

\subsection{H$\alpha$ equivalent width} \label{sec:HaEW}

Figure \ref{fig:cum_dist_EW} shows the cumulative distributions of the H$\alpha$ EW for the different CCSN types.
The median and the average of each distribution, their KS test and AD test values are in Tables \ref{tab:EW_stats},  \ref{tab:EW_KS}, and \ref{tab:EW_AD}, respectively. 
In the top panel of Figure \ref{fig:cum_dist_EW}, we see that SNe Ic have the highest median H$\alpha$ EW ($221 \pm 9$~\AA), followed by SNe IIb and Ib ($139.8 \pm 8$~\AA \ and $87.2 \pm 12$~\AA, respectively). 
{SNe Ibn show the lowest median H$\alpha$ EW ($45 \pm 11$~\AA), although only three events are considered in the analysis and chance alignment with older stellar populations cannot be excluded.}
The KS test values between the different distributions are not statistically significant, with the resulting KS test p-value between the SNe II and Ic being $\sim 0.1$. However, the AD test shows a significance value of $\sim 0.03$ between the SNe II and Ic, suggesting a statistically significant difference between the distributions.

When we group the different types as SNe II, SESNe, and IIn/Ibn, in the bottom panel of Figure \ref{fig:cum_dist_EW}, SESNe show a higher median of H$\alpha$ EW ($109 \pm 5$~\AA) than the other subtypes. 
The KS test shows a statistically significant difference between SNe II and SESNe, with a p-value $\sim 0.03$. The AD test also shows a statistically significant difference between SNe II and SESNe, with a significance value of $\sim 0.005$. 

As can be seen in Figure \ref{fig:cum_dist_EW}, one event has a very large H$\alpha$ EW: SN 2016cdd, a SN Ib, with an H$\alpha$ EW of $995 \pm 20$~\AA. 
SN 2016cdd is a regular SN Ib, with He I absorption features and is similar to the prototypical SN 1999dn \citep[][]{1999AAS...195.3810Q, 2016ATel.9065....1H}. 
{Although this appears to be a common example of a SN Ib, it is associated to a very high value of H$\alpha$ EW, suggesting an association with very young stars and very recent episodes of star formation. Finally, a chance alignment of the SN with a very recent star-forming region cannot be excluded.}

\begin{figure*}[t!]
\centering
\includegraphics[width=0.40\textwidth]{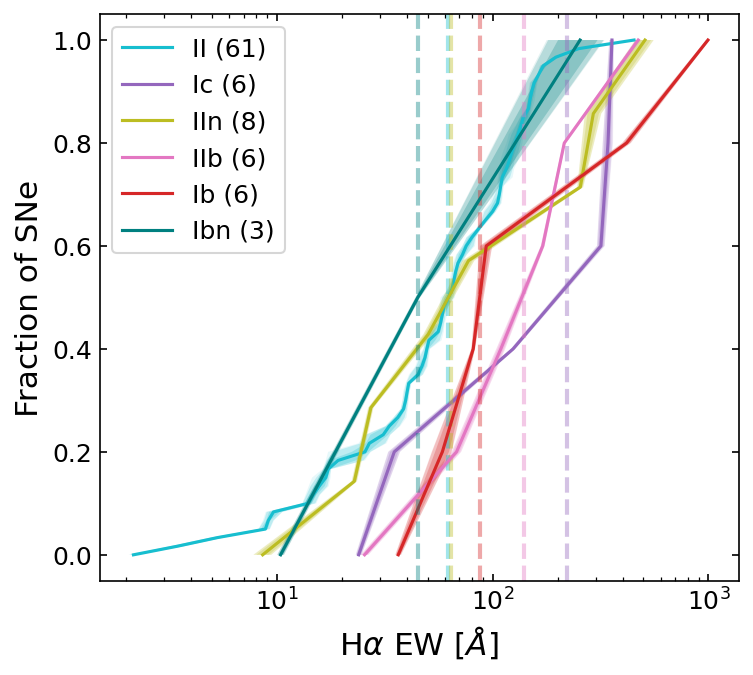}
\includegraphics[width=0.42\textwidth]{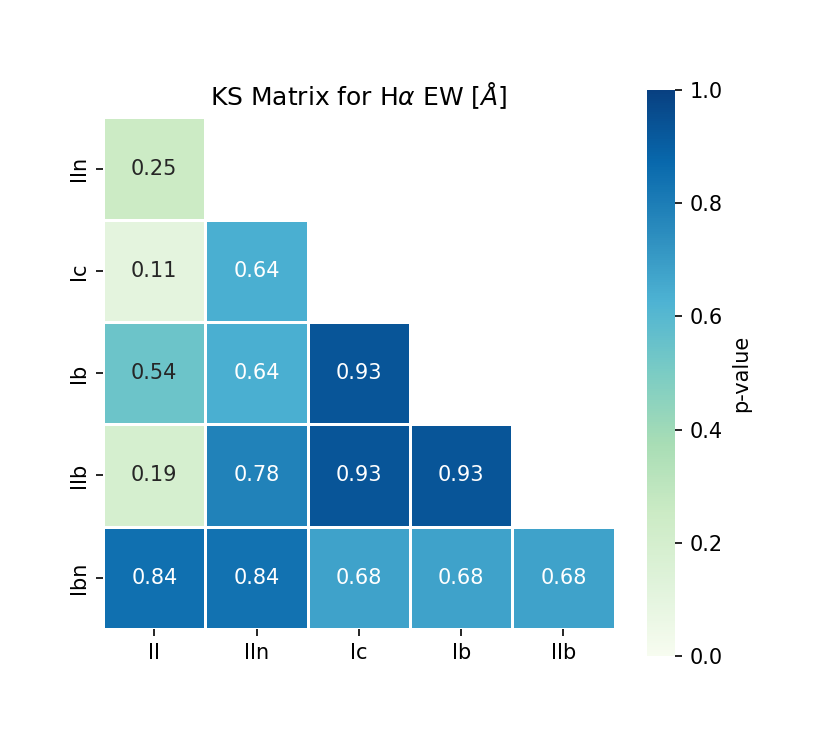}
\includegraphics[width=0.40\textwidth]{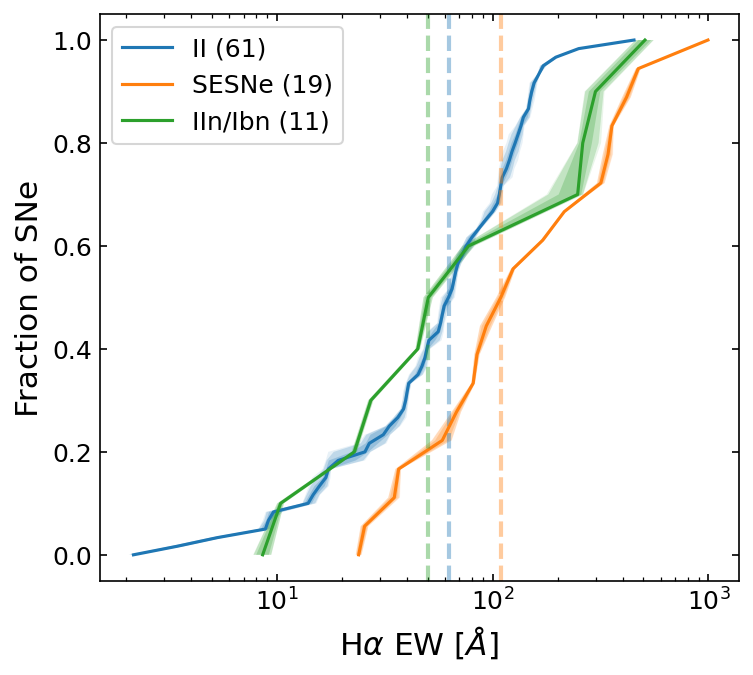}
\includegraphics[width=0.42\textwidth]{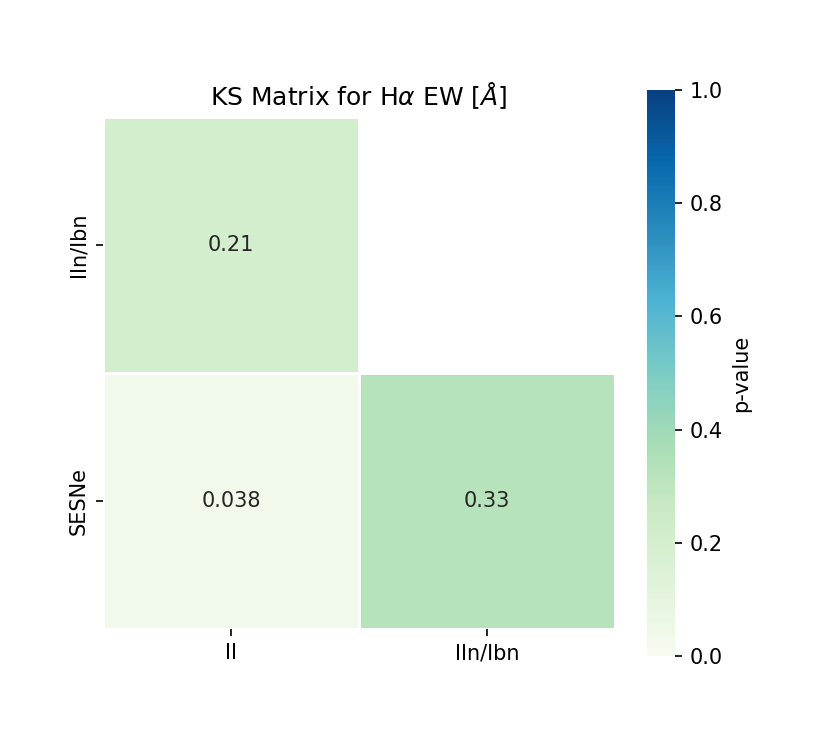}
\caption{{Cumulative distributions of the H$\alpha$ EW of the \ion{H}{ii} regions at the SN locations.} The top row shows the distribution for all the SN types, while the bottom row shows the distribution for the types grouped into SNe II, SESNe, and IIn/Ibn. The Kolmogorov-Smirnov (KS) statistic matrix is shown in the right column. }
\label{fig:cum_dist_EW}
\end{figure*}

\subsection{Star formation rate} \label{sec:SFR}

The top panel of Figure \ref{fig:cum_dist_SFR} shows the cumulative distributions of $\Sigma$SFR for the different CCSN types.
The median and the average of each distribution, their KS test and AD test values are reported in Tables \ref{tab:SFR_stats}, \ref{tab:SFR_KS}, and \ref{tab:SFR_AD}, respectively. 
The SNe Ic have the highest median $\textrm{log}_{10} \Sigma$SFR ($-1.11 \pm 0.50$~M$_\odot$~yr$^{-1}$~kpc$^{-2}$), with no associated \ion{H}{ii} regions having $\textrm{log}_{10} \Sigma \textrm{SFR} \leq -3$~M$_\odot$~yr$^{-1}$~kpc$^{-2}$. 
SNe IIn have the smallest median value ($-2.13 \pm 0.17$~M$_\odot$~yr$^{-1}$~kpc$^{-2}$), followed by {SNe Ibn ($-1.97 \pm 1.0$~M$_\odot$~yr$^{-1}$~kpc$^{-2}$).} 
Based on the KS and AD tests the differences are not statistically significant. 

The differences between SNe II, IIn/Ibn, and SESNe are clearer in the bottom panel of Figure \ref{fig:cum_dist_SFR}. The SESNe have the highest values of $\Sigma$SFR, with a median of $\textrm{log}_{10} \Sigma \textrm{SFR} = -1.27 \pm 0.23$~M$_\odot$~yr$^{-1}$~kpc$^{-2}$. 
The KS and AD tests show no statistically significant difference.

\begin{figure*}[t!]
\centering
\includegraphics[width=0.40\textwidth]{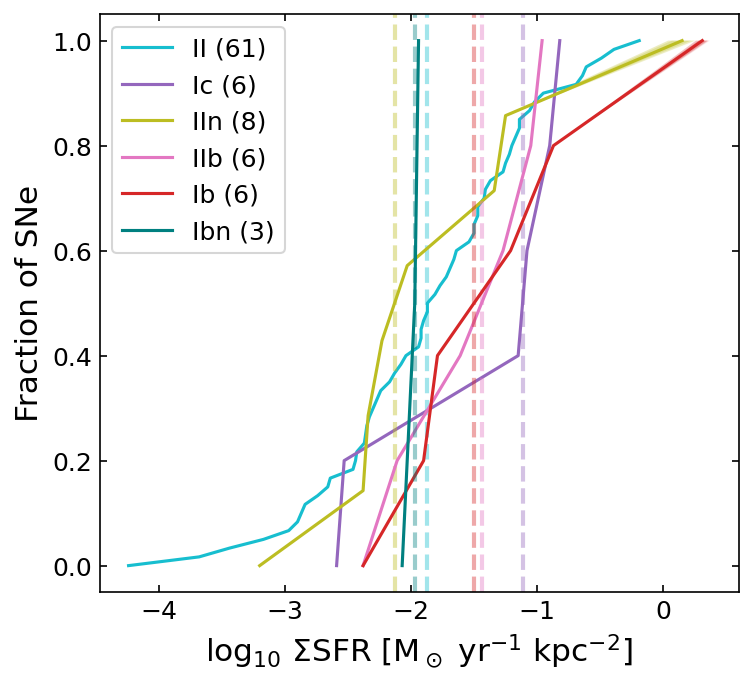}
\includegraphics[width=0.42\textwidth]{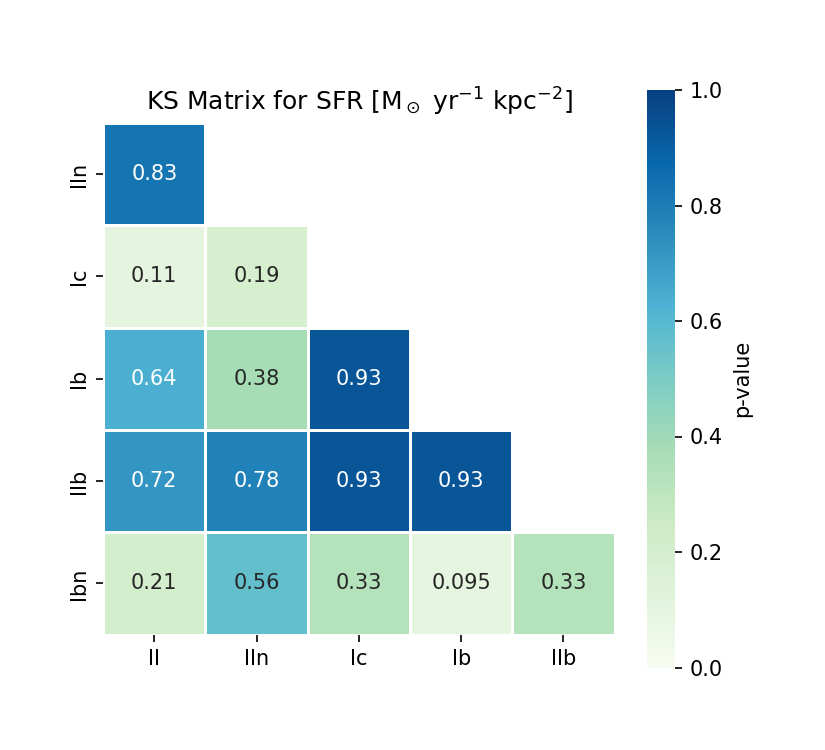}
\includegraphics[width=0.40\textwidth]{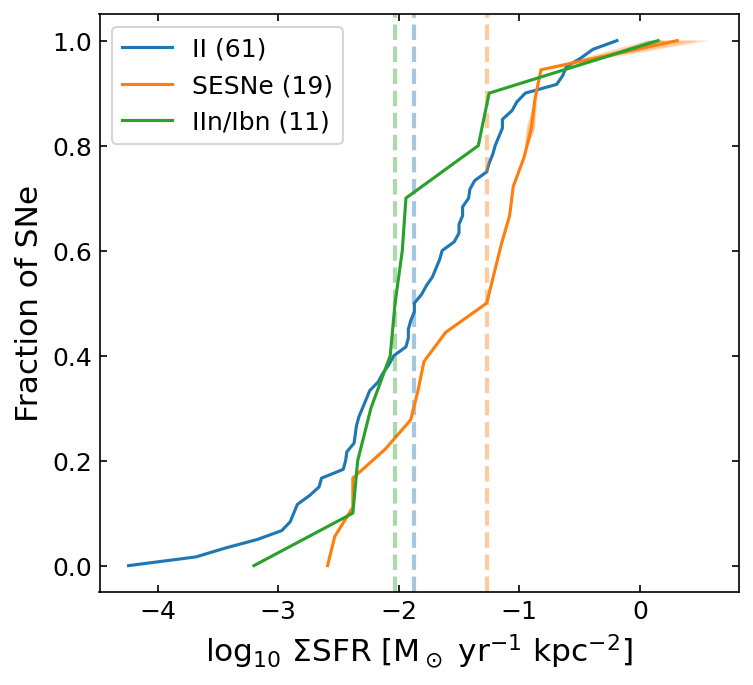}
\includegraphics[width=0.42\textwidth]{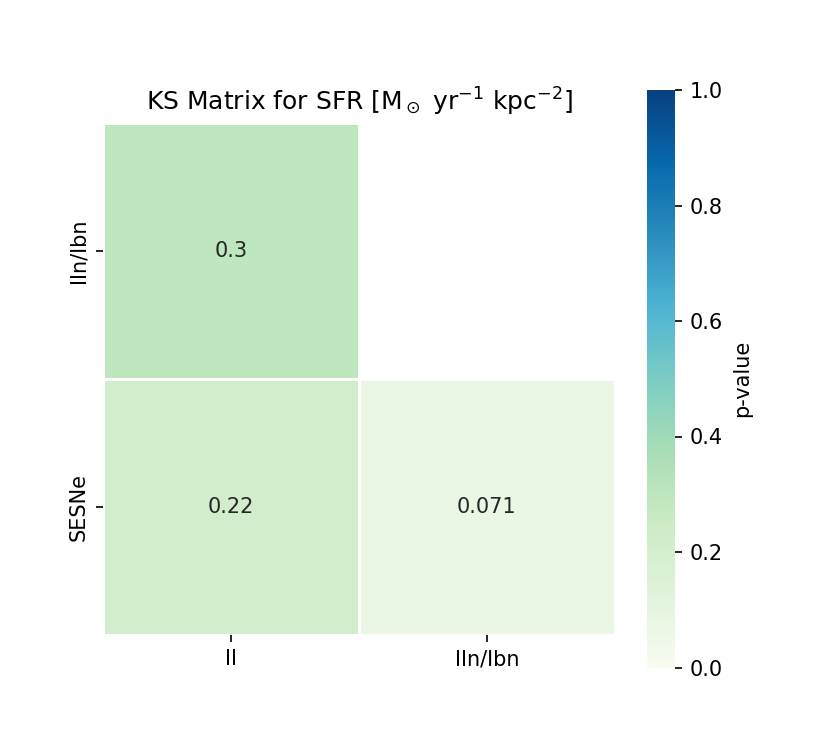}
\caption{{Cumulative distributions of the star formation rate surface density ($\Sigma$SFR) of \ion{H}{ii} regions at the SN locations, for the different SN types, and for the types grouped into SNe II, SESNe, and IIn/Ibn.} The right column shows the Kolmogorov-Smirnov (KS) statistic matrix.  \label{fig:cum_dist_SFR}}
\end{figure*}

\subsection{Extinction} \label{sec:extinc}

In the top panel of Figure \ref{fig:cum_dist_extinction} we show the cumulative distributions for the host extinction for the different CCSNe, and report the median and average of each distribution, the KS and AD test values in Tables \ref{tab:ebv_stats},  \ref{tab:ebv_KS}, and \ref{tab:ebv_AD}, respectively. 
The extinctions have values up to $\sim 1$~mag, with a SN Ic having the highest value and the SNe IIb having the highest median of $0.32 \pm 0.3$~mag (although it has a large uncertainty associated).
{SNe Ib and Ic have the same median value for their distribution ($\sim 0.20$~mag) and SNe II have a similar median ($0.23 \pm 0.1$~mag).}
However, the differences are not statistically significant.

Grouping the different SN types does not produce a more statistically significant result, as it is shown in the bottom of Figure \ref{fig:cum_dist_extinction}. {The SNe II, SESNe, and IIn/Ibn have very similar median values, $\sim 0.23$, $\sim 0.22$, and $\sim 0.26$~mag, respectively. }

\begin{figure*}[t!]
\centering
\includegraphics[width=0.40\textwidth]{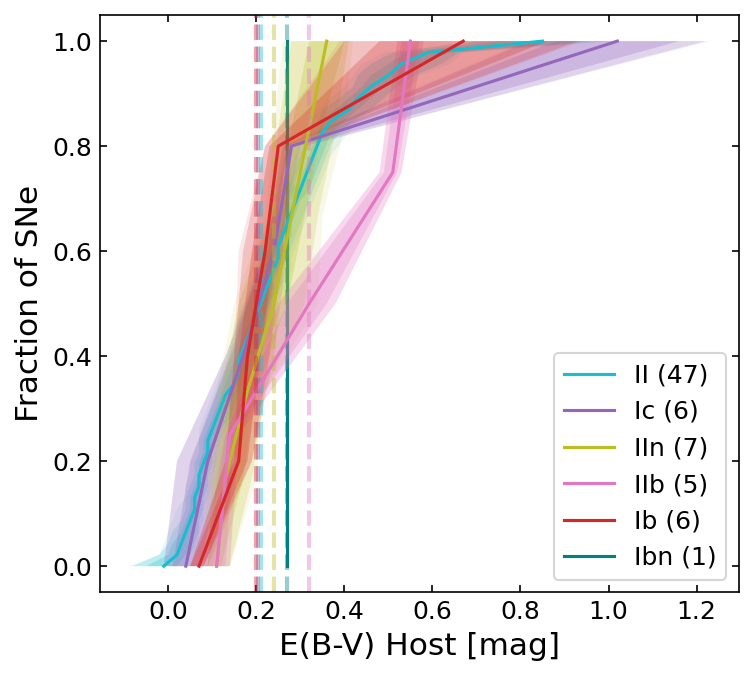}
\includegraphics[width=0.42\textwidth]{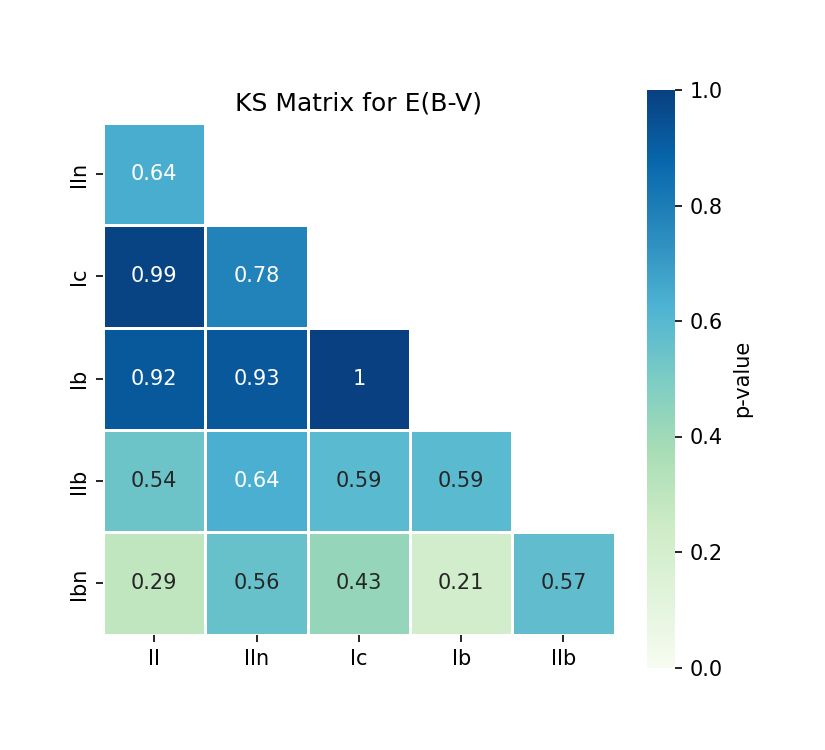}
\includegraphics[width=0.40\textwidth]{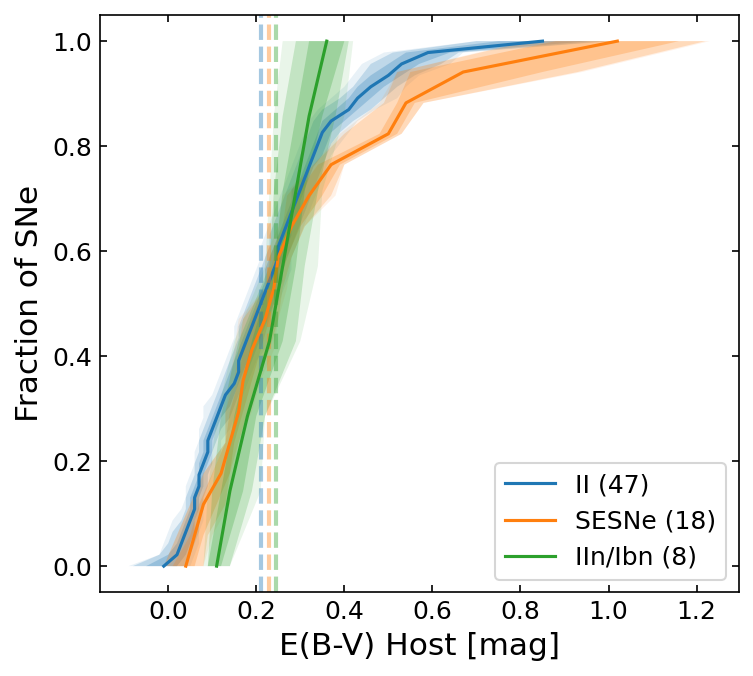}
\includegraphics[width=0.42\textwidth]{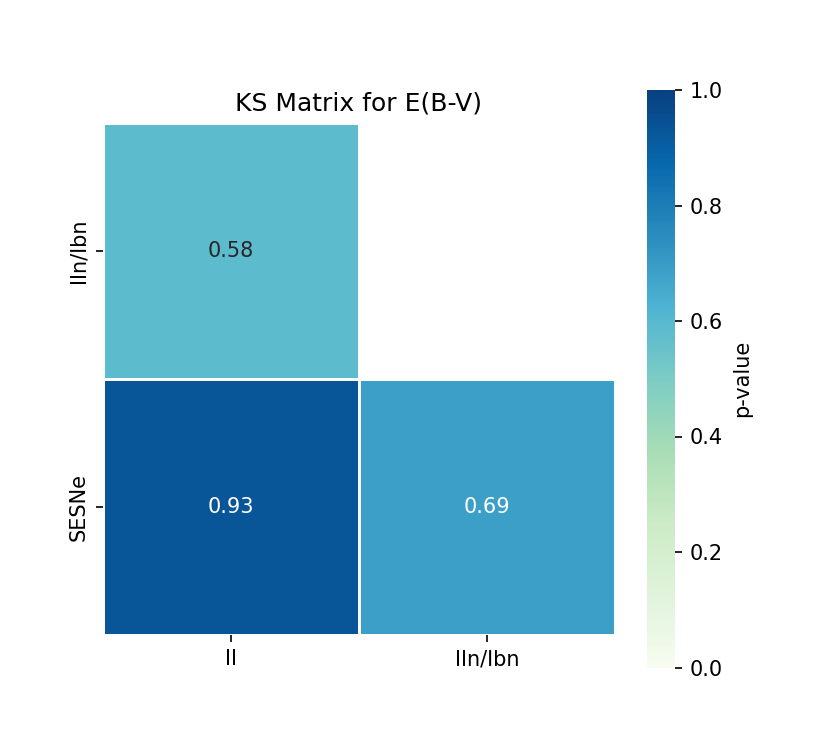}
\caption{Cumulative distributions of the line of sight extinction $E(B-V)$, estimated from the Balmer decrement. The top row shows the distributions for the different types of CCSNe, while the bottom row shows the distributions for the CCSNe grouped into SNe II, SESNe, and IIn/Ibn.
The right column shows the Kolmogorov-Smirnov (KS) statistic matrix.   \label{fig:cum_dist_extinction}} 
\end{figure*}

\section{Discussion} \label{sec:dis}

\subsection{Constraints on progenitor properties}

As discussed in the introduction, other studies of SN environments with larger samples have been achieved, but most of them have the major drawback of selecting SNe from targeted surveys, or selecting a sample with little or no control for systematics and biases. This might have an important effect in the final results, since these studies are focused on more massive and more metal rich galaxies, while also combining selection effects from many different SN search surveys.

We find in Section \ref{sec:res} that SESNe are associated with larger median H$\alpha$ EW, $\Sigma$SFR, and oxygen abundances than SNe II and IIn/Ibn, and that SNe Ic have larger values of these parameters than SNe Ib and IIb.
{Although this could indicate that SESNe come from younger, more massive and more metal rich stars than SNe II and IIn/Ibn, the differences between the distributions are not statistically significant in most cases, suggesting that any difference in progenitor properties is not large.}

For the different CCSN subtypes, we see a decreasing trend {in median values of H$\alpha$ EW and $\Sigma$SFR as Ic $\rightarrow$ IIb/Ib $\rightarrow$ II/Ibn/IIn. 
We found a statistically significant difference only between SNe~Ic and II for H$\alpha$ EW.}
This suggests that SNe Ic are associated to the youngest and more star-forming stellar populations, while SNe II and IIn are related to the oldest and less star-forming populations between the CCSNe.
This relation also shows that SNe Ic are associated to younger and more star-forming stellar population than SNe IIb and Ib, pointing to differences in progenitor age and mass-loss mechanisms. 
Similar results were also found by other previous analyses \citep[see, e.g.,][]{2008MNRAS.390.1527A, 2012ApJ...758..132S, 2012MNRAS.424.1372A, 2014A&A...572A..38G, 2014MNRAS.441.2230H, 2018ApJ...855..107G, 2018A&A...613A..35K}. 
{Again, we note that the differences in H$\alpha$ EW and $\Sigma$SFR between the CCSN subtypes are not large, as no statistically significant result was found for most cases. We also note the large intrinsic errors in the measured $\Sigma$SFR.}

In the analysis of oxygen abundance, we see a decreasing trend of {median values as Ic/IIb/Ibn $\rightarrow$ Ib $\rightarrow$ II/IIn in the D16 and N2 indexes, and Ic/IIb/Ibn $\rightarrow$ II/IIn $\rightarrow$ Ib in the O3N2 index.} 
These results suggests that SNe Ic are related to the more metal-rich environments, while SNe II and IIn are associated to the less metal-rich regions within CCSNe. This also suggests that SNe Ic and IIb are associated to stellar populations with higher metal content than SNe Ib. 
{However, the only statistically significant differences found were in the D16 index between SNe~II and Ic. The lack of statistically significant differences in the N2 and O3N2 indexes suggests that the metallicity differences in the environments of the different CCSN subtypes is not large. We also note that the large intrinsic errors associated to the oxygen abundance calibrators might input uncertainties in this analysis.
Although previous studies found significant differences in the environment metallicity of SNe~II and SESNe \citep[see, e.g.,][]{2003A&A...406..259P, 2009A&A...503..137B, 2008ApJ...673..999P, 2010ApJ...721..777A, 2011ApJ...731L...4M}, recent analyses have shown that such differences might not be large.}
\citet[][]{2016A&A...591A..48G} and \citet[][]{2018ApJ...855..107G} showed that no difference in the metal content of SN environments is seen when using a targeted sample, but that significant differences arise when taking host galaxies from untargeted searches. 
By using the integrated properties of SN host galaxies, \citet{2021ApJS..255...29S} found no significant differences in the galaxy mass (related to metallicity) and SFR for the different CCSN types, although they show that
SNe~Ibc prefer galaxies with slightly higher masses and SFR than SNe~II and IIb. 

Rotating single star models predict that SNe Ib should be generated by stars in the range of $20 - 40$~M$_\odot$, while SNe Ic should arise from progenitors with masses greater than $40$~M$_\odot$, at solar metallicity \citep[][]{2009A&A...502..611G}. 
Binary population models predict that SNe originating in such systems would require less massive progenitor stars for the envelope stripping \citep[e.g.,][]{2008MNRAS.384.1109E}. 
However, recent binary models show that a thin layer of hydrogen is still present after Roche-lobe overflow in these systems, making it hard to the stars to lose their deeper helium layer \citep[][]{2017MNRAS.470.3970Y}.
{Other recent results propose a hybrid mass-loss mechanism, with both binary interaction and strong winds playing a role \citep{2022ApJ...928..151F, 2023MNRAS.521.2860S}.}
Our results suggest that SNe Ic have a higher chance to be originated by more massive and more metal-rich stars than SNe Ib, either in single or binary systems.

{We find that SNe IIb are related to slightly younger and star-forming regions than SNe II and Ib, and with values of oxygen abundance higher than these types. 
Previous studies either report SNe IIb environments older than \citep[e.g.,][]{2012MNRAS.424.1372A, 2018A&A...613A..35K, 2018MNRAS.476.2629M} or similar to  \citep[e.g.,][]{2013MNRAS.436.3464K, 2023MNRAS.521.2860S} SNe Ib.
The fact that the progenitors of SNe IIb are in a similar age and mass range to SNe II might suggest that their progenitors are predominantly in binary systems, as they would not be massive enough to lose their envelopes through single-star winds.
% Such differences might arise from the small number of events in our sample ($\leq 10$).
Another possibility is that SNe IIb come from a mixture of progenitors with different metallicities and mass ranges.}

We have shown that SNe IIn/Ibn have very similar H$\alpha$ EW, SFR, and oxygen abundances to SNe II. 
{Given the large amount of material ejected by these events, a massive progenitor (LBVs in the case of SNe~IIn and massive WRs for SNe~Ibn) is expetected to generate such strongly interacting transients.} However, lower mass progenitors could also generate SNe IIn, such as RSG with superwinds or static CSM shells close to the star \citep[e.g.,][]{2002ApJ...572..350F, 2009AJ....137.3558S}.
Other analyses of the environments of SNe IIn also found no association to more metal rich or more younger stellar populations, showing very similar properties to normal SNe II \citep[][]{2013A&A...555A..10T, 2012MNRAS.424.1372A, 2017A&A...597A..92K, 2018A&A...613A..35K}. 
\citet[][]{2018ApJ...855..107G}, however, showed that the age distributions of SNe IIn had a bimodal distribution, and \citet[][]{2022MNRAS.513.3564R} also found a bimodal distribution of progenitors associated to younger and older stellar populations. 
{Recent studies of SNe~Ibn also show that they could be associated to older stellar populations \citep[e.g.,][]{2017MNRAS.471.4381S, 2020MNRAS.491.6000S}, in agreement with our results (although we note the low number of SNe~Ibn in our sample).}

Finally, we have shown that the CCSNe have similar values of host galaxy extinctions, with a median extinction $\sim 0.2$~mag for most of the subtypes, although SNe Ib have a slightly larger median.
This is similar to what was found by \citet[][]{2011ApJ...741...97D} and \citet[][]{2012ApJ...758..132S}. \citet[][]{2012ApJ...759..107K} found that SNe Ib and Ic have significantly higher values of host extinction than SNe II and IIb.
\citet[][]{2017MNRAS.468..628G} also demonstrated that SESNe have a larger dust reddening and \ion{H}{ii}  column density than SNe Ia and II.
\citet[][]{2012ApJ...758..132S} showed that the hosts of SESNe do not have any significant difference in extinction between themselves, but \citet[][]{2012ApJ...759..107K} found that SNe Ib and Ic have significantly higher values of host extinction than SNe II and IIb.

\subsection{Environment versus light curve properties} \label{sec:lc_prop}

CCSNe display a variety of properties in their LCs that are directly connected to the evolution of their progenitor stars and explosion properties. Properties such as the hydrogen envelope mass, progenitor radius, explosion energy and ejected $^{56}$Ni affect the observed luminosity, duration and shapes of their LCs \citep[e.g., ][]{2009ApJ...703.2205K, 2011ApJ...729...61B}. By analyzing 116 $V$-band LCs of SNe II, \citet[][]{2014ApJ...786...67A} showed that they display a large range of decline rates, and that the decline rate is directly proportional to the magnitude at peak luminosity. Their results also suggest that fast-decliner SNe II have lower ejected masses. Recently, \citet[][]{2022A&A...660A..42M} found that SNe~II with higher luminosity have higher $^{56}$Ni masses and that faster declining events have a more centrally concentrated $^{56}$Ni distribution. \citet[][]{2022A&A...660A..42M} also concluded that the explosion energy was the property that dominates the SN LC diversity. G18 looked for correlations between the LC parameters and the environments of SNe~II in low-luminosity host galaxies, and concluded that this sample had slower declining LCs than events in higher-luminosity galaxies. G18 however found no correlation between the plateau duration or the explosion energy of the Type II SNe with the metallicity at their environments.

Recent analyses of SESNe LCs have shown that they are characterized by relatively small ejecta masses and explosion energies \citep[][]{2013MNRAS.434.1098C, 2015A&A...574A..60T, 2016MNRAS.457..328L, 2016MNRAS.458.2973P}.
Since the LCs of these events are powered by radioactive decay, the amount of $^{56}$Ni synthesized in the explosion is directly proportional to the observed peak luminosity. In a study of 34 SESNe, \citet[][]{2018A&A...609A.136T} showed that the $\Delta m_{15}$ parameter, the brightness difference between the peak and 15 days later, is correlated with the slope of the LC
during its linear decay and with the peak absolute $B$-band magnitude. 

{Observations of SNe IIn have shown that they have very heterogeneous spectra and LCs \citep[][]{2013A&A...555A..10T}. Due to interaction with CSM, SNe IIn can stay bright for a much longer period than normal SNe, and the geometry of the CSM may drive further LC diversity.} \citet[][]{2015A&A...580A.131T} proposed a grouping of SNe IIn based on the similarities with the prototypical SN 1988Z \citep[][]{1993MNRAS.262..128T}, SN 1994W \citep[][]{1998ApJ...493..933S}, and SN 1998S \citep[][]{2000MNRAS.318.1093F}, and found that 1998S-like SNe show higher metallicities than the other events. \citet[][]{2020A&A...637A..73N} analyzed the LC parameters of a sample of 42 SNe IIn and found that they could be divided into two groups with distinct rise times, and that more luminous events are also the more long-lasting.

Here, we look for correlations between the environment and the explosion properties of the SNe in our sample. We compare the LC properties to the derived physical properties of H$\alpha$ EW, $\Sigma$SFR, and oxygen abundance. We expand the study presented in \citet[][]{2018ApJ...855..107G} and make this analysis using the LCs of 64 SNe II, eight SESNe, and ten SNe IIn {(because only one SN Ibn in our sample has good photometry coverage, this type is not included in this work)}. 
For most SNe, we use $BVr$ photometry obtained by ASAS-SN or the follow-up observations obtained with the Las Cumbres Observatory (LCO) telescopes \citep[for a in depth discussion on ASASSN photometry, see][]{2017PASP..129j4502K}. 
In a few cases {for which ASAS-SN photometry is not available} we use other published photometry, such as for ASASSN-14jb \citep{2019yCat..36290057M}, SN 2014cy, SN 2014dw, SN 2015W \citep[][]{2016MNRAS.459.3939V}, and SN 2017gmr \citep[][]{2021yCat..18850043A}. We characterize the LCs of SNe II using the magnitude at peak luminosity and the decline slope, $s$, similar to what was used by \citet[][]{2014ApJ...786...67A} and \citet{2016AJ....151...33G}. 
For SESNe, together with the magnitude at peak luminosity, we also use the $\Delta m_{15}$ parameter, first defined by \citet[][]{1993ApJ...413L.105P} in the study of SNe Ia.
For SNe IIn, we only characterize their LCs with the magnitude at peak luminosity.
As noted before, our sample is composed of intrinsically bright SNe, due to the magnitude-limited nature of ASAS-SN. Therefore, this analysis is biased toward including brighter events at the expense of fainter events, and fainter SNe could present distinct correlations between their LCs and environments. 

The LC parameters are given in Appendix \ref{app:LC_params}, in Tables \ref{tab:phot_II}, \ref{tab:phot_SESNe}, and \ref{tab:phot_int} for SNe II, SESNe, and IIn, respectively.
We show how the LCs and environment parameters are related to each other in Figures \ref{fig:II_LCs}, \ref{fig:SESNe_LCs}, and \ref{fig:int_LCs}. 
To define the possible correlation between the parameters analyzed, we use the Pearson test factor and $p-$value of the distributions.
We find a correlation for the postmaximum decline slope, $s$, in the $B$ and $r$ bands with the H$\alpha$ EW for SNe II.
The resulting Pearson test factors for these correlations were ($B,V,r$)=(0.63, 0.33, 0.86) ($p$-value=0.06, 0.07, 0.004).
However, only a few SNe II could have the $s$ parameters derived from their LCs in the $B$ and $r$ bands, leading to low-number statistics. 
In contrast to \citet[][]{2018ApJ...855..107G}, no significant correlation is found between the SFR and the $V$-band postmaximum decline slope for SNe II.

For SESNe, correlations are found between $B$-band magnitude at maximum luminosity and H$\alpha$ EW, with a Pearson correlation factor of $\sim 0.99$ ($p$-value $\sim 0.06$) and small correlations are found between the $B$-band $\Delta m_{15}$ parameter to $\Sigma$SFR and H$\alpha$ EW, with Pearson correlation factors $\sim 0.97$ (although with a nonstatistically significant $p$-value $\sim 0.1$) and $0.98$ ($p$-value $\sim 0.09$), respectively.
Small correlations are also found between M$_{peak}$ and the oxygen abundance, with Pearson test factors of ($B,V,r$)=(0.5, 0.5, 0.76) ($p$-value=0.1, 0.4, 0.2) for the D16 index.
{This suggests that the production of synthesized $^{56}$Ni in these explosions might be connected to the age of their parent stellar population, with events that produce more $^{56}$Ni occurring in the oldest and less star-forming regions.}
This also suggests a connection of their metal abundance to $^{56}$Ni production, with events in lower metallicity regions producing more $^{56}$Ni than events in higher metallicity environments.
{We also note that low-number statistics bring a significant caveat to this analysis.}
A similar result was found for SNe Ia, as higher metallicity leads to more production of $^{58}$Ni instead of $^{56}$Ni \citep[for a recent description on SNe Ia modeling, see e.g.,][]{2019MNRAS.482.4346B}.
However, we note that all of these results suffer from low-number statistics, and thus additional events are required to test the robustness of any correlations observed.

For SNe IIn, no correlation is found between M$_{peak}$ and the environment. Again, such an analysis might be suffering from small number statistics, and more observations of SNe IIn are required to enable stronger conclusions.

\begin{figure*}[t!]
\centering
\includegraphics[width=0.9\textwidth]{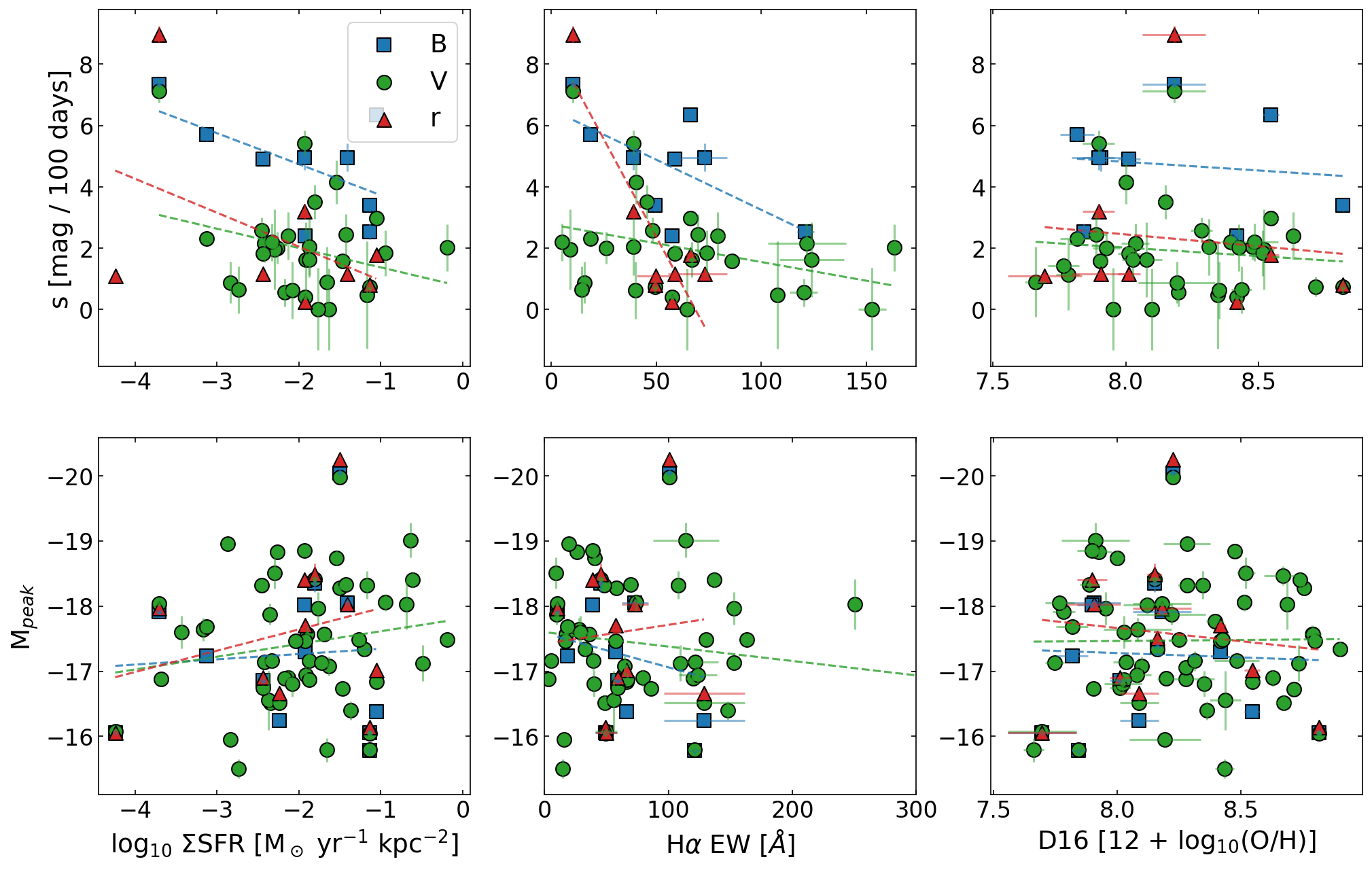}
\caption{SNe II postmaximum brightness decline rate (s) and absolute magnitude at peak (M$_{peak}$) in $B$, $V$, and $r$ bands, as a function of $\Sigma$SFR (left), H$\alpha$ EW (middle), and the oxygen abundance given by the D16 indicator (right).  \label{fig:II_LCs}} 
\end{figure*}

\begin{figure*}[t!]
\centering
\includegraphics[width=0.9\textwidth]{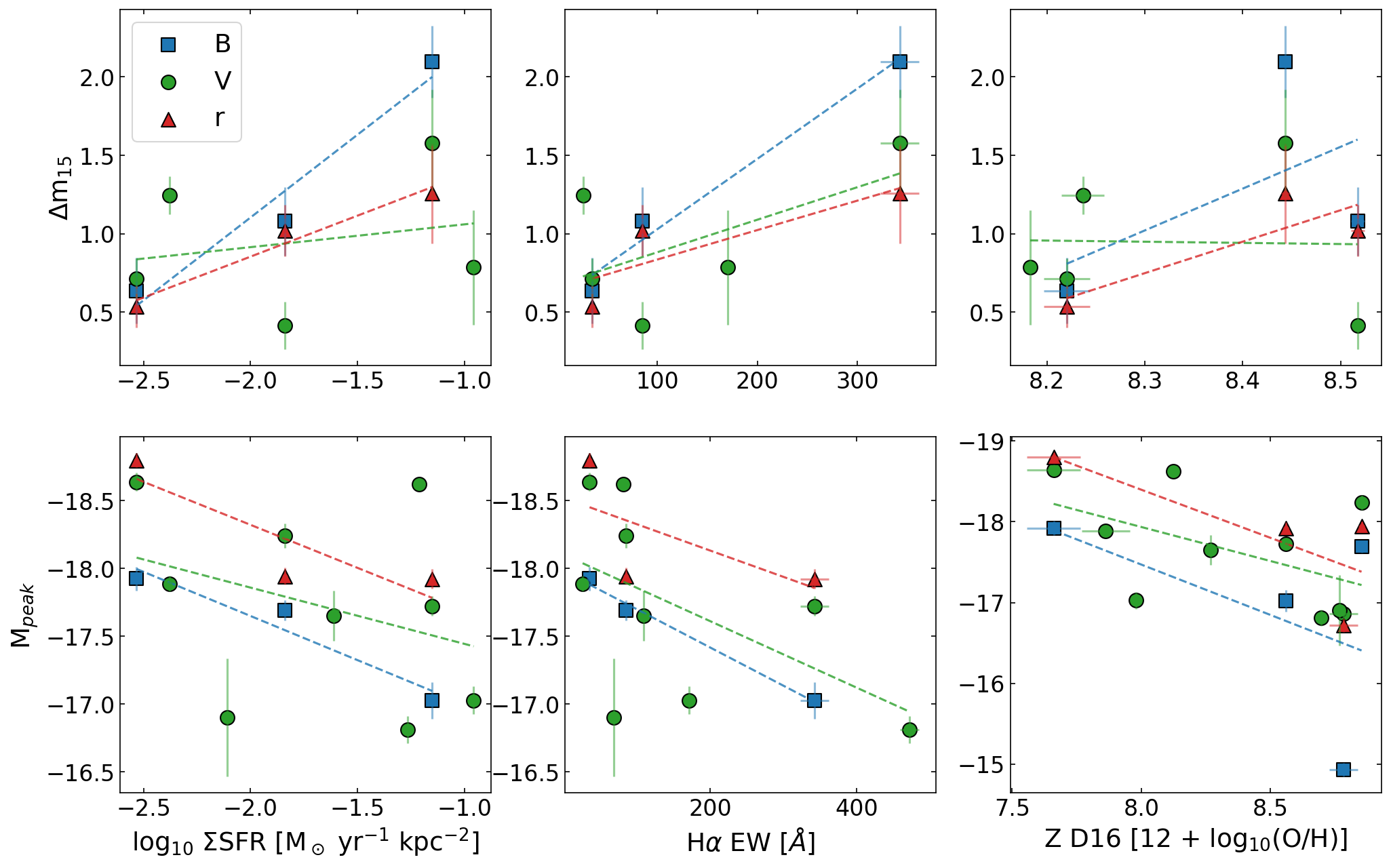}
\caption{Brightness difference between peak and 15 days later and absolute magnitude at peak (M$_{peak}$) for SESNe in $B$, $V$, and $r$ bands, as a function of $\Sigma$SFR (left), H$\alpha$ EW (middle), and the oxygen abundance given by the D16 indicator (right).  \label{fig:SESNe_LCs}} 
\end{figure*}

\begin{figure*}[t!]
\centering
\includegraphics[width=0.9\textwidth]{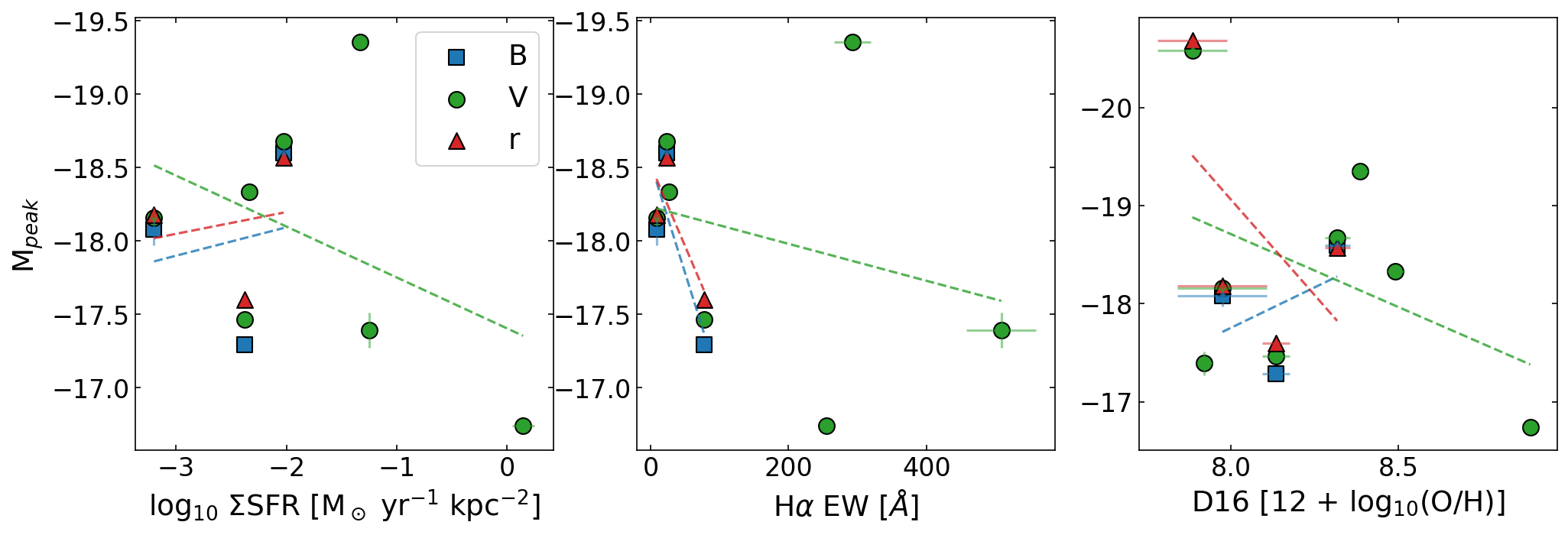}
\caption{SNe IIn absolute magnitude at peak (M$_{peak}$) in $B$, $V$, and $r$ bands, as a function of $\Sigma$SFR (left), H$\alpha$ EW (middle), and the oxygen abundance given by the D16 indicator (right).  \label{fig:int_LCs}} 
\end{figure*}

\section{Conclusions} \label{sec:conc}

In this work, we analyzed 112 CCSN environments observed with the MUSE instrument on the VLT. 
By using a untargeted sample of galaxies that hosted CCSNe detected by the ASAS-SN survey, we were able to make a minimally target-biased analysis of their environments, with a larger control of survey selection systematics than previous studies.  
We used the {\ion{H}{ii} regions centered at the SN positions} to derive physical properties of their environments, such as H$\alpha$ EW, $\Sigma$SFR, oxygen abundance, and host extinction. Our main results can be summarized as follows:

\begin{itemize}

    \item SESNe occur in environments with higher median H$\alpha$ EW, $\Sigma$SFR, and oxygen abundance than SNe II and IIn/Ibn. However, a statistically significant difference between the cumulative distributions is found only for H$\alpha$ EW between SNe~II and SESNe and for oxygen abundances {in the D16 index} between SNe~II and Ic.

    \item Within SESNe, SNe Ic show higher median SFR, H$\alpha$ EW, and oxygen abundance than SNe Ib, although no statistically significant difference between the distributions is found. This could suggests that SNe Ic are produced by higher mass and more metal rich progenitors than SNe Ib.
    
    \item The environments of SNe IIb have similar {median} H$\alpha$ EW and SFR to SNe Ib SNe, and similar oxygen abundance {medians} to SNe Ic. 
    
    \item The properties of SNe Ic are consistent with higher mass and higher metallicity progenitors, while SNe Ib might be connected to lower mass and lower metallicity populations. SNe IIb are consistent with relatively lower mass progenitors in higher metallicity environments, but a mixture of single stars and binary systems might be possible.
    
    \item {SNe IIn} have similar distributions to SNe II, suggesting that they occur in environments with similar physical properties. SNe Ibn have a median oxygen abundance similar to SNe IIb and Ic, and $\Sigma$SFR similar to SNe IIn.
    
    \item The distributions of host-galaxy extinctions are very similar for the different CCSN types, suggesting that different CCSN types suffer from similar degrees of host extinction.

    \item For SNe II, a correlation was found between the postmaximum decline rate in $B$ and $r$ bands to H$\alpha$ EW, with faster-declining events happening in environments with lower H$\alpha$ EW. However only a few points are used in this analysis.
    No further correlation of LC properties of SNe II to their environments was found.

    \item For SESNe, weak correlations are found between $B$-band peak magnitude and $\Delta m_{15}$ to H$\alpha$ EW and $\Sigma$SFR. Brighter events seem to occur in environments with lower H$\alpha$ EW and $\Sigma$SFR, and SESNe with higher $\Delta m_{15}$ are connected to higher H$\alpha$ EW and $\Sigma$SFR.
    The $BVr$ peak luminosity of SESNe is also weakly correlated to the oxygen abundance at their environments, with brighter events happening in locations with less oxygen content. 
    This suggests an intrinsic relation between the metal content, the star formation intensity, and stellar age at the environments of SESNe and the amount of synthesized $^{56}$Ni.

\end{itemize}

Although some contrasts to previous results are found in this work, this represents the most homogeneous sample of CCSN host galaxies to date. The low number statistics in our sample (specially for SESNe) might also have contributed to such differences. We aim to complement this analysis with new observations of SESN and SNe IIn/Ibn host galaxies obtained with MUSE, which will give more statistical significance to the study of the environments of these events.

As a legacy of this work, all the resultant physical parameter maps derived for the analyzed galaxies are available online\footnote{{\url{https://sites.google.com/view/theamusingasassnsample}}} or under request to the corresponding author.

Following the results presented here, the Paper II will present a global analysis and characterization of all star-forming regions within the CCSN host galaxies within our sample. 
Comparing SN host \ion{H}{ii}  region environment properties to all other \ion{H}{ii}  regions within their galaxies - and indeed the full sample of galaxies - one can ask whether different SN types prefer to explode from, for example, lower or higher metallicity star forming regions.
As a future work, we aim to use the data and physical quantities derived here to estimate the relative rates of the different CCSN types as a function of their environment properties, which will help in constraining the mass ranges of their progenitor stars.

\begin{acknowledgements}

      Based on observations collected at the European Southern Observatory under ESO programme(s): 
      096.D-0296 (A), 
      0103.D-0440 (A),
      096.D-0263 (A), 
         097.B-0165 (A),
         097.D-0408 (A),
        0104.D-0503 (A),
         60.A-9301 (A),
         096.D-0786 (A),
         097.D-1054 (B),
         0101.C-0329 (D),
         0100.D-0341 (A),
         1100.B-0651 (A),
         094.B-0298 (A),
         097.B-0640 (A),
         0101.D-0748 (A),
         095.D-0172 (A),
         1100.B-0651 (A),
         0100.D-0649 (F),
         096.B-0309 (A).
         
     This work makes use of the following softwares: ESO reduction pipeline \citep{2014ASPC..485..451W}, \textsc{EsoReflex} \citep{2013A&A...559A..96F}, \textsc{CosmicFlows} \citep{2015MNRAS.450..317C}, \textsc{Astropy} \citep{astropy:2013}, \textsc{SciPy} \citep{2020SciPy-NMeth}, \textsc{HyperLEDA} \citep{2014A&A...570A..13M}, \textsc{VizieR Queries} \citep{2019AJ....157...98G}, and \textsc{IFUanal} \citep{2018MNRAS.473.1359L}.   
     
     This research has made use of the NASA/IPAC Extragalactic Database (NED), which is funded by the National Aeronautics and Space Administration and operated by the California Institute of Technology.
     T.P. acknowledges the support by ANID through the Beca Doctorado Nacional 202221222222.
    J.L.P. acknowledges support by ANID through the Fondecyt regular grant 1191038 and through the Millennium Science Initiative grant ICN12\_009, awarded to The Millennium Institute of Astrophysics, MAS. L.G. acknowledges financial support from the Spanish Ministerio de Ciencia e Innovaci\'on (MCIN), the Agencia Estatal de Investigaci\'on (AEI) 10.13039/501100011033, and the European Social Fund (ESF) "Investing in your future" under the 2019 Ram\'on y Cajal program RYC2019-027683-I and the PID2020-115253GA-I00 HOSTFLOWS project, from Centro Superior de Investigaciones Cient\'ificas (CSIC) under the PIE project 20215AT016, and the program Unidad de Excelencia Mar\'ia de Maeztu CEX2020-001058-M.
      E.J.J. acknowledges support from FONDECYT Iniciación en investigación 2020 Project 11200263.
      J.D.L. acknowledges support from a UK Research and Innovation Fellowship(MR/T020784/1).
      S.D. acknowledge support from the National Natural Science Foundation of China (grant No. 12133005) and the XPLORER PRIZE.
      H.K. was funded by the Academy of Finland projects 324504 and 328898.
      F.F. acknowledges the Chilean Ministry of Economy, Development, and Tourism’s Millennium Science Initiative through grant ICN12 12009, awarded to the Millennium Institute of Astrophysics National Agency for Research and Development (ANID) grants: BASAL Center of Mathematical Modelling Grant PAI AFB-170001 FONDECYT Regular 1200710.
      Support for T.W.-S.H. was provided by NASA through the NASA Hubble Fellowship grant HST-HF2-51458.001-A awarded by the Space Telescope Science Institute (STScI), which is operated by the Association of Universities for Research in Astronomy, Inc., for NASA, under contract NAS5-26555.
      {We thank the referee for the very detailed and constructive report.}
      
\end{acknowledgements}

% WARNING
%-------------------------------------------------------------------
% Please note that we have included the references to the file aa.dem in
% order to compile it, but we ask you to:
%
% - use BibTeX with the regular commands:
  % \bibliographystyle{aa} % style aa.bst
%   \bibliography{Yourfile} % your references Yourfile.bib
%
% - join the .bib files when you upload your source files
%-------------------------------------------------------------------

% \begin{thebibliography}{}

\bibliographystyle{aa}
\bibliography{ref}

\begin{thebibliography}{172}
\expandafter\ifx\csname natexlab\endcsname\relax\def\natexlab#1{#1}\fi

\bibitem[{{Alloin} {et~al.}(1979){Alloin}, {Collin-Souffrin}, {Joly}, \&
  {Vigroux}}]{1979A&A....78..200A}
{Alloin}, D., {Collin-Souffrin}, S., {Joly}, M., \& {Vigroux}, L. 1979, \aap,
  78, 200

\bibitem[{{Anderson} {et~al.}(2010){Anderson}, {Covarrubias}, {James}, {Hamuy},
  \& {Habergham}}]{2010MNRAS.407.2660A}
{Anderson}, J.~P., {Covarrubias}, R.~A., {James}, P.~A., {Hamuy}, M., \&
  {Habergham}, S.~M. 2010, \mnras, 407, 2660

\bibitem[{{Anderson} {et~al.}(2014){Anderson}, {Gonz{\'a}lez-Gait{\'a}n},
  {Hamuy}, {Guti{\'e}rrez}, {Stritzinger}, {Olivares E.}, {Phillips},
  {Schulze}, {Antezana}, {Bolt}, {Campillay}, {Castell{\'o}n}, {Contreras}, {de
  Jaeger}, {Folatelli}, {F{\"o}rster}, {Freedman}, {Gonz{\'a}lez}, {Hsiao},
  {Krzemi{\'n}ski}, {Krisciunas}, {Maza}, {McCarthy}, {Morrell}, {Persson},
  {Roth}, {Salgado}, {Suntzeff}, \& {Thomas-Osip}}]{2014ApJ...786...67A}
{Anderson}, J.~P., {Gonz{\'a}lez-Gait{\'a}n}, S., {Hamuy}, M., {et~al.} 2014,
  \apj, 786, 67

\bibitem[{{Anderson} {et~al.}(2012){Anderson}, {Habergham}, {James}, \&
  {Hamuy}}]{2012MNRAS.424.1372A}
{Anderson}, J.~P., {Habergham}, S.~M., {James}, P.~A., \& {Hamuy}, M. 2012,
  \mnras, 424, 1372

\bibitem[{{Anderson} \& {James}(2008)}]{2008MNRAS.390.1527A}
{Anderson}, J.~P. \& {James}, P.~A. 2008, \mnras, 390, 1527

\bibitem[{{Anderson} {et~al.}(2015){Anderson}, {James}, {Habergham}, {Galbany},
  \& {Kuncarayakti}}]{2015PASA...32...19A}
{Anderson}, J.~P., {James}, P.~A., {Habergham}, S.~M., {Galbany}, L., \&
  {Kuncarayakti}, H. 2015, \pasa, 32, e019

\bibitem[{{Andrews} {et~al.}(2021){Andrews}, {Sand}, {Valenti}, {Smith},
  {Dastidar}, {Sahu}, {Misra}, {Singh}, {Hiramatsu}, {Brown}, {Hosseinzadeh},
  {Wyatt}, {Vinko}, {Anupama}, {Arcavi}, {Ashall}, {Benetti}, {Berton},
  {Bostroem}, {Bulla}, {Burke}, {Chen}, {Chomiuk}, {Cikota}, {Congiu}, {Cseh},
  {Davis}, {Elias-Rosa}, {Faran}, {Fraser}, {Galbany}, {Gall}, {Gal-Yam},
  {Gangopadhyay}, {Gromadzki}, {Haislip}, {Howell}, {Hsiao}, {Inserra},
  {Kankare}, {Kuncarayakti}, {Kouprianov}, {Kumar}, {Li}, {Lin}, {Maguire},
  {Mazzali}, {McCully}, {Milne}, {Mo}, {Morrell}, {Nicholl}, {Ochner},
  {Olivares}, {Pastorello}, {Patat}, {Phillips}, {Pignata}, {Prentice},
  {Reguitti}, {Reichart}, {Rodriguez}, {Rui}, {Sanwal}, {Sarneczky},
  {Shahbandeh}, {Singh}, {Smartt}, {Strader}, {Stritzinger}, {Szakats},
  {Tartaglia}, {Wang}, {Wang}, {Wang}, {Wheeler}, {Xiang}, {Yaron}, {Young}, \&
  {Zhang}}]{2021yCat..18850043A}
{Andrews}, J.~E., {Sand}, D.~J., {Valenti}, S., {et~al.} 2021, VizieR Online
  Data Catalog, J/ApJ/885/43

\bibitem[{{Arcavi} {et~al.}(2010){Arcavi}, {Gal-Yam}, {Kasliwal}, {Quimby},
  {Ofek}, {Kulkarni}, {Nugent}, {Cenko}, {Bloom}, {Sullivan}, {Howell},
  {Poznanski}, {Filippenko}, {Law}, {Hook}, {J{\"o}nsson}, {Blake}, {Cooke},
  {Dekany}, {Rahmer}, {Hale}, {Smith}, {Zolkower}, {Velur}, {Walters},
  {Henning}, {Bui}, {McKenna}, \& {Jacobsen}}]{2010ApJ...721..777A}
{Arcavi}, I., {Gal-Yam}, A., {Kasliwal}, M.~M., {et~al.} 2010, \apj, 721, 777

\bibitem[{{Arnett} {et~al.}(1989){Arnett}, {Bahcall}, {Kirshner}, \&
  {Woosley}}]{1989ARA&A..27..629A}
{Arnett}, W.~D., {Bahcall}, J.~N., {Kirshner}, R.~P., \& {Woosley}, S.~E. 1989,
  \araa, 27, 629

\bibitem[{{Astropy Collaboration} {et~al.}(2013){Astropy Collaboration},
  {Robitaille}, {Tollerud}, {Greenfield}, {Droettboom}, {Bray}, {Aldcroft},
  {Davis}, {Ginsburg}, {Price-Whelan}, {Kerzendorf}, {Conley}, {Crighton},
  {Barbary}, {Muna}, {Ferguson}, {Grollier}, {Parikh}, {Nair}, {Unther},
  {Deil}, {Woillez}, {Conseil}, {Kramer}, {Turner}, {Singer}, {Fox}, {Weaver},
  {Zabalza}, {Edwards}, {Azalee Bostroem}, {Burke}, {Casey}, {Crawford},
  {Dencheva}, {Ely}, {Jenness}, {Labrie}, {Lim}, {Pierfederici}, {Pontzen},
  {Ptak}, {Refsdal}, {Servillat}, \& {Streicher}}]{astropy:2013}
{Astropy Collaboration}, {Robitaille}, T.~P., {Tollerud}, E.~J., {et~al.} 2013,
  \aap, 558, A33

\bibitem[{{Bacon} {et~al.}(2014){Bacon}, {Vernet}, {Borisova}, {Bouch{\'e}},
  {Brinchmann}, {Carollo}, {Carton}, {Caruana}, {Cerda}, {Contini}, {Franx},
  {Girard}, {Guerou}, {Haddad}, {Hau}, {Herenz}, {Herrera}, {Husemann},
  {Husser}, {Jarno}, {Kamann}, {Krajnovic}, {Lilly}, {Mainieri}, {Martinsson},
  {Palsa}, {Patricio}, {P{\'e}contal}, {Pello}, {Piqueras}, {Richard},
  {Sandin}, {Schroetter}, {Selman}, {Shirazi}, {Smette}, {Soto}, {Streicher},
  {Urrutia}, {Weilbacher}, {Wisotzki}, \& {Zins}}]{2014Msngr.157...13B}
{Bacon}, R., {Vernet}, J., {Borisova}, E., {et~al.} 2014, The Messenger, 157,
  13

\bibitem[{{Baldwin} {et~al.}(1981){Baldwin}, {Phillips}, \&
  {Terlevich}}]{1981PASP...93....5B}
{Baldwin}, J.~A., {Phillips}, M.~M., \& {Terlevich}, R. 1981, \pasp, 93, 5

\bibitem[{{Ben-Ami} {et~al.}(2023){Ben-Ami}, {Arcavi}, {Newsome}, {Farah},
  {Pellegrino}, {Terreran}, {Burke}, {Hosseinzadeh}, {McCully}, {Hiramatsu},
  {Gonzalez}, \& {Howell}}]{2023ApJ...946...30B}
{Ben-Ami}, T., {Arcavi}, I., {Newsome}, M., {et~al.} 2023, \apj, 946, 30

\bibitem[{{Bersten} {et~al.}(2011){Bersten}, {Benvenuto}, \&
  {Hamuy}}]{2011ApJ...729...61B}
{Bersten}, M.~C., {Benvenuto}, O., \& {Hamuy}, M. 2011, \apj, 729, 61

\bibitem[{{Bethe}(1990)}]{1990RvMP...62..801B}
{Bethe}, H.~A. 1990, Reviews of Modern Physics, 62, 801

\bibitem[{{Bethe} {et~al.}(1979){Bethe}, {Brown}, {Applegate}, \&
  {Lattimer}}]{1979NuPhA.324..487B}
{Bethe}, H.~A., {Brown}, G.~E., {Applegate}, J., \& {Lattimer}, J.~M. 1979,
  \nphysa, 324, 487

\bibitem[{{Boissier} \& {Prantzos}(2009)}]{2009A&A...503..137B}
{Boissier}, S. \& {Prantzos}, N. 2009, \aap, 503, 137

\bibitem[{{Bravo} {et~al.}(2019){Bravo}, {Badenes}, \&
  {Mart{\'\i}nez-Rodr{\'\i}guez}}]{2019MNRAS.482.4346B}
{Bravo}, E., {Badenes}, C., \& {Mart{\'\i}nez-Rodr{\'\i}guez}, H. 2019, \mnras,
  482, 4346

\bibitem[{{Brown} {et~al.}(2019){Brown}, {Stanek}, {Holoien}, {Kochanek},
  {Shappee}, {Prieto}, {Dong}, {Chen}, {Thompson}, {Beacom}, {Stritzinger},
  {Bersier}, \& {Brimacombe}}]{2019MNRAS.484.3785B}
{Brown}, J.~S., {Stanek}, K.~Z., {Holoien}, T.~W.~S., {et~al.} 2019, \mnras,
  484, 3785

\bibitem[{{Bruzual} \& {Charlot}(2003)}]{2003MNRAS.344.1000B}
{Bruzual}, G. \& {Charlot}, S. 2003, \mnras, 344, 1000

\bibitem[{{Burrows} \& {Vartanyan}(2021)}]{2021Natur.589...29B}
{Burrows}, A. \& {Vartanyan}, D. 2021, \nat, 589, 29

\bibitem[{{Cano}(2013)}]{2013MNRAS.434.1098C}
{Cano}, Z. 2013, \mnras, 434, 1098

\bibitem[{{Cardelli} {et~al.}(1989){Cardelli}, {Clayton}, \&
  {Mathis}}]{1989ApJ...345..245C}
{Cardelli}, J.~A., {Clayton}, G.~C., \& {Mathis}, J.~S. 1989, \apj, 345, 245

\bibitem[{{Carrick} {et~al.}(2015){Carrick}, {Turnbull}, {Lavaux}, \&
  {Hudson}}]{2015MNRAS.450..317C}
{Carrick}, J., {Turnbull}, S.~J., {Lavaux}, G., \& {Hudson}, M.~J. 2015,
  \mnras, 450, 317

\bibitem[{{Chabrier}(2003)}]{2003PASP..115..763C}
{Chabrier}, G. 2003, \pasp, 115, 763

\bibitem[{{Chambers} {et~al.}(2016){Chambers}, {Magnier}, {Metcalfe},
  {Flewelling}, {Huber}, {Waters}, {Denneau}, {Draper}, {Farrow}, {Finkbeiner},
  {Holmberg}, {Koppenhoefer}, {Price}, {Rest}, {Saglia}, {Schlafly}, {Smartt},
  {Sweeney}, {Wainscoat}, {Burgett}, {Chastel}, {Grav}, {Heasley}, {Hodapp},
  {Jedicke}, {Kaiser}, {Kudritzki}, {Luppino}, {Lupton}, {Monet}, {Morgan},
  {Onaka}, {Shiao}, {Stubbs}, {Tonry}, {White}, {Ba{\~n}ados}, {Bell},
  {Bender}, {Bernard}, {Boegner}, {Boffi}, {Botticella}, {Calamida},
  {Casertano}, {Chen}, {Chen}, {Cole}, {Deacon}, {Frenk}, {Fitzsimmons},
  {Gezari}, {Gibbs}, {Goessl}, {Goggia}, {Gourgue}, {Goldman}, {Grant},
  {Grebel}, {Hambly}, {Hasinger}, {Heavens}, {Heckman}, {Henderson}, {Henning},
  {Holman}, {Hopp}, {Ip}, {Isani}, {Jackson}, {Keyes}, {Koekemoer}, {Kotak},
  {Le}, {Liska}, {Long}, {Lucey}, {Liu}, {Martin}, {Masci}, {McLean}, {Mindel},
  {Misra}, {Morganson}, {Murphy}, {Obaika}, {Narayan}, {Nieto-Santisteban},
  {Norberg}, {Peacock}, {Pier}, {Postman}, {Primak}, {Rae}, {Rai}, {Riess},
  {Riffeser}, {Rix}, {R{\"o}ser}, {Russel}, {Rutz}, {Schilbach}, {Schultz},
  {Scolnic}, {Strolger}, {Szalay}, {Seitz}, {Small}, {Smith}, {Soderblom},
  {Taylor}, {Thomson}, {Taylor}, {Thakar}, {Thiel}, {Thilker}, {Unger},
  {Urata}, {Valenti}, {Wagner}, {Walder}, {Walter}, {Watters}, {Werner},
  {Wood-Vasey}, \& {Wyse}}]{2016arXiv161205560C}
{Chambers}, K.~C., {Magnier}, E.~A., {Metcalfe}, N., {et~al.} 2016, arXiv
  e-prints, arXiv:1612.05560

\bibitem[{{Cid Fernandes} {et~al.}(2005){Cid Fernandes}, {Mateus}, {Sodr{\'e}},
  {Stasi{\'n}ska}, \& {Gomes}}]{2005MNRAS.358..363C}
{Cid Fernandes}, R., {Mateus}, A., {Sodr{\'e}}, L., {Stasi{\'n}ska}, G., \&
  {Gomes}, J.~M. 2005, \mnras, 358, 363

\bibitem[{{Cronin} {et~al.}(2021){Cronin}, {Utomo}, {Leroy}, {Behrens},
  {Chastenet}, {Holland-Ashford}, {Koch}, {Lopez}, {Sandstrom}, \&
  {Williams}}]{2021ApJ...923...86C}
{Cronin}, S.~A., {Utomo}, D., {Leroy}, A.~K., {et~al.} 2021, \apj, 923, 86

\bibitem[{{Crowther}(2013)}]{2013MNRAS.428.1927C}
{Crowther}, P.~A. 2013, \mnras, 428, 1927

\bibitem[{{Dom{\'\i}nguez} {et~al.}(2013){Dom{\'\i}nguez}, {Siana}, {Henry},
  {Scarlata}, {Bedregal}, {Malkan}, {Atek}, {Ross}, {Colbert}, {Teplitz},
  {Rafelski}, {McCarthy}, {Bunker}, {Hathi}, {Dressler}, {Martin}, \&
  {Masters}}]{2013ApJ...763..145D}
{Dom{\'\i}nguez}, A., {Siana}, B., {Henry}, A.~L., {et~al.} 2013, \apj, 763,
  145

\bibitem[{{Dopita} {et~al.}(2016){Dopita}, {Kewley}, {Sutherland}, \&
  {Nicholls}}]{2016Ap&SS.361...61D}
{Dopita}, M.~A., {Kewley}, L.~J., {Sutherland}, R.~S., \& {Nicholls}, D.~C.
  2016, \apss, 361, 61

\bibitem[{{Drout} {et~al.}(2011){Drout}, {Soderberg}, {Gal-Yam}, {Cenko},
  {Fox}, {Leonard}, {Sand}, {Moon}, {Arcavi}, \& {Green}}]{2011ApJ...741...97D}
{Drout}, M.~R., {Soderberg}, A.~M., {Gal-Yam}, A., {et~al.} 2011, \apj, 741, 97

\bibitem[{{Eldridge} {et~al.}(2008){Eldridge}, {Izzard}, \&
  {Tout}}]{2008MNRAS.384.1109E}
{Eldridge}, J.~J., {Izzard}, R.~G., \& {Tout}, C.~A. 2008, \mnras, 384, 1109

\bibitem[{{Eldridge} {et~al.}(2017){Eldridge}, {Stanway}, {Xiao}, {McClelland},
  {Taylor}, {Ng}, {Greis}, \& {Bray}}]{2017PASA...34...58E}
{Eldridge}, J.~J., {Stanway}, E.~R., {Xiao}, L., {et~al.} 2017, \pasa, 34, e058

\bibitem[{{Elias-Rosa} {et~al.}(2016){Elias-Rosa}, {Pastorello}, {Benetti},
  {Cappellaro}, {Taubenberger}, {Terreran}, {Fraser}, {Brown}, {Tartaglia},
  {Morales-Garoffolo}, {Harmanen}, {Richardson}, {Artigau}, {Tomasella},
  {Margutti}, {Smartt}, {Dennefeld}, {Turatto}, {Anupama}, {Arbour}, {Berton},
  {Bjorkman}, {Boles}, {Briganti}, {Chornock}, {Ciabattari}, {Cortini},
  {Dimai}, {Gerhartz}, {Itagaki}, {Kotak}, {Mancini}, {Martinelli},
  {Milisavljevic}, {Misra}, {Ochner}, {Patnaude}, {Polshaw}, {Sahu}, \&
  {Zaggia}}]{2016MNRAS.463.3894E}
{Elias-Rosa}, N., {Pastorello}, A., {Benetti}, S., {et~al.} 2016, \mnras, 463,
  3894

\bibitem[{{Fang} {et~al.}(2022){Fang}, {Maeda}, {Kuncarayakti}, {Tanaka},
  {Kawabata}, {Hattori}, {Aoki}, {Moriya}, \& {Yamanaka}}]{2022ApJ...928..151F}
{Fang}, Q., {Maeda}, K., {Kuncarayakti}, H., {et~al.} 2022, \apj, 928, 151

\bibitem[{{Fassia} {et~al.}(2000){Fassia}, {Meikle}, {Vacca}, {Kemp}, {Walton},
  {Pollacco}, {Smartt}, {Oscoz}, {Arag{\'o}n-Salamanca}, {Bennett}, {Hawarden},
  {Alonso}, {Alcalde}, {Pedrosa}, {Telting}, {Arevalo}, {Deeg}, {Garz{\'o}n},
  {G{\'o}mez-Rold{\'a}n}, {G{\'o}mez}, {Guti{\'e}rrez}, {L{\'o}pez}, {Rozas},
  {Serra-Ricart}, \& {Zapatero-Osorio}}]{2000MNRAS.318.1093F}
{Fassia}, A., {Meikle}, W.~P.~S., {Vacca}, W.~D., {et~al.} 2000, \mnras, 318,
  1093

\bibitem[{{Filippenko}(1997)}]{1997ARA&A..35..309F}
{Filippenko}, A.~V. 1997, \araa, 35, 309

\bibitem[{{Filippenko} {et~al.}(1994){Filippenko}, {Matheson}, \&
  {Barth}}]{1994AJ....108.2220F}
{Filippenko}, A.~V., {Matheson}, T., \& {Barth}, A.~J. 1994, \aj, 108, 2220

\bibitem[{{Flaugher}(2005)}]{2005IJMPA..20.3121F}
{Flaugher}, B. 2005, International Journal of Modern Physics A, 20, 3121

\bibitem[{{Flewelling} {et~al.}(2020){Flewelling}, {Magnier}, {Chambers},
  {Heasley}, {Holmberg}, {Huber}, {Sweeney}, {Waters}, {Calamida}, {Casertano},
  {Chen}, {Farrow}, {Hasinger}, {Henderson}, {Long}, {Metcalfe}, {Narayan},
  {Nieto-Santisteban}, {Norberg}, {Rest}, {Saglia}, {Szalay}, {Thakar},
  {Tonry}, {Valenti}, {Werner}, {White}, {Denneau}, {Draper}, {Hodapp},
  {Jedicke}, {Kaiser}, {Kudritzki}, {Price}, {Wainscoat}, {Chastel}, {McLean},
  {Postman}, \& {Shiao}}]{2020ApJS..251....7F}
{Flewelling}, H.~A., {Magnier}, E.~A., {Chambers}, K.~C., {et~al.} 2020, \apjs,
  251, 7

\bibitem[{{Folatelli} {et~al.}(2015){Folatelli}, {Bersten}, {Kuncarayakti},
  {Benvenuto}, {Maeda}, \& {Nomoto}}]{2015ApJ...811..147F}
{Folatelli}, G., {Bersten}, M.~C., {Kuncarayakti}, H., {et~al.} 2015, \apj,
  811, 147

\bibitem[{{Fox} {et~al.}(2014){Fox}, {Azalee Bostroem}, {Van Dyk},
  {Filippenko}, {Fransson}, {Matheson}, {Cenko}, {Chandra}, {Dwarkadas}, {Li},
  {Parker}, \& {Smith}}]{2014ApJ...790...17F}
{Fox}, O.~D., {Azalee Bostroem}, K., {Van Dyk}, S.~D., {et~al.} 2014, \apj,
  790, 17

\bibitem[{{Fox} {et~al.}(2022){Fox}, {Van Dyk}, {Williams}, {Drout},
  {Zapartas}, {Smith}, {Milisavljevic}, {Andrews}, {Bostroem}, {Filippenko},
  {Gomez}, {Kelly}, {de Mink}, {Pierel}, {Rest}, {Ryder}, {Sravan}, {Strolger},
  {Wang}, \& {Weil}}]{2022ApJ...929L..15F}
{Fox}, O.~D., {Van Dyk}, S.~D., {Williams}, B.~F., {et~al.} 2022, \apjl, 929,
  L15

\bibitem[{{Fransson} {et~al.}(2002){Fransson}, {Chevalier}, {Filippenko},
  {Leibundgut}, {Barth}, {Fesen}, {Kirshner}, {Leonard}, {Li}, {Lundqvist},
  {Sollerman}, \& {Van Dyk}}]{2002ApJ...572..350F}
{Fransson}, C., {Chevalier}, R.~A., {Filippenko}, A.~V., {et~al.} 2002, \apj,
  572, 350

\bibitem[{{Fraser}(2016)}]{2016MNRAS.456L..16F}
{Fraser}, M. 2016, \mnras, 456, L16

\bibitem[{{Freudling} {et~al.}(2013){Freudling}, {Romaniello}, {Bramich},
  {Ballester}, {Forchi}, {Garc{\'{\i}}a-Dabl{\'o}}, {Moehler}, \&
  {Neeser}}]{2013A&A...559A..96F}
{Freudling}, W., {Romaniello}, M., {Bramich}, D.~M., {et~al.} 2013, \aap, 559,
  A96

\bibitem[{{Gaia Collaboration} {et~al.}(2018){Gaia Collaboration}, {Brown},
  {Vallenari}, {Prusti}, {de Bruijne}, {Babusiaux}, {Bailer-Jones}, {Biermann},
  {Evans}, {Eyer}, {Jansen}, {Jordi}, {Klioner}, {Lammers}, {Lindegren},
  {Luri}, {Mignard}, {Panem}, {Pourbaix}, {Randich}, {Sartoretti}, {Siddiqui},
  {Soubiran}, {van Leeuwen}, {Walton}, {Arenou}, {Bastian}, {Cropper},
  {Drimmel}, {Katz}, {Lattanzi}, {Bakker}, {Cacciari}, {Casta{\~n}eda},
  {Chaoul}, {Cheek}, {De Angeli}, {Fabricius}, {Guerra}, {Holl}, {Masana},
  {Messineo}, {Mowlavi}, {Nienartowicz}, {Panuzzo}, {Portell}, {Riello},
  {Seabroke}, {Tanga}, {Th{\'e}venin}, {Gracia-Abril}, {Comoretto},
  {Garcia-Reinaldos}, {Teyssier}, {Altmann}, {Andrae}, {Audard},
  {Bellas-Velidis}, {Benson}, {Berthier}, {Blomme}, {Burgess}, {Busso},
  {Carry}, {Cellino}, {Clementini}, {Clotet}, {Creevey}, {Davidson}, {De
  Ridder}, {Delchambre}, {Dell'Oro}, {Ducourant},
  {Fern{\'a}ndez-Hern{\'a}ndez}, {Fouesneau}, {Fr{\'e}mat}, {Galluccio},
  {Garc{\'\i}a-Torres}, {Gonz{\'a}lez-N{\'u}{\~n}ez}, {Gonz{\'a}lez-Vidal},
  {Gosset}, {Guy}, {Halbwachs}, {Hambly}, {Harrison}, {Hern{\'a}ndez},
  {Hestroffer}, {Hodgkin}, {Hutton}, {Jasniewicz}, {Jean-Antoine-Piccolo},
  {Jordan}, {Korn}, {Krone-Martins}, {Lanzafame}, {Lebzelter}, {L{\"o}ffler},
  {Manteiga}, {Marrese}, {Mart{\'\i}n-Fleitas}, {Moitinho}, {Mora}, {Muinonen},
  {Osinde}, {Pancino}, {Pauwels}, {Petit}, {Recio-Blanco}, {Richards},
  {Rimoldini}, {Robin}, {Sarro}, {Siopis}, {Smith}, {Sozzetti}, {S{\"u}veges},
  {Torra}, {van Reeven}, {Abbas}, {Abreu Aramburu}, {Accart}, {Aerts},
  {Altavilla}, {{\'A}lvarez}, {Alvarez}, {Alves}, {Anderson}, {Andrei},
  {Anglada Varela}, {Antiche}, {Antoja}, {Arcay}, {Astraatmadja}, {Bach},
  {Baker}, {Balaguer-N{\'u}{\~n}ez}, {Balm}, {Barache}, {Barata}, {Barbato},
  {Barblan}, {Barklem}, {Barrado}, {Barros}, {Barstow}, {Bartholom{\'e}
  Mu{\~n}oz}, {Bassilana}, {Becciani}, {Bellazzini}, {Berihuete}, {Bertone},
  {Bianchi}, {Bienaym{\'e}}, {Blanco-Cuaresma}, {Boch}, {Boeche}, {Bombrun},
  {Borrachero}, {Bossini}, {Bouquillon}, {Bourda}, {Bragaglia}, {Bramante},
  {Breddels}, {Bressan}, {Brouillet}, {Br{\"u}semeister}, {Brugaletta},
  {Bucciarelli}, {Burlacu}, {Busonero}, {Butkevich}, {Buzzi}, {Caffau},
  {Cancelliere}, {Cannizzaro}, {Cantat-Gaudin}, {Carballo}, {Carlucci},
  {Carrasco}, {Casamiquela}, {Castellani}, {Castro-Ginard}, {Charlot},
  {Chemin}, {Chiavassa}, {Cocozza}, {Costigan}, {Cowell}, {Crifo}, {Crosta},
  {Crowley}, {Cuypers}, {Dafonte}, {Damerdji}, {Dapergolas}, {David}, {David},
  {de Laverny}, {De Luise}, {De March}, {de Martino}, {de Souza}, {de Torres},
  {Debosscher}, {del Pozo}, {Delbo}, {Delgado}, {Delgado}, {Di Matteo},
  {Diakite}, {Diener}, {Distefano}, {Dolding}, {Drazinos}, {Dur{\'a}n},
  {Edvardsson}, {Enke}, {Eriksson}, {Esquej}, {Eynard Bontemps}, {Fabre},
  {Fabrizio}, {Faigler}, {Falc{\~a}o}, {Farr{\`a}s Casas}, {Federici},
  {Fedorets}, {Fernique}, {Figueras}, {Filippi}, {Findeisen}, {Fonti},
  {Fraile}, {Fraser}, {Fr{\'e}zouls}, {Gai}, {Galleti}, {Garabato},
  {Garc{\'\i}a-Sedano}, {Garofalo}, {Garralda}, {Gavel}, {Gavras}, {Gerssen},
  {Geyer}, {Giacobbe}, {Gilmore}, {Girona}, {Giuffrida}, {Glass}, {Gomes},
  {Granvik}, {Gueguen}, {Guerrier}, {Guiraud}, {Guti{\'e}rrez-S{\'a}nchez},
  {Haigron}, {Hatzidimitriou}, {Hauser}, {Haywood}, {Heiter}, {Helmi}, {Heu},
  {Hilger}, {Hobbs}, {Hofmann}, {Holland}, {Huckle}, {Hypki}, {Icardi},
  {Jan{\ss}en}, {Jevardat de Fombelle}, {Jonker}, {Juh{\'a}sz}, {Julbe},
  {Karampelas}, {Kewley}, {Klar}, {Kochoska}, {Kohley}, {Kolenberg},
  {Kontizas}, {Kontizas}, {Koposov}, {Kordopatis}, {Kostrzewa-Rutkowska},
  {Koubsky}, {Lambert}, {Lanza}, {Lasne}, {Lavigne}, {Le Fustec}, {Le
  Poncin-Lafitte}, {Lebreton}, {Leccia}, {Leclerc}, {Lecoeur-Taibi},
  {Lenhardt}, {Leroux}, {Liao}, {Licata}, {Lindstr{\o}m}, {Lister}, {Livanou},
  {Lobel}, {L{\'o}pez}, {Managau}, {Mann}, {Mantelet}, {Marchal}, {Marchant},
  {Marconi}, {Marinoni}, {Marschalk{\'o}}, {Marshall}, {Martino}, {Marton},
  {Mary}, {Massari}, {Matijevi{\v{c}}}, {Mazeh}, {McMillan}, {Messina},
  {Michalik}, {Millar}, {Molina}, {Molinaro}, {Moln{\'a}r}, {Montegriffo},
  {Mor}, {Morbidelli}, {Morel}, {Morris}, {Mulone}, {Muraveva}, {Musella},
  {Nelemans}, {Nicastro}, {Noval}, {O'Mullane}, {Ord{\'e}novic},
  {Ord{\'o}{\~n}ez-Blanco}, {Osborne}, {Pagani}, {Pagano}, {Pailler},
  {Palacin}, {Palaversa}, {Panahi}, {Pawlak}, {Piersimoni}, {Pineau}, {Plachy},
  {Plum}, {Poggio}, {Poujoulet}, {Pr{\v{s}}a}, {Pulone}, {Racero}, {Ragaini},
  {Rambaux}, {Ramos-Lerate}, {Regibo}, {Reyl{\'e}}, {Riclet}, {Ripepi}, {Riva},
  {Rivard}, {Rixon}, {Roegiers}, {Roelens}, {Romero-G{\'o}mez}, {Rowell},
  {Royer}, {Ruiz-Dern}, {Sadowski}, {Sagrist{\`a} Sell{\'e}s}, {Sahlmann},
  {Salgado}, {Salguero}, {Sanna}, {Santana-Ros}, {Sarasso}, {Savietto},
  {Schultheis}, {Sciacca}, {Segol}, {Segovia}, {S{\'e}gransan}, {Shih},
  {Siltala}, {Silva}, {Smart}, {Smith}, {Solano}, {Solitro}, {Sordo}, {Soria
  Nieto}, {Souchay}, {Spagna}, {Spoto}, {Stampa}, {Steele},
  {Steidelm{\"u}ller}, {Stephenson}, {Stoev}, {Suess}, {Surdej}, {Szabados},
  {Szegedi-Elek}, {Tapiador}, {Taris}, {Tauran}, {Taylor}, {Teixeira},
  {Terrett}, {Teyssandier}, {Thuillot}, {Titarenko}, {Torra Clotet}, {Turon},
  {Ulla}, {Utrilla}, {Uzzi}, {Vaillant}, {Valentini}, {Valette}, {van Elteren},
  {Van Hemelryck}, {van Leeuwen}, {Vaschetto}, {Vecchiato}, {Veljanoski},
  {Viala}, {Vicente}, {Vogt}, {von Essen}, {Voss}, {Votruba}, {Voutsinas},
  {Walmsley}, {Weiler}, {Wertz}, {Wevers}, {Wyrzykowski}, {Yoldas},
  {{\v{Z}}erjal}, {Ziaeepour}, {Zorec}, {Zschocke}, {Zucker}, {Zurbach}, \&
  {Zwitter}}]{2018A&A...616A...1G}
{Gaia Collaboration}, {Brown}, A.~G.~A., {Vallenari}, A., {et~al.} 2018, \aap,
  616, A1

\bibitem[{{Gal-Yam}(2017)}]{2017hsn..book..195G}
{Gal-Yam}, A. 2017, in Handbook of Supernovae, ed. A.~W. {Alsabti} \&
  P.~{Murdin}, 195

\bibitem[{{Gal-Yam} \& {Leonard}(2009)}]{2009Natur.458..865G}
{Gal-Yam}, A. \& {Leonard}, D.~C. 2009, \nat, 458, 865

\bibitem[{{Gal-Yam} {et~al.}(2021){Gal-Yam}, {Yaron}, {Pastorello},
  {Taubenberger}, {Fraser}, \& {Perley}}]{2021TNSAN..76....1G}
{Gal-Yam}, A., {Yaron}, O., {Pastorello}, A., {et~al.} 2021, Transient Name
  Server AstroNote, 76, 1

\bibitem[{{Galbany} {et~al.}(2016{\natexlab{a}}){Galbany}, {Anderson},
  {Rosales-Ortega}, {Kuncarayakti}, {Kr{\"u}hler}, {S{\'a}nchez},
  {Falc{\'o}n-Barroso}, {P{\'e}rez}, {Maureira}, {Hamuy},
  {Gonz{\'a}lez-Gait{\'a}n}, {F{\"o}rster}, \& {Moral}}]{2016MNRAS.455.4087G}
{Galbany}, L., {Anderson}, J.~P., {Rosales-Ortega}, F.~F., {et~al.}
  2016{\natexlab{a}}, \mnras, 455, 4087

\bibitem[{{Galbany} {et~al.}(2018){Galbany}, {Anderson}, {S{\'a}nchez},
  {Kuncarayakti}, {Pedraz}, {Gonz{\'a}lez-Gait{\'a}n}, {Stanishev},
  {Dom{\'\i}nguez}, {Moreno-Raya}, {Wood-Vasey}, {Mour{\~a}o}, {Ponder},
  {Badenes}, {Moll{\'a}}, {L{\'o}pez-S{\'a}nchez}, {Rosales-Ortega},
  {V{\'\i}lchez}, {Garc{\'\i}a-Benito}, \& {Marino}}]{2018ApJ...855..107G}
{Galbany}, L., {Anderson}, J.~P., {S{\'a}nchez}, S.~F., {et~al.} 2018, \apj,
  855, 107

\bibitem[{{Galbany} {et~al.}(2016{\natexlab{b}}){Galbany}, {Hamuy}, {Phillips},
  {Suntzeff}, {Maza}, {de Jaeger}, {Moraga}, {Gonz{\'a}lez-Gait{\'a}n},
  {Krisciunas}, {Morrell}, {Thomas-Osip}, {Krzeminski}, {Gonz{\'a}lez},
  {Antezana}, {Wishnjewski}, {McCarthy}, {Anderson}, {Guti{\'e}rrez},
  {Stritzinger}, {Folatelli}, {Anguita}, {Galaz}, {Green}, {Impey}, {Kim},
  {Kirhakos}, {Malkan}, {Mulchaey}, {Phillips}, {Pizzella}, {Prosser},
  {Schmidt}, {Schommer}, {Sherry}, {Strolger}, {Wells}, \&
  {Williger}}]{2016AJ....151...33G}
{Galbany}, L., {Hamuy}, M., {Phillips}, M.~M., {et~al.} 2016{\natexlab{b}},
  \aj, 151, 33

\bibitem[{{Galbany} {et~al.}(2017){Galbany}, {Mora}, {Gonz{\'a}lez-Gait{\'a}n},
  {Bolatto}, {Dannerbauer}, {L{\'o}pez-S{\'a}nchez}, {Maeda}, {P{\'e}rez},
  {P{\'e}rez-Torres}, {S{\'a}nchez}, {Wong}, {Badenes}, {Blitz}, {Marino},
  {Utomo}, \& {Van de Ven}}]{2017MNRAS.468..628G}
{Galbany}, L., {Mora}, L., {Gonz{\'a}lez-Gait{\'a}n}, S., {et~al.} 2017,
  \mnras, 468, 628

\bibitem[{{Galbany} {et~al.}(2014){Galbany}, {Stanishev}, {Mour{\~a}o},
  {Rodrigues}, {Flores}, {Garc{\'\i}a-Benito}, {Mast}, {Mendoza},
  {S{\'a}nchez}, {Badenes}, {Barrera-Ballesteros}, {Bland-Hawthorn},
  {Falc{\'o}n-Barroso}, {Garc{\'\i}a-Lorenzo}, {Gomes}, {Gonz{\'a}lez Delgado},
  {Kehrig}, {Lyubenova}, {L{\'o}pez-S{\'a}nchez}, {de Lorenzo-C{\'a}ceres},
  {Marino}, {Meidt}, {Moll{\'a}}, {Papaderos}, {P{\'e}rez-Torres},
  {Rosales-Ortega}, \& {van de Ven}}]{2014A&A...572A..38G}
{Galbany}, L., {Stanishev}, V., {Mour{\~a}o}, A.~M., {et~al.} 2014, \aap, 572,
  A38

\bibitem[{{Galbany} {et~al.}(2016{\natexlab{c}}){Galbany}, {Stanishev},
  {Mour{\~a}o}, {Rodrigues}, {Flores}, {Walcher}, {S{\'a}nchez},
  {Garc{\'\i}a-Benito}, {Mast}, {Badenes}, {Gonz{\'a}lez Delgado}, {Kehrig},
  {Lyubenova}, {Marino}, {Moll{\'a}}, {Meidt}, {P{\'e}rez}, {van de Ven}, \&
  {V{\'\i}lchez}}]{2016A&A...591A..48G}
{Galbany}, L., {Stanishev}, V., {Mour{\~a}o}, A.~M., {et~al.}
  2016{\natexlab{c}}, \aap, 591, A48

\bibitem[{{Georgy} {et~al.}(2009){Georgy}, {Meynet}, {Walder}, {Folini}, \&
  {Maeder}}]{2009A&A...502..611G}
{Georgy}, C., {Meynet}, G., {Walder}, R., {Folini}, D., \& {Maeder}, A. 2009,
  \aap, 502, 611

\bibitem[{{Ginsburg} {et~al.}(2019){Ginsburg}, {Sip{\H{o}}cz}, {Brasseur},
  {Cowperthwaite}, {Craig}, {Deil}, {Guillochon}, {Guzman}, {Liedtke}, {Lian
  Lim}, {Lockhart}, {Mommert}, {Morris}, {Norman}, {Parikh}, {Persson},
  {Robitaille}, {Segovia}, {Singer}, {Tollerud}, {de Val-Borro}, {Valtchanov},
  {Woillez}, {Astroquery Collaboration}, \& {a subset of astropy
  Collaboration}}]{2019AJ....157...98G}
{Ginsburg}, A., {Sip{\H{o}}cz}, B.~M., {Brasseur}, C.~E., {et~al.} 2019, \aj,
  157, 98

\bibitem[{{Guti{\'e}rrez} {et~al.}(2017){Guti{\'e}rrez}, {Anderson}, {Hamuy},
  {Morrell}, {Gonz{\'a}lez-Gaitan}, {Stritzinger}, {Phillips}, {Galbany},
  {Folatelli}, {Dessart}, {Contreras}, {Della Valle}, {Freedman}, {Hsiao},
  {Krisciunas}, {Madore}, {Maza}, {Suntzeff}, {Prieto}, {Gonz{\'a}lez},
  {Cappellaro}, {Navarrete}, {Pizzella}, {Ruiz}, {Smith}, \&
  {Turatto}}]{2017ApJ...850...89G}
{Guti{\'e}rrez}, C.~P., {Anderson}, J.~P., {Hamuy}, M., {et~al.} 2017, \apj,
  850, 89

\bibitem[{{Guti{\'e}rrez} {et~al.}(2018){Guti{\'e}rrez}, {Anderson},
  {Sullivan}, {Dessart}, {Gonz{\'a}lez-Gaitan}, {Galbany}, {Dimitriadis},
  {Arcavi}, {Bufano}, {Chen}, {Dennefeld}, {Gromadzki}, {Haislip},
  {Hosseinzadeh}, {Howell}, {Inserra}, {Kankare}, {Leloudas}, {Maguire},
  {McCully}, {Morrell}, {Olivares E}, {Pignata}, {Reichart}, {Reynolds},
  {Smartt}, {Sollerman}, {Taddia}, {Tak{\'a}ts}, {Terreran}, {Valenti}, \&
  {Young}}]{2018MNRAS.479.3232G}
{Guti{\'e}rrez}, C.~P., {Anderson}, J.~P., {Sullivan}, M., {et~al.} 2018,
  \mnras, 479, 3232

\bibitem[{{Habergham} {et~al.}(2014){Habergham}, {Anderson}, {James}, \&
  {Lyman}}]{2014MNRAS.441.2230H}
{Habergham}, S.~M., {Anderson}, J.~P., {James}, P.~A., \& {Lyman}, J.~D. 2014,
  \mnras, 441, 2230

\bibitem[{{Hakobyan} {et~al.}(2009){Hakobyan}, {Mamon}, {Petrosian}, {Kunth},
  \& {Turatto}}]{2009A&A...508.1259H}
{Hakobyan}, A.~A., {Mamon}, G.~A., {Petrosian}, A.~R., {Kunth}, D., \&
  {Turatto}, M. 2009, \aap, 508, 1259

\bibitem[{{Holoien} {et~al.}(2017{\natexlab{a}}){Holoien}, {Brown}, {Stanek},
  {Kochanek}, {Shappee}, {Prieto}, {Dong}, {Brimacombe}, {Bishop}, {Basu},
  {Beacom}, {Bersier}, {Chen}, {Danilet}, {Falco}, {Godoy-Rivera}, {Goss},
  {Pojmanski}, {Simonian}, {Skowron}, {Thompson}, {Wo{\'z}niak}, {{\'A}vila},
  {Bock}, {Carballo}, {Conseil}, {Contreras}, {Cruz}, {And{\'u}jar}, {Guo},
  {Hsiao}, {Kiyota}, {Koff}, {Krannich}, {Madore}, {Marples}, {Masi},
  {Morrell}, {Monard}, {Munoz-Mateos}, {Nicholls}, {Nicolas}, {Wagner}, \&
  {Wiethoff}}]{2017MNRAS.467.1098H}
{Holoien}, T.~W.~S., {Brown}, J.~S., {Stanek}, K.~Z., {et~al.}
  2017{\natexlab{a}}, \mnras, 467, 1098

\bibitem[{{Holoien} {et~al.}(2017{\natexlab{b}}){Holoien}, {Brown}, {Stanek},
  {Kochanek}, {Shappee}, {Prieto}, {Dong}, {Brimacombe}, {Bishop}, {Bose},
  {Beacom}, {Bersier}, {Chen}, {Chomiuk}, {Falco}, {Godoy-Rivera}, {Morrell},
  {Pojmanski}, {Shields}, {Strader}, {Stritzinger}, {Thompson}, {Wo{\'z}niak},
  {Bock}, {Cacella}, {Conseil}, {Cruz}, {Fernandez}, {Kiyota}, {Koff},
  {Krannich}, {Marples}, {Masi}, {Monard}, {Nicholls}, {Nicolas}, {Post},
  {Stone}, \& {Wiethoff}}]{2017MNRAS.471.4966H}
{Holoien}, T.~W.~S., {Brown}, J.~S., {Stanek}, K.~Z., {et~al.}
  2017{\natexlab{b}}, \mnras, 471, 4966

\bibitem[{{Holoien} {et~al.}(2019){Holoien}, {Brown}, {Vallely}, {Stanek},
  {Kochanek}, {Shappee}, {Prieto}, {Dong}, {Brimacombe}, {Bishop}, {Bose},
  {Beacom}, {Bersier}, {Chen}, {Chomiuk}, {Falco}, {Holmbo}, {Jayasinghe},
  {Morrell}, {Pojmanski}, {Shields}, {Strader}, {Stritzinger}, {Thompson},
  {Wo{\'z}niak}, {Bock}, {Cacella}, {Carballo}, {Cruz}, {Conseil}, {Farfan},
  {Fernandez}, {Kiyota}, {Koff}, {Krannich}, {Marples}, {Masi}, {Monard},
  {Mu{\~n}oz}, {Nicholls}, {Post}, {Stone}, {Trappett}, \&
  {Wiethoff}}]{2019MNRAS.484.1899H}
{Holoien}, T.~W.~S., {Brown}, J.~S., {Vallely}, P.~J., {et~al.} 2019, \mnras,
  484, 1899

\bibitem[{{Holoien} {et~al.}(2017{\natexlab{c}}){Holoien}, {Stanek},
  {Kochanek}, {Shappee}, {Prieto}, {Brimacombe}, {Bersier}, {Bishop}, {Dong},
  {Brown}, {Danilet}, {Simonian}, {Basu}, {Beacom}, {Falco}, {Pojmanski},
  {Skowron}, {Wo{\'z}niak}, {{\'A}vila}, {Conseil}, {Contreras}, {Cruz},
  {Fern{\'a}ndez}, {Koff}, {Guo}, {Herczeg}, {Hissong}, {Hsiao}, {Jose},
  {Kiyota}, {Long}, {Monard}, {Nicholls}, {Nicolas}, \&
  {Wiethoff}}]{2017MNRAS.464.2672H}
{Holoien}, T.~W.~S., {Stanek}, K.~Z., {Kochanek}, C.~S., {et~al.}
  2017{\natexlab{c}}, \mnras, 464, 2672

\bibitem[{{Hosseinzadeh} {et~al.}(2016){Hosseinzadeh}, {Arcavi}, {Howell},
  {McCully}, \& {Valenti}}]{2016ATel.9065....1H}
{Hosseinzadeh}, G., {Arcavi}, I., {Howell}, D.~A., {McCully}, C., \& {Valenti},
  S. 2016, The Astronomer's Telegram, 9065, 1

\bibitem[{{Hosseinzadeh} {et~al.}(2019){Hosseinzadeh}, {McCully}, {Zabludoff},
  {Arcavi}, {French}, {Howell}, {Berger}, \& {Hiramatsu}}]{2019ApJ...871L...9H}
{Hosseinzadeh}, G., {McCully}, C., {Zabludoff}, A.~I., {et~al.} 2019, \apjl,
  871, L9

\bibitem[{{Jha}(2017)}]{2017TNSCR.870....1J}
{Jha}, S. 2017, Transient Name Server Classification Report, 2017-870, 1

\bibitem[{{Kaiser} {et~al.}(2002){Kaiser}, {Aussel}, {Burke}, {Boesgaard},
  {Chambers}, {Chun}, {Heasley}, {Hodapp}, {Hunt}, {Jedicke}, {Jewitt},
  {Kudritzki}, {Luppino}, {Maberry}, {Magnier}, {Monet}, {Onaka}, {Pickles},
  {Rhoads}, {Simon}, {Szalay}, {Szapudi}, {Tholen}, {Tonry}, {Waterson}, \&
  {Wick}}]{2002SPIE.4836..154K}
{Kaiser}, N., {Aussel}, H., {Burke}, B.~E., {et~al.} 2002, in Society of
  Photo-Optical Instrumentation Engineers (SPIE) Conference Series, Vol. 4836,
  Survey and Other Telescope Technologies and Discoveries, ed. J.~A. {Tyson} \&
  S.~{Wolff}, 154--164

\bibitem[{{Kangas} {et~al.}(2013){Kangas}, {Mattila}, {Kankare}, {Kotilainen},
  {V{\"a}is{\"a}nen}, {Greimel}, \& {Takalo}}]{2013MNRAS.436.3464K}
{Kangas}, T., {Mattila}, S., {Kankare}, E., {et~al.} 2013, \mnras, 436, 3464

\bibitem[{{Kangas} {et~al.}(2017){Kangas}, {Portinari}, {Mattila}, {Fraser},
  {Kankare}, {Izzard}, {James}, {Gonz{\'a}lez-Fern{\'a}ndez}, {Maund}, \&
  {Thompson}}]{2017A&A...597A..92K}
{Kangas}, T., {Portinari}, L., {Mattila}, S., {et~al.} 2017, \aap, 597, A92

\bibitem[{{Kasen} \& {Woosley}(2009)}]{2009ApJ...703.2205K}
{Kasen}, D. \& {Woosley}, S.~E. 2009, \apj, 703, 2205

\bibitem[{{Kauffmann} {et~al.}(2003){Kauffmann}, {Heckman}, {Tremonti},
  {Brinchmann}, {Charlot}, {White}, {Ridgway}, {Brinkmann}, {Fukugita}, {Hall},
  {Ivezi{\'c}}, {Richards}, \& {Schneider}}]{2003MNRAS.346.1055K}
{Kauffmann}, G., {Heckman}, T.~M., {Tremonti}, C., {et~al.} 2003, \mnras, 346,
  1055

\bibitem[{{Kelly} \& {Kirshner}(2012)}]{2012ApJ...759..107K}
{Kelly}, P.~L. \& {Kirshner}, R.~P. 2012, \apj, 759, 107

\bibitem[{{Kennicutt}(1998)}]{1998ApJ...498..541K}
{Kennicutt}, Robert~C., J. 1998, \apj, 498, 541

\bibitem[{{Kewley} {et~al.}(2001){Kewley}, {Dopita}, {Sutherland}, {Heisler},
  \& {Trevena}}]{2001ApJ...556..121K}
{Kewley}, L.~J., {Dopita}, M.~A., {Sutherland}, R.~S., {Heisler}, C.~A., \&
  {Trevena}, J. 2001, \apj, 556, 121

\bibitem[{{Kilpatrick} {et~al.}(2021){Kilpatrick}, {Drout}, {Auchettl},
  {Dimitriadis}, {Foley}, {Jones}, {DeMarchi}, {French}, {Gall}, {Hjorth},
  {Jacobson-Gal{\'a}n}, {Margutti}, {Piro}, {Ramirez-Ruiz}, {Rest}, \&
  {Rojas-Bravo}}]{2021MNRAS.504.2073K}
{Kilpatrick}, C.~D., {Drout}, M.~R., {Auchettl}, K., {et~al.} 2021, \mnras,
  504, 2073

\bibitem[{{Kochanek} {et~al.}(2017){Kochanek}, {Shappee}, {Stanek}, {Holoien},
  {Thompson}, {Prieto}, {Dong}, {Shields}, {Will}, {Britt}, {Perzanowski}, \&
  {Pojma{\'n}ski}}]{2017PASP..129j4502K}
{Kochanek}, C.~S., {Shappee}, B.~J., {Stanek}, K.~Z., {et~al.} 2017, \pasp,
  129, 104502

\bibitem[{{Kuncarayakti} {et~al.}(2018){Kuncarayakti}, {Anderson}, {Galbany},
  {Maeda}, {Hamuy}, {Aldering}, {Arimoto}, {Doi}, {Morokuma}, \&
  {Usuda}}]{2018A&A...613A..35K}
{Kuncarayakti}, H., {Anderson}, J.~P., {Galbany}, L., {et~al.} 2018, \aap, 613,
  A35

\bibitem[{{Kuncarayakti} {et~al.}(2013{\natexlab{a}}){Kuncarayakti}, {Doi},
  {Aldering}, {Arimoto}, {Maeda}, {Morokuma}, {Pereira}, {Usuda}, \&
  {Hashiba}}]{2013AJ....146...30K}
{Kuncarayakti}, H., {Doi}, M., {Aldering}, G., {et~al.} 2013{\natexlab{a}},
  \aj, 146, 30

\bibitem[{{Kuncarayakti} {et~al.}(2013{\natexlab{b}}){Kuncarayakti}, {Doi},
  {Aldering}, {Arimoto}, {Maeda}, {Morokuma}, {Pereira}, {Usuda}, \&
  {Hashiba}}]{2013AJ....146...31K}
{Kuncarayakti}, H., {Doi}, M., {Aldering}, G., {et~al.} 2013{\natexlab{b}},
  \aj, 146, 31

\bibitem[{{Le F{\`e}vre} {et~al.}(2003){Le F{\`e}vre}, {Saisse}, {Mancini},
  {Brau-Nogue}, {Caputi}, {Castinel}, {D'Odorico}, {Garilli}, {Kissler-Patig},
  {Lucuix}, {Mancini}, {Pauget}, {Sciarretta}, {Scodeggio}, {Tresse}, \&
  {Vettolani}}]{2003SPIE.4841.1670L}
{Le F{\`e}vre}, O., {Saisse}, M., {Mancini}, D., {et~al.} 2003, in Society of
  Photo-Optical Instrumentation Engineers (SPIE) Conference Series, Vol. 4841,
  Instrument Design and Performance for Optical/Infrared Ground-based
  Telescopes, ed. M.~{Iye} \& A.~F.~M. {Moorwood}, 1670--1681

\bibitem[{{Leitherer} {et~al.}(1999){Leitherer}, {Schaerer}, {Goldader},
  {Delgado}, {Robert}, {Kune}, {de Mello}, {Devost}, \&
  {Heckman}}]{1999ApJS..123....3L}
{Leitherer}, C., {Schaerer}, D., {Goldader}, J.~D., {et~al.} 1999, \apjs, 123,
  3

\bibitem[{{Leloudas} {et~al.}(2011){Leloudas}, {Gallazzi}, {Sollerman},
  {Stritzinger}, {Fynbo}, {Hjorth}, {Malesani}, {Micha{\l}owski},
  {Milvang-Jensen}, \& {Smith}}]{2011A&A...530A..95L}
{Leloudas}, G., {Gallazzi}, A., {Sollerman}, J., {et~al.} 2011, \aap, 530, A95

\bibitem[{{Leloudas} {et~al.}(2015){Leloudas}, {Schulze}, {Kr{\"u}hler},
  {Gorosabel}, {Christensen}, {Mehner}, {de Ugarte Postigo}, {Amor{\'\i}n},
  {Th{\"o}ne}, {Anderson}, {Bauer}, {Gallazzi}, {He{\l}miniak}, {Hjorth},
  {Ibar}, {Malesani}, {Morell}, {Vinko}, \& {Wheeler}}]{2015MNRAS.449..917L}
{Leloudas}, G., {Schulze}, S., {Kr{\"u}hler}, T., {et~al.} 2015, \mnras, 449,
  917

\bibitem[{{Li} {et~al.}(2011){Li}, {Leaman}, {Chornock}, {Filippenko},
  {Poznanski}, {Ganeshalingam}, {Wang}, {Modjaz}, {Jha}, {Foley}, \&
  {Smith}}]{2011MNRAS.412.1441L}
{Li}, W., {Leaman}, J., {Chornock}, R., {et~al.} 2011, \mnras, 412, 1441

\bibitem[{{Lineweaver}(1997)}]{1997mba..conf...69L}
{Lineweaver}, C.~H. 1997, in Microwave Background Anisotropies, Vol.~16, 69--75

\bibitem[{{L{\'o}pez-Cob{\'a}} {et~al.}(2020){L{\'o}pez-Cob{\'a}},
  {S{\'a}nchez}, {Anderson}, {Cruz-Gonz{\'a}lez}, {Galbany}, {Ruiz-Lara},
  {Barrera-Ballesteros}, {Prieto}, \& {Kuncarayakti}}]{2020AJ....159..167L}
{L{\'o}pez-Cob{\'a}}, C., {S{\'a}nchez}, S.~F., {Anderson}, J.~P., {et~al.}
  2020, \aj, 159, 167

\bibitem[{{Lugo-Aranda} {et~al.}(2022){Lugo-Aranda}, {S{\'a}nchez},
  {Espinosa-Ponce}, {L{\'o}pez-Cob{\'a}}, {Galbany}, {Barrera-Ballesteros},
  {S{\'a}nchez-Menguiano}, \& {Anderson}}]{2022RASTI...1....3L}
{Lugo-Aranda}, A.~Z., {S{\'a}nchez}, S.~F., {Espinosa-Ponce}, C., {et~al.}
  2022, RAS Techniques and Instruments, 1, 3

\bibitem[{{Lyman} {et~al.}(2016){Lyman}, {Bersier}, {James}, {Mazzali},
  {Eldridge}, {Fraser}, \& {Pian}}]{2016MNRAS.457..328L}
{Lyman}, J.~D., {Bersier}, D., {James}, P.~A., {et~al.} 2016, \mnras, 457, 328

\bibitem[{{Lyman} {et~al.}(2018){Lyman}, {Taddia}, {Stritzinger}, {Galbany},
  {Leloudas}, {Anderson}, {Eldridge}, {James}, {Kr{\"u}hler}, {Levan},
  {Pignata}, \& {Stanway}}]{2018MNRAS.473.1359L}
{Lyman}, J.~D., {Taddia}, F., {Stritzinger}, M.~D., {et~al.} 2018, \mnras, 473,
  1359

\bibitem[{{Makarov} {et~al.}(2014){Makarov}, {Prugniel}, {Terekhova},
  {Courtois}, \& {Vauglin}}]{2014A&A...570A..13M}
{Makarov}, D., {Prugniel}, P., {Terekhova}, N., {Courtois}, H., \& {Vauglin},
  I. 2014, \aap, 570, A13

\bibitem[{{Marino} {et~al.}(2013){Marino}, {Rosales-Ortega}, {S{\'a}nchez},
  {Gil de Paz}, {V{\'\i}lchez}, {Miralles-Caballero}, {Kehrig},
  {P{\'e}rez-Montero}, {Stanishev}, {Iglesias-P{\'a}ramo}, {D{\'\i}az},
  {Castillo-Morales}, {Kennicutt}, {L{\'o}pez-S{\'a}nchez}, {Galbany},
  {Garc{\'\i}a-Benito}, {Mast}, {Mendez-Abreu}, {Monreal-Ibero}, {Husemann},
  {Walcher}, {Garc{\'\i}a-Lorenzo}, {Masegosa}, {Del Olmo Orozco},
  {Mour{\~a}o}, {Ziegler}, {Moll{\'a}}, {Papaderos},
  {S{\'a}nchez-Bl{\'a}zquez}, {Gonz{\'a}lez Delgado}, {Falc{\'o}n-Barroso},
  {Roth}, {van de Ven}, \& {Califa Team}}]{2013A&A...559A.114M}
{Marino}, R.~A., {Rosales-Ortega}, F.~F., {S{\'a}nchez}, S.~F., {et~al.} 2013,
  \aap, 559, A114

\bibitem[{{Martig} \& {Bournaud}(2010)}]{2010ApJ...714L.275M}
{Martig}, M. \& {Bournaud}, F. 2010, \apjl, 714, L275

\bibitem[{{Martinez} {et~al.}(2022){Martinez}, {Anderson}, {Bersten}, {Hamuy},
  {Gonz{\'a}lez-Gait{\'a}n}, {Orellana}, {Stritzinger}, {Phillips},
  {Guti{\'e}rrez}, {Burns}, {de Jaeger}, {Ertini}, {Folatelli}, {F{\"o}rster},
  {Galbany}, {Hoeflich}, {Hsiao}, {Morrell}, {Pessi}, \&
  {Suntzeff}}]{2022A&A...660A..42M}
{Martinez}, L., {Anderson}, J.~P., {Bersten}, M.~C., {et~al.} 2022, \aap, 660,
  A42

\bibitem[{{Mateus} {et~al.}(2006){Mateus}, {Sodr{\'e}}, {Cid Fernandes},
  {Stasi{\'n}ska}, {Schoenell}, \& {Gomes}}]{2006MNRAS.370..721M}
{Mateus}, A., {Sodr{\'e}}, L., {Cid Fernandes}, R., {et~al.} 2006, \mnras, 370,
  721

\bibitem[{{Matteucci} \& {Greggio}(1986)}]{1986A&A...154..279M}
{Matteucci}, F. \& {Greggio}, L. 1986, \aap, 154, 279

\bibitem[{{Mauerhan} \& {Smith}(2012)}]{2012MNRAS.424.2659M}
{Mauerhan}, J. \& {Smith}, N. 2012, \mnras, 424, 2659

\bibitem[{{Mauerhan} {et~al.}(2013){Mauerhan}, {Smith}, {Filippenko},
  {Blanchard}, {Blanchard}, {Casper}, {Cenko}, {Clubb}, {Cohen}, {Fuller},
  {Li}, \& {Silverman}}]{2013MNRAS.430.1801M}
{Mauerhan}, J.~C., {Smith}, N., {Filippenko}, A.~V., {et~al.} 2013, \mnras,
  430, 1801

\bibitem[{{Maund}(2018)}]{2018MNRAS.476.2629M}
{Maund}, J.~R. 2018, \mnras, 476, 2629

\bibitem[{{Maund} {et~al.}(2011){Maund}, {Fraser}, {Ergon}, {Pastorello},
  {Smartt}, {Sollerman}, {Benetti}, {Botticella}, {Bufano}, {Danziger},
  {Kotak}, {Magill}, {Stephens}, \& {Valenti}}]{2011ApJ...739L..37M}
{Maund}, J.~R., {Fraser}, M., {Ergon}, M., {et~al.} 2011, \apjl, 739, L37

\bibitem[{{Maund} {et~al.}(2015){Maund}, {Fraser}, {Reilly}, {Ergon}, \&
  {Mattila}}]{2015MNRAS.447.3207M}
{Maund}, J.~R., {Fraser}, M., {Reilly}, E., {Ergon}, M., \& {Mattila}, S. 2015,
  \mnras, 447, 3207

\bibitem[{{Maund} {et~al.}(2014){Maund}, {Reilly}, \&
  {Mattila}}]{2014MNRAS.438..938M}
{Maund}, J.~R., {Reilly}, E., \& {Mattila}, S. 2014, \mnras, 438, 938

\bibitem[{{Maund} \& {Smartt}(2009)}]{2009Sci...324..486M}
{Maund}, J.~R. \& {Smartt}, S.~J. 2009, Science, 324, 486

\bibitem[{{Mayker Chen} {et~al.}(2023){Mayker Chen}, {Leroy}, {Lopez},
  {Benincasa}, {Chevance}, {Glover}, {Hughes}, {Kreckel}, {Sarbadhicary},
  {Sun}, {Thompson}, {Utomo}, {Bigiel}, {Blanc}, {Dale}, {Grasha}, {Kruijssen},
  {Pan}, {Querejeta}, {Schinnerer}, {Watkins}, \&
  {Williams}}]{2023ApJ...944..110M}
{Mayker Chen}, N., {Leroy}, A.~K., {Lopez}, L.~A., {et~al.} 2023, \apj, 944,
  110

\bibitem[{{Mazzali} {et~al.}(2002){Mazzali}, {Deng}, {Maeda}, {Nomoto},
  {Umeda}, {Hatano}, {Iwamoto}, {Yoshii}, {Kobayashi}, {Minezaki}, {Doi},
  {Enya}, {Tomita}, {Smartt}, {Kinugasa}, {Kawakita}, {Ayani}, {Kawabata},
  {Yamaoka}, {Qiu}, {Motohara}, {Gerardy}, {Fesen}, {Kawabata}, {Iye},
  {Kashikawa}, {Kosugi}, {Ohyama}, {Takada-Hidai}, {Zhao}, {Chornock},
  {Filippenko}, {Benetti}, \& {Turatto}}]{2002ApJ...572L..61M}
{Mazzali}, P.~A., {Deng}, J., {Maeda}, K., {et~al.} 2002, \apjl, 572, L61

\bibitem[{{Meza} {et~al.}(2019{\natexlab{a}}){Meza}, {Prieto}, {Clocchiatti},
  {Galbany}, {Anderson}, {Falco}, {Kochanek}, {Kuncarayakti}, {S{\'a}nchez},
  {Brimacombe}, {Holoien}, {Shappee}, {Stanek}, \&
  {Thompson}}]{2019A&A...629A..57M}
{Meza}, N., {Prieto}, J.~L., {Clocchiatti}, A., {et~al.} 2019{\natexlab{a}},
  \aap, 629, A57

\bibitem[{{Meza} {et~al.}(2019{\natexlab{b}}){Meza}, {Prieto}, {Clocchiatti},
  {Galbany}, {Anderson}, {Falco}, {Kochanek}, {Kuncarayakti}, {Sanchez},
  {Brimacombe}, {Holoien}, {Shappee}, {Stanek}, \&
  {Thompson}}]{2019yCat..36290057M}
{Meza}, N., {Prieto}, J.~L., {Clocchiatti}, A., {et~al.} 2019{\natexlab{b}},
  VizieR Online Data Catalog, J/A+A/629/A57

\bibitem[{{Mezzacappa}(2005)}]{2005ARNPS..55..467M}
{Mezzacappa}, A. 2005, Annual Review of Nuclear and Particle Science, 55, 467

\bibitem[{{Modjaz} {et~al.}(2011){Modjaz}, {Kewley}, {Bloom}, {Filippenko},
  {Perley}, \& {Silverman}}]{2011ApJ...731L...4M}
{Modjaz}, M., {Kewley}, L., {Bloom}, J.~S., {et~al.} 2011, \apjl, 731, L4

\bibitem[{{Neumann} {et~al.}(2023){Neumann}, {Holoien}, {Kochanek}, {Stanek},
  {Vallely}, {Shappee}, {Prieto}, {Pessi}, {Jayasinghe}, {Brimacombe},
  {Bersier}, {Aydi}, {Basinger}, {Beacom}, {Bose}, {Brown}, {Chen},
  {Clocchiatti}, {Desai}, {Dong}, {Falco}, {Holmbo}, {Morrell}, {Shields},
  {Sokolovsky}, {Strader}, {Stritzinger}, {Swihart}, {Thompson}, {Way},
  {Aslan}, {Bishop}, {Bock}, {Bradshaw}, {Cacella}, {Castro-Morales},
  {Conseil}, {Cornect}, {Cruz}, {Farfan}, {Fernandez}, {Gabuya},
  {Gonzalez-Carballo}, {Kendurkar}, {Kiyota}, {Koff}, {Krannich}, {Marples},
  {Masi}, {Monard}, {Mu{\~n}oz}, {Nicholls}, {Post}, {Pujic}, {Stone},
  {Tomasella}, {Trappett}, \& {Wiethoff}}]{2022arXiv221006492N}
{Neumann}, K.~D., {Holoien}, T.~W.~S., {Kochanek}, C.~S., {et~al.} 2023,
  \mnras, 520, 4356

\bibitem[{{Nicholl} {et~al.}(2016){Nicholl}, {Berger}, {Smartt}, {Margutti},
  {Kamble}, {Alexander}, {Chen}, {Inserra}, {Arcavi}, {Blanchard}, {Cartier},
  {Chambers}, {Childress}, {Chornock}, {Cowperthwaite}, {Drout}, {Flewelling},
  {Fraser}, {Gal-Yam}, {Galbany}, {Harmanen}, {Holoien}, {Hosseinzadeh},
  {Howell}, {Huber}, {Jerkstrand}, {Kankare}, {Kochanek}, {Lin}, {Lunnan},
  {Magnier}, {Maguire}, {McCully}, {McDonald}, {Metzger}, {Milisavljevic},
  {Mitra}, {Reynolds}, {Saario}, {Shappee}, {Smith}, {Valenti}, {Villar},
  {Waters}, \& {Young}}]{2016ApJ...826...39N}
{Nicholl}, M., {Berger}, E., {Smartt}, S.~J., {et~al.} 2016, \apj, 826, 39

\bibitem[{{Nicholl} {et~al.}(2018){Nicholl}, {Blanchard}, {Berger},
  {Alexander}, {Metzger}, {Bhirombhakdi}, {Chornock}, {Coppejans}, {Gomez},
  {Margalit}, {Margutti}, \& {Terreran}}]{2018ApJ...866L..24N}
{Nicholl}, M., {Blanchard}, P.~K., {Berger}, E., {et~al.} 2018, \apjl, 866, L24

\bibitem[{{Nyholm} {et~al.}(2020){Nyholm}, {Sollerman}, {Tartaglia}, {Taddia},
  {Fremling}, {Blagorodnova}, {Filippenko}, {Gal-Yam}, {Howell},
  {Karamehmetoglu}, {Kulkarni}, {Laher}, {Leloudas}, {Masci}, {Kasliwal},
  {Mor{\r{a}}}, {Moriya}, {Ofek}, {Papadogiannakis}, {Quimby}, {Rebbapragada},
  \& {Schulze}}]{2020A&A...637A..73N}
{Nyholm}, A., {Sollerman}, J., {Tartaglia}, L., {et~al.} 2020, \aap, 637, A73

\bibitem[{{Pastorello} {et~al.}(2013){Pastorello}, {Cappellaro}, {Inserra},
  {Smartt}, {Pignata}, {Benetti}, {Valenti}, {Fraser}, {Tak{\'a}ts}, {Benitez},
  {Botticella}, {Brimacombe}, {Bufano}, {Cellier-Holzem}, {Costado}, {Cupani},
  {Curtis}, {Elias-Rosa}, {Ergon}, {Fynbo}, {Hambsch}, {Hamuy}, {Harutyunyan},
  {Ivarson}, {Kankare}, {Martin}, {Kotak}, {LaCluyze}, {Maguire}, {Mattila},
  {Maza}, {McCrum}, {Miluzio}, {Norgaard-Nielsen}, {Nysewander}, {Ochner},
  {Pan}, {Pumo}, {Reichart}, {Tan}, {Taubenberger}, {Tomasella}, {Turatto}, \&
  {Wright}}]{2013ApJ...767....1P}
{Pastorello}, A., {Cappellaro}, E., {Inserra}, C., {et~al.} 2013, \apj, 767, 1

\bibitem[{{Pastorello} {et~al.}(2008){Pastorello}, {Mattila}, {Zampieri},
  {Della Valle}, {Smartt}, {Valenti}, {Agnoletto}, {Benetti}, {Benn}, {Branch},
  {Cappellaro}, {Dennefeld}, {Eldridge}, {Gal-Yam}, {Harutyunyan}, {Hunter},
  {Kjeldsen}, {Lipkin}, {Mazzali}, {Milne}, {Navasardyan}, {Ofek}, {Pian},
  {Shemmer}, {Spiro}, {Stathakis}, {Taubenberger}, {Turatto}, \&
  {Yamaoka}}]{2008MNRAS.389..113P}
{Pastorello}, A., {Mattila}, S., {Zampieri}, L., {et~al.} 2008, \mnras, 389,
  113

\bibitem[{{Pastorello} {et~al.}(2007){Pastorello}, {Smartt}, {Mattila},
  {Eldridge}, {Young}, {Itagaki}, {Yamaoka}, {Navasardyan}, {Valenti}, {Patat},
  {Agnoletto}, {Augusteijn}, {Benetti}, {Cappellaro}, {Boles}, {Bonnet-Bidaud},
  {Botticella}, {Bufano}, {Cao}, {Deng}, {Dennefeld}, {Elias-Rosa},
  {Harutyunyan}, {Keenan}, {Iijima}, {Lorenzi}, {Mazzali}, {Meng}, {Nakano},
  {Nielsen}, {Smoker}, {Stanishev}, {Turatto}, {Xu}, \&
  {Zampieri}}]{2007Natur.447..829P}
{Pastorello}, A., {Smartt}, S.~J., {Mattila}, S., {et~al.} 2007, \nat, 447, 829

\bibitem[{{Patat} {et~al.}(2001){Patat}, {Cappellaro}, {Danziger}, {Mazzali},
  {Sollerman}, {Augusteijn}, {Brewer}, {Doublier}, {Gonzalez}, {Hainaut},
  {Lidman}, {Leibundgut}, {Nomoto}, {Nakamura}, {Spyromilio}, {Rizzi},
  {Turatto}, {Walsh}, {Galama}, {van Paradijs}, {Kouveliotou}, {Vreeswijk},
  {Frontera}, {Masetti}, {Palazzi}, \& {Pian}}]{2001ApJ...555..900P}
{Patat}, F., {Cappellaro}, E., {Danziger}, J., {et~al.} 2001, \apj, 555, 900

\bibitem[{{Perley} {et~al.}(2016){Perley}, {Quimby}, {Yan}, {Vreeswijk}, {De
  Cia}, {Lunnan}, {Gal-Yam}, {Yaron}, {Filippenko}, {Graham}, {Laher}, \&
  {Nugent}}]{2016ApJ...830...13P}
{Perley}, D.~A., {Quimby}, R.~M., {Yan}, L., {et~al.} 2016, \apj, 830, 13

\bibitem[{{Perley} {et~al.}(2022){Perley}, {Sollerman}, {Schulze}, {Yao},
  {Fremling}, {Gal-Yam}, {Ho}, {Yang}, {Kool}, {Irani}, {Yan}, {Andreoni},
  {Baade}, {Bellm}, {Brink}, {Chen}, {Cikota}, {Coughlin}, {Dahiwale},
  {Dekany}, {Duev}, {Filippenko}, {Hoeflich}, {Kasliwal}, {Kulkarni}, {Lunnan},
  {Masci}, {Maund}, {Medford}, {Riddle}, {Rosnet}, {Shupe}, {Strotjohann},
  {Tzanidakis}, \& {Zheng}}]{2022ApJ...927..180P}
{Perley}, D.~A., {Sollerman}, J., {Schulze}, S., {et~al.} 2022, \apj, 927, 180

\bibitem[{{Pettini} \& {Pagel}(2004)}]{2004MNRAS.348L..59P}
{Pettini}, M. \& {Pagel}, B. E.~J. 2004, \mnras, 348, L59

\bibitem[{{Phillips}(1993)}]{1993ApJ...413L.105P}
{Phillips}, M.~M. 1993, \apjl, 413, L105

\bibitem[{{Prantzos} \& {Boissier}(2003)}]{2003A&A...406..259P}
{Prantzos}, N. \& {Boissier}, S. 2003, \aap, 406, 259

\bibitem[{{Prentice} {et~al.}(2016){Prentice}, {Mazzali}, {Pian}, {Gal-Yam},
  {Kulkarni}, {Rubin}, {Corsi}, {Fremling}, {Sollerman}, {Yaron}, {Arcavi},
  {Zheng}, {Kasliwal}, {Filippenko}, {Cenko}, {Cao}, \&
  {Nugent}}]{2016MNRAS.458.2973P}
{Prentice}, S.~J., {Mazzali}, P.~A., {Pian}, E., {et~al.} 2016, \mnras, 458,
  2973

\bibitem[{{Prieto} {et~al.}(2008){Prieto}, {Stanek}, \&
  {Beacom}}]{2008ApJ...673..999P}
{Prieto}, J.~L., {Stanek}, K.~Z., \& {Beacom}, J.~F. 2008, \apj, 673, 999

\bibitem[{{Qiu} {et~al.}(1999){Qiu}, {Hatano}, {Branch}, \&
  {Baron}}]{1999AAS...195.3810Q}
{Qiu}, Y.~L., {Hatano}, K., {Branch}, D., \& {Baron}, E. 1999, in American
  Astronomical Society Meeting Abstracts, Vol. 195, American Astronomical
  Society Meeting Abstracts, 38.10

\bibitem[{{Raimann} {et~al.}(2000){Raimann}, {Storchi-Bergmann}, {Bica},
  {Melnick}, \& {Schmitt}}]{2000MNRAS.316..559R}
{Raimann}, D., {Storchi-Bergmann}, T., {Bica}, E., {Melnick}, J., \& {Schmitt},
  H. 2000, \mnras, 316, 559

\bibitem[{{Ransome} {et~al.}(2022){Ransome}, {Habergham-Mawson}, {Darnley},
  {James}, \& {Percival}}]{2022MNRAS.513.3564R}
{Ransome}, C.~L., {Habergham-Mawson}, S.~M., {Darnley}, M.~J., {James}, P.~A.,
  \& {Percival}, S.~M. 2022, \mnras, 513, 3564

\bibitem[{{Ryder} {et~al.}(1993){Ryder}, {Staveley-Smith}, {Dopita}, {Petre},
  {Colbert}, {Malin}, \& {Schlegel}}]{1993ApJ...416..167R}
{Ryder}, S., {Staveley-Smith}, L., {Dopita}, M., {et~al.} 1993, \apj, 416, 167

\bibitem[{{S{\'a}nchez-Bl{\'a}zquez} {et~al.}(2006){S{\'a}nchez-Bl{\'a}zquez},
  {Peletier}, {Jim{\'e}nez-Vicente}, {Cardiel}, {Cenarro},
  {Falc{\'o}n-Barroso}, {Gorgas}, {Selam}, \& {Vazdekis}}]{2006MNRAS.371..703S}
{S{\'a}nchez-Bl{\'a}zquez}, P., {Peletier}, R.~F., {Jim{\'e}nez-Vicente}, J.,
  {et~al.} 2006, \mnras, 371, 703

\bibitem[{{Sanders} {et~al.}(2012){Sanders}, {Soderberg}, {Levesque}, {Foley},
  {Chornock}, {Milisavljevic}, {Margutti}, {Berger}, {Drout}, {Czekala}, \&
  {Dittmann}}]{2012ApJ...758..132S}
{Sanders}, N.~E., {Soderberg}, A.~M., {Levesque}, E.~M., {et~al.} 2012, \apj,
  758, 132

\bibitem[{{Schady} {et~al.}(2019){Schady}, {Eldridge}, {Anderson}, {Chen},
  {Galbany}, {Kuncarayakti}, \& {Xiao}}]{2019MNRAS.490.4515S}
{Schady}, P., {Eldridge}, J.~J., {Anderson}, J., {et~al.} 2019, \mnras, 490,
  4515

\bibitem[{{Schlafly} \& {Finkbeiner}(2011)}]{2011ApJ...737..103S}
{Schlafly}, E.~F. \& {Finkbeiner}, D.~P. 2011, \apj, 737, 103

\bibitem[{{Schlegel}(1990)}]{1990MNRAS.244..269S}
{Schlegel}, E.~M. 1990, \mnras, 244, 269

\bibitem[{{Schulze} {et~al.}(2021){Schulze}, {Yaron}, {Sollerman}, {Leloudas},
  {Gal}, {Wright}, {Lunnan}, {Gal-Yam}, {Ofek}, {Perley}, {Filippenko},
  {Kasliwal}, {Kulkarni}, {Neill}, {Nugent}, {Quimby}, {Sullivan},
  {Strotjohann}, {Arcavi}, {Ben-Ami}, {Bianco}, {Bloom}, {De}, {Fraser},
  {Fremling}, {Horesh}, {Johansson}, {Kelly}, {Kne{\v{z}}evi{\'c}},
  {Kne{\v{z}}evi{\'c}}, {Maguire}, {Nyholm}, {Papadogiannakis}, {Petrushevska},
  {Rubin}, {Yan}, {Yang}, {Adams}, {Bufano}, {Clubb}, {Foley}, {Green},
  {Harmanen}, {Ho}, {Hook}, {Hosseinzadeh}, {Howell}, {Kong}, {Kotak},
  {Matheson}, {McCully}, {Milisavljevic}, {Pan}, {Poznanski}, {Shivvers}, {van
  Velzen}, \& {Verbeek}}]{2021ApJS..255...29S}
{Schulze}, S., {Yaron}, O., {Sollerman}, J., {et~al.} 2021, \apjs, 255, 29

\bibitem[{{Shappee} {et~al.}(2014){Shappee}, {Prieto}, {Grupe}, {Kochanek},
  {Stanek}, {De Rosa}, {Mathur}, {Zu}, {Peterson}, {Pogge}, {Komossa}, {Im},
  {Jencson}, {Holoien}, {Basu}, {Beacom}, {Szczygie{\l}}, {Brimacombe},
  {Adams}, {Campillay}, {Choi}, {Contreras}, {Dietrich}, {Dubberley},
  {Elphick}, {Foale}, {Giustini}, {Gonzalez}, {Hawkins}, {Howell}, {Hsiao},
  {Koss}, {Leighly}, {Morrell}, {Mudd}, {Mullins}, {Nugent}, {Parrent},
  {Phillips}, {Pojmanski}, {Rosing}, {Ross}, {Sand}, {Terndrup}, {Valenti},
  {Walker}, \& {Yoon}}]{2014ApJ...788...48S}
{Shappee}, B.~J., {Prieto}, J.~L., {Grupe}, D., {et~al.} 2014, \apj, 788, 48

\bibitem[{{Shivvers} {et~al.}(2017){Shivvers}, {Zheng}, {Van Dyk}, {Mauerhan},
  {Filippenko}, {Smith}, {Foley}, {Mazzali}, {Kamble}, {Kilpatrick},
  {Margutti}, {Yuk}, {Graham}, {Kelly}, {Andrews}, {Matheson}, {Wood-Vasey},
  {Ponder}, {Brown}, {Chevalier}, {Milisavljevic}, {Drout}, {Parrent},
  {Soderberg}, {Ashall}, {Piascik}, \& {Prentice}}]{2017MNRAS.471.4381S}
{Shivvers}, I., {Zheng}, W., {Van Dyk}, S.~D., {et~al.} 2017, \mnras, 471, 4381

\bibitem[{{Smartt}(2015)}]{2015PASA...32...16S}
{Smartt}, S.~J. 2015, \pasa, 32, e016

\bibitem[{{Smith}(2014)}]{2014ARA&A..52..487S}
{Smith}, N. 2014, \araa, 52, 487

\bibitem[{{Smith}(2019)}]{2019MNRAS.489.4378S}
{Smith}, N. 2019, \mnras, 489, 4378

\bibitem[{{Smith} {et~al.}(2009){Smith}, {Hinkle}, \&
  {Ryde}}]{2009AJ....137.3558S}
{Smith}, N., {Hinkle}, K.~H., \& {Ryde}, N. 2009, \aj, 137, 3558

\bibitem[{{Sollerman} {et~al.}(1998){Sollerman}, {Cumming}, \&
  {Lundqvist}}]{1998ApJ...493..933S}
{Sollerman}, J., {Cumming}, R.~J., \& {Lundqvist}, P. 1998, \apj, 493, 933

\bibitem[{{Storchi-Bergmann} {et~al.}(1994){Storchi-Bergmann}, {Calzetti}, \&
  {Kinney}}]{1994ApJ...429..572S}
{Storchi-Bergmann}, T., {Calzetti}, D., \& {Kinney}, A.~L. 1994, \apj, 429, 572

\bibitem[{{Sun} {et~al.}(2023){Sun}, {Maund}, \&
  {Crowther}}]{2023MNRAS.521.2860S}
{Sun}, N.-C., {Maund}, J.~R., \& {Crowther}, P.~A. 2023, \mnras, 521, 2860

\bibitem[{{Sun} {et~al.}(2021){Sun}, {Maund}, {Crowther}, {Fang}, \&
  {Zapartas}}]{2021MNRAS.504.2253S}
{Sun}, N.-C., {Maund}, J.~R., {Crowther}, P.~A., {Fang}, X., \& {Zapartas}, E.
  2021, \mnras, 504, 2253

\bibitem[{{Sun} {et~al.}(2022){Sun}, {Maund}, {Crowther}, {Hirai}, {Kashapov},
  {Liu}, {Liu}, \& {Zapartas}}]{2022MNRAS.510.3701S}
{Sun}, N.-C., {Maund}, J.~R., {Crowther}, P.~A., {et~al.} 2022, \mnras, 510,
  3701

\bibitem[{{Sun} {et~al.}(2020){Sun}, {Maund}, {Hirai}, {Crowther}, \&
  {Podsiadlowski}}]{2020MNRAS.491.6000S}
{Sun}, N.-C., {Maund}, J.~R., {Hirai}, R., {Crowther}, P.~A., \&
  {Podsiadlowski}, P. 2020, \mnras, 491, 6000

\bibitem[{{Taddia} {et~al.}(2015{\natexlab{a}}){Taddia}, {Sollerman},
  {Fremling}, {Pastorello}, {Leloudas}, {Fransson}, {Nyholm}, {Stritzinger},
  {Ergon}, {Roy}, \& {Migotto}}]{2015A&A...580A.131T}
{Taddia}, F., {Sollerman}, J., {Fremling}, C., {et~al.} 2015{\natexlab{a}},
  \aap, 580, A131

\bibitem[{{Taddia} {et~al.}(2015{\natexlab{b}}){Taddia}, {Sollerman},
  {Leloudas}, {Stritzinger}, {Valenti}, {Galbany}, {Kessler}, {Schneider}, \&
  {Wheeler}}]{2015A&A...574A..60T}
{Taddia}, F., {Sollerman}, J., {Leloudas}, G., {et~al.} 2015{\natexlab{b}},
  \aap, 574, A60

\bibitem[{{Taddia} {et~al.}(2018){Taddia}, {Stritzinger}, {Bersten}, {Baron},
  {Burns}, {Contreras}, {Holmbo}, {Hsiao}, {Morrell}, {Phillips}, {Sollerman},
  \& {Suntzeff}}]{2018A&A...609A.136T}
{Taddia}, F., {Stritzinger}, M.~D., {Bersten}, M., {et~al.} 2018, \aap, 609,
  A136

\bibitem[{{Taddia} {et~al.}(2013){Taddia}, {Stritzinger}, {Sollerman},
  {Phillips}, {Anderson}, {Boldt}, {Campillay}, {Castell{\'o}n}, {Contreras},
  {Folatelli}, {Hamuy}, {Heinrich-Josties}, {Krzeminski}, {Morrell}, {Burns},
  {Freedman}, {Madore}, {Persson}, \& {Suntzeff}}]{2013A&A...555A..10T}
{Taddia}, F., {Stritzinger}, M.~D., {Sollerman}, J., {et~al.} 2013, \aap, 555,
  A10

\bibitem[{{Taggart} \& {Perley}(2021)}]{2021MNRAS.503.3931T}
{Taggart}, K. \& {Perley}, D.~A. 2021, \mnras, 503, 3931

\bibitem[{{Taibi} {et~al.}(2022){Taibi}, {Battaglia}, {Leaman}, {Brooks},
  {Riggs}, {Munshi}, {Revaz}, \& {Jablonka}}]{2022A&A...665A..92T}
{Taibi}, S., {Battaglia}, G., {Leaman}, R., {et~al.} 2022, \aap, 665, A92

\bibitem[{{The Dark Energy Survey Collaboration}(2005)}]{2005astro.ph.10346T}
{The Dark Energy Survey Collaboration}. 2005, arXiv e-prints, astro

\bibitem[{{Thone} {et~al.}(2015){Thone}, {de Ugarte Postigo}, {Garcia-Benito},
  {Leloudas}, {Schulze}, \& {Amorin}}]{2015MNRAS.451L..65T}
{Thone}, C.~C., {de Ugarte Postigo}, A., {Garcia-Benito}, R., {et~al.} 2015,
  \mnras, 451, L65

\bibitem[{{Turatto} {et~al.}(1993){Turatto}, {Cappellaro}, {Danziger},
  {Benetti}, {Gouiffes}, \& {della Valle}}]{1993MNRAS.262..128T}
{Turatto}, M., {Cappellaro}, E., {Danziger}, I.~J., {et~al.} 1993, \mnras, 262,
  128

\bibitem[{{Valenti} {et~al.}(2016){Valenti}, {Howell}, {Stritzinger}, {Graham},
  {Hosseinzadeh}, {Arcavi}, {Bildsten}, {Jerkstrand}, {McCully}, {Pastorello},
  {Piro}, {Sand}, {Smartt}, {Terreran}, {Baltay}, {Benetti}, {Brown},
  {Filippenko}, {Fraser}, {Rabinowitz}, {Sullivan}, \&
  {Yuan}}]{2016MNRAS.459.3939V}
{Valenti}, S., {Howell}, D.~A., {Stritzinger}, M.~D., {et~al.} 2016, \mnras,
  459, 3939

\bibitem[{{Van Dyk} {et~al.}(2012){Van Dyk}, {Davidge}, {Elias-Rosa},
  {Taubenberger}, {Li}, {Levesque}, {Howerton}, {Pignata}, {Morrell}, {Hamuy},
  \& {Filippenko}}]{2012AJ....143...19V}
{Van Dyk}, S.~D., {Davidge}, T.~J., {Elias-Rosa}, N., {et~al.} 2012, \aj, 143,
  19

\bibitem[{{Van Dyk} {et~al.}(2014){Van Dyk}, {Zheng}, {Fox}, {Cenko}, {Clubb},
  {Filippenko}, {Foley}, {Miller}, {Smith}, {Kelly}, {Lee}, {Ben-Ami}, \&
  {Gal-Yam}}]{2014AJ....147...37V}
{Van Dyk}, S.~D., {Zheng}, W., {Fox}, O.~D., {et~al.} 2014, \aj, 147, 37

\bibitem[{{van Zee} {et~al.}(1998){van Zee}, {Salzer}, \&
  {Haynes}}]{1998ApJ...497L...1V}
{van Zee}, L., {Salzer}, J.~J., \& {Haynes}, M.~P. 1998, \apjl, 497, L1

\bibitem[{Virtanen {et~al.}(2020)Virtanen, Gommers, Oliphant, Haberland, Reddy,
  Cournapeau, Burovski, Peterson, Weckesser, Bright, {van der Walt}, Brett,
  Wilson, Millman, Mayorov, Nelson, Jones, Kern, Larson, Carey, Polat, Feng,
  Moore, {VanderPlas}, Laxalde, Perktold, Cimrman, Henriksen, Quintero, Harris,
  Archibald, Ribeiro, Pedregosa, {van Mulbregt}, \& {SciPy 1.0
  Contributors}}]{2020SciPy-NMeth}
Virtanen, P., Gommers, R., Oliphant, T.~E., {et~al.} 2020, Nature Methods, 17,
  261

\bibitem[{{Weilbacher} {et~al.}(2014){Weilbacher}, {Streicher}, {Urrutia},
  {P{\'e}contal-Rousset}, {Jarno}, \& {Bacon}}]{2014ASPC..485..451W}
{Weilbacher}, P.~M., {Streicher}, O., {Urrutia}, T., {et~al.} 2014, in
  Astronomical Society of the Pacific Conference Series, Vol. 485, Astronomical
  Data Analysis Software and Systems XXIII, ed. N.~{Manset} \& P.~{Forshay},
  451

\bibitem[{{Woosley} \& {Bloom}(2006)}]{2006ARA&A..44..507W}
{Woosley}, S.~E. \& {Bloom}, J.~S. 2006, \araa, 44, 507

\bibitem[{{Woosley} \& {Weaver}(1986)}]{1986ARA&A..24..205W}
{Woosley}, S.~E. \& {Weaver}, T.~A. 1986, \araa, 24, 205

\bibitem[{{Xiao} {et~al.}(2019){Xiao}, {Galbany}, {Eldridge}, \&
  {Stanway}}]{2019MNRAS.482..384X}
{Xiao}, L., {Galbany}, L., {Eldridge}, J.~J., \& {Stanway}, E.~R. 2019, \mnras,
  482, 384

\bibitem[{{Xiao} {et~al.}(2018){Xiao}, {Stanway}, \&
  {Eldridge}}]{2018MNRAS.477..904X}
{Xiao}, L., {Stanway}, E.~R., \& {Eldridge}, J.~J. 2018, \mnras, 477, 904

\bibitem[{{Yoon}(2017)}]{2017MNRAS.470.3970Y}
{Yoon}, S.-C. 2017, \mnras, 470, 3970

\bibitem[{{Yoon} {et~al.}(2010){Yoon}, {Woosley}, \&
  {Langer}}]{2010ApJ...725..940Y}
{Yoon}, S.~C., {Woosley}, S.~E., \& {Langer}, N. 2010, \apj, 725, 940

\bibitem[{{York} {et~al.}(2000){York}, {Adelman}, {Anderson}, {Anderson},
  {Annis}, {Bahcall}, {Bakken}, {Barkhouser}, {Bastian}, {Berman}, {Boroski},
  {Bracker}, {Briegel}, {Briggs}, {Brinkmann}, {Brunner}, {Burles}, {Carey},
  {Carr}, {Castander}, {Chen}, {Colestock}, {Connolly}, {Crocker}, {Csabai},
  {Czarapata}, {Davis}, {Doi}, {Dombeck}, {Eisenstein}, {Ellman}, {Elms},
  {Evans}, {Fan}, {Federwitz}, {Fiscelli}, {Friedman}, {Frieman}, {Fukugita},
  {Gillespie}, {Gunn}, {Gurbani}, {de Haas}, {Haldeman}, {Harris}, {Hayes},
  {Heckman}, {Hennessy}, {Hindsley}, {Holm}, {Holmgren}, {Huang}, {Hull},
  {Husby}, {Ichikawa}, {Ichikawa}, {Ivezi{\'c}}, {Kent}, {Kim}, {Kinney},
  {Klaene}, {Kleinman}, {Kleinman}, {Knapp}, {Korienek}, {Kron}, {Kunszt},
  {Lamb}, {Lee}, {Leger}, {Limmongkol}, {Lindenmeyer}, {Long}, {Loomis},
  {Loveday}, {Lucinio}, {Lupton}, {MacKinnon}, {Mannery}, {Mantsch}, {Margon},
  {McGehee}, {McKay}, {Meiksin}, {Merelli}, {Monet}, {Munn}, {Narayanan},
  {Nash}, {Neilsen}, {Neswold}, {Newberg}, {Nichol}, {Nicinski}, {Nonino},
  {Okada}, {Okamura}, {Ostriker}, {Owen}, {Pauls}, {Peoples}, {Peterson},
  {Petravick}, {Pier}, {Pope}, {Pordes}, {Prosapio}, {Rechenmacher}, {Quinn},
  {Richards}, {Richmond}, {Rivetta}, {Rockosi}, {Ruthmansdorfer}, {Sandford},
  {Schlegel}, {Schneider}, {Sekiguchi}, {Sergey}, {Shimasaku}, {Siegmund},
  {Smee}, {Smith}, {Snedden}, {Stone}, {Stoughton}, {Strauss}, {Stubbs},
  {SubbaRao}, {Szalay}, {Szapudi}, {Szokoly}, {Thakar}, {Tremonti}, {Tucker},
  {Uomoto}, {Vanden Berk}, {Vogeley}, {Waddell}, {Wang}, {Watanabe},
  {Weinberg}, {Yanny}, {Yasuda}, \& {SDSS Collaboration}}]{2000AJ....120.1579Y}
{York}, D.~G., {Adelman}, J., {Anderson}, John~E., J., {et~al.} 2000, \aj, 120,
  1579

\bibitem[{{Zapartas} {et~al.}(2017){Zapartas}, {de Mink}, {Izzard}, {Yoon},
  {Badenes}, {G{\"o}tberg}, {de Koter}, {Neijssel}, {Renzo}, {Schootemeijer},
  \& {Shrotriya}}]{2017A&A...601A..29Z}
{Zapartas}, E., {de Mink}, S.~E., {Izzard}, R.~G., {et~al.} 2017, \aap, 601,
  A29

\end{thebibliography}

% \end{thebibliography}

\appendix

\section{BPT analysis} \label{app:BPT}

Figure \ref{fig:BPT} shows a BPT diagram of all \ion{H}{ii} regions (blue circles) extracted from the galaxies in our sample. The dashed line marks the active galactic nucleus (AGN) ionization region defined by \citet{2001ApJ...556..121K} while the solid line shows the composite region limit defined by \citet{2003MNRAS.346.1055K}. The orange circles in Figure \ref{fig:BPT} show the position of the CCSNe in our sample and the green circles mark the SNe that fall above the composite region line: ASASSN-14ha (NGC 1566), SN 2014eh (NGC 6907), and SN 2016bev (ESO 560-G013). Since we are interested in the properties of regions that are ionized only by star formation, we exclude these three SNe of our analysis as they might be contaminated by other ionizing sources. In Paper II we will present an analysis of all \ion{H}{ii}  regions in detail, and compare the SN host \ion{H}{ii}  regions to all the other star-forming regions within their galaxies. 

\begin{figure}[t!]
\centering
\includegraphics[width=0.45\textwidth]{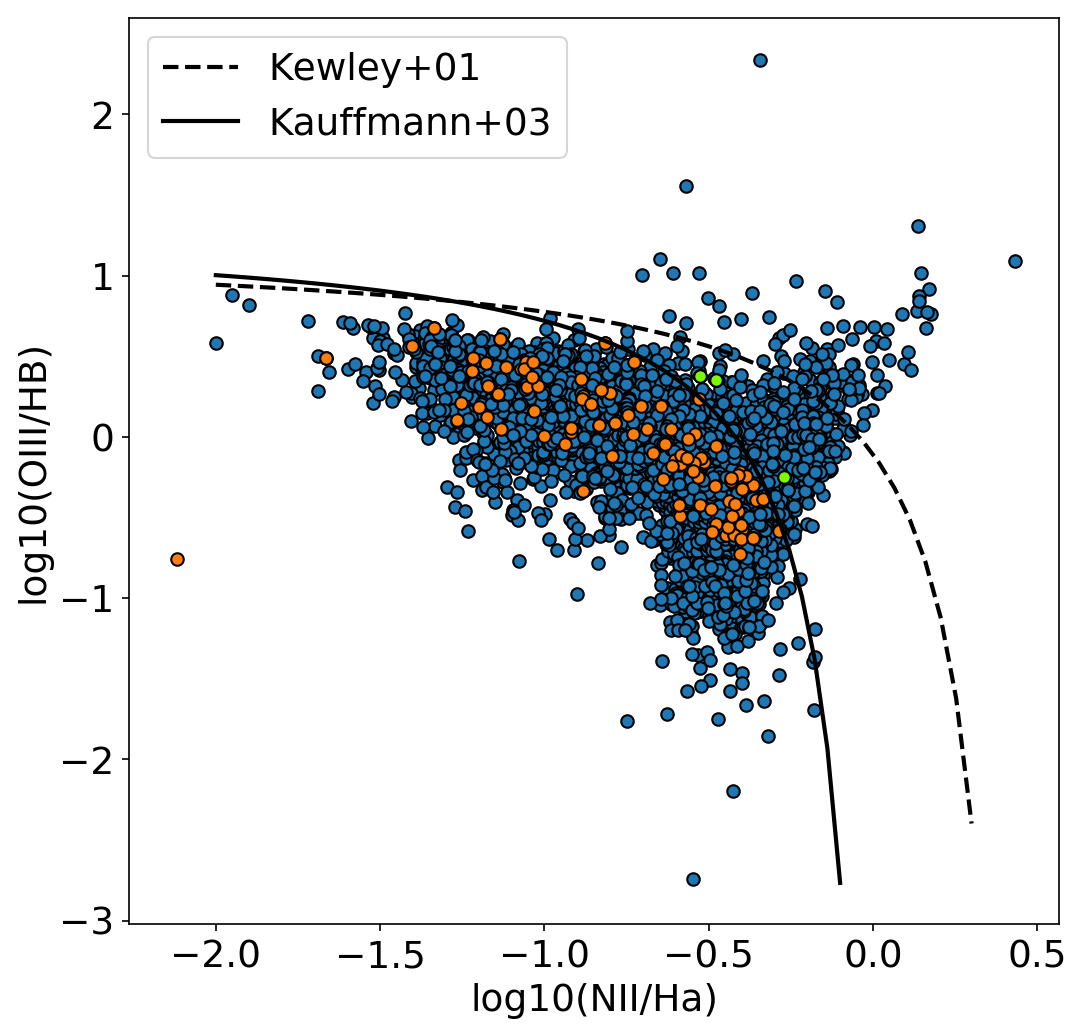}
\caption{BPT diagram of the \ion{H}{ii} regions in our sample. The blue circles show all the \ion{H}{ii} regions, while the orange circles mark the position of the \ion{H}{ii} regions associated to the CCSNe. The green circles show the SNe that are above the composite region (solid line) defined by \citet{2003MNRAS.346.1055K}. We also show the AGN ionization region as the dashed line, as defined by \citet{2001ApJ...556..121K}. \label{fig:BPT}}
\end{figure}

\section{Parameter tests} \label{app:param_tests}

We tested \textsc{IFUanal} in three galaxies in order to estimate how the selected \ion{H}{ii} segmentation parameters could affect the final result for the physical parameters. The galaxies (ESO 381-IG048/ASASSN-15bb, NGC 0988/SN 2017gmr, and NGC 2466/ASASSN-14dd) were selected to be representative of the full sample: one galaxy with a small projected size; one nearby spiral galaxy; and one spiral galaxy with a large projected size. We performed five runs of the code for each galaxy, varying the maximum bin radius, the minimum H$\alpha$ flux threshold to define the seed peaks, and the maximum flux threshold to where the region will expand. We compare the extracted results for all the selected bins in each galaxy in Figure \ref{fig:ASASSN-15bb_param_tests}, where we show the normalized cumulative distributions for H$\alpha$ flux and EW, the oxygen abundance given by the D16 indicator, and the estimated host extinction.

Table \ref{tab:param_tests} lists the different sets of parameters used for the tests. Sets 1 and 2 have fixed values for the minimum and maximum H$\alpha$ flux thresholds, relative to the standard deviation of the H$\alpha$ flux in a empty region of the cube, while the maximum radii are set to 6 or 10 pixels, respectively. The sets 3 and 4 have a maximum radius of 6 pixels, while the maximum H$\alpha$ flux threshold varies between $8 \sigma$ and $12 \sigma$, respectively, and the minimum H$\alpha$ flux threshold varies between $1.5 \sigma$ and $4.5 \sigma$, respectively.
We also used the \ion{H}{ii} region segmentation obtained through  \textsc{py\ion{H}{ii} extractor} \citep{2022RASTI...1....3L}, which allows the bin regions to expand with no predefined maximum radius or threshold in the flux. We show the results from this segmentation as the set of parameters 5 in Figure \ref{fig:ASASSN-15bb_param_tests}, and report them as varying parameters in Table \ref{tab:param_tests}.

Figure \ref{fig:ASASSN-15bb_param_tests} shows the cumulative distributions for H$\alpha$ flux. The galaxy ESO 381-IG048 presents the largest scatter between the distributions, with the set of parameters 3 showing the lowest median. This is also true for the oxygen abundance and the H$\alpha$ EW, shown in the second and third panels, respectively. Lowering the H$\alpha$ flux thresholds includes more pixels with lower values of H$\alpha$ flux, EW, and oxygen abundance. 
The results for NGC 0988 and NGC 2466, on the other hand, do not show significant differences between the distinct tests. 
These tests show that the input parameters in \textsc{IFUanal} might have a larger effect on the final analysis in smaller galaxies, where less \ion{H}{ii}  regions are selected. This should be taken as a caveat for the analyses presented in this work and for the conclusions taken from them. 

We set an intermediate minimum threshold of 10 times the standard deviation of the H$\alpha$ flux in a  background region of the datacube, and a limiting threshold of three times the standard deviation of the H$\alpha$ flux to where the region can expand. Given our results, varying these parameters around the chosen values do not produce significant differences in the output for the majority of the galaxies.

\begin{figure*}[t!]
\centering
\includegraphics[width=0.9\textwidth]{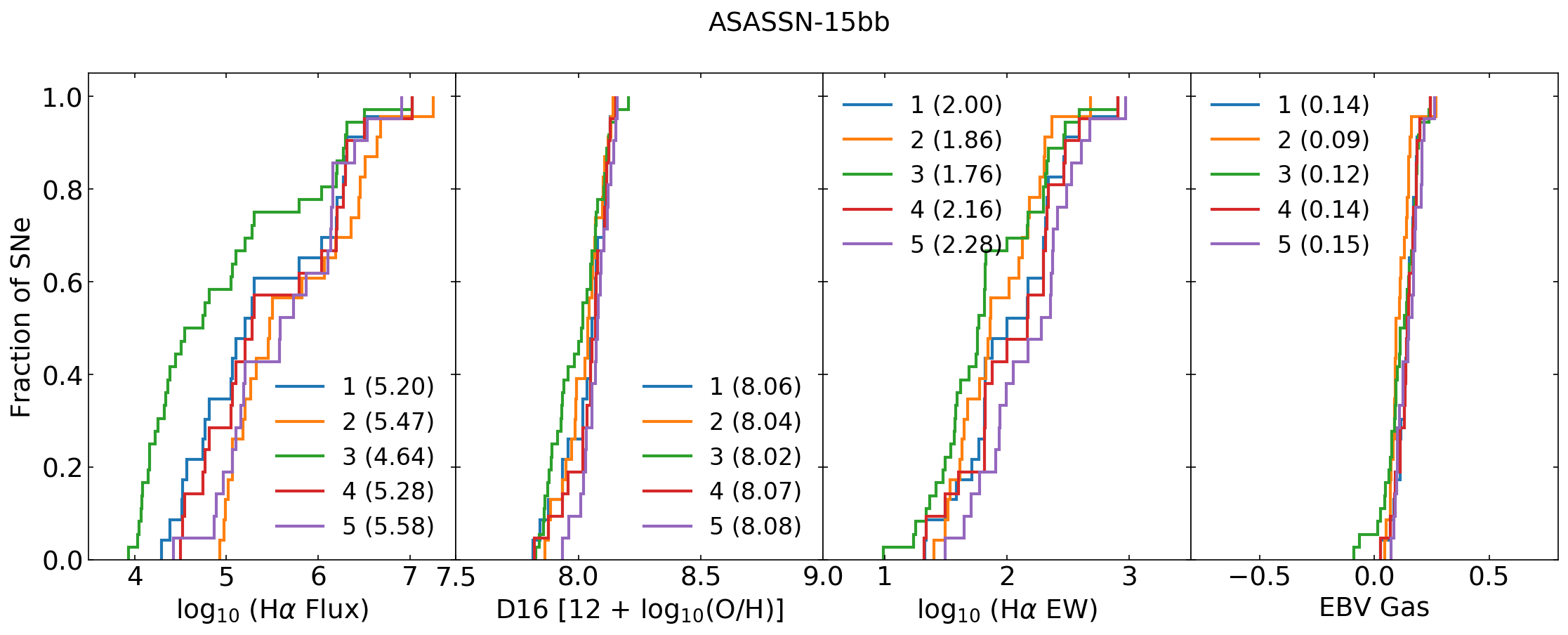}
\includegraphics[width=0.9\textwidth]{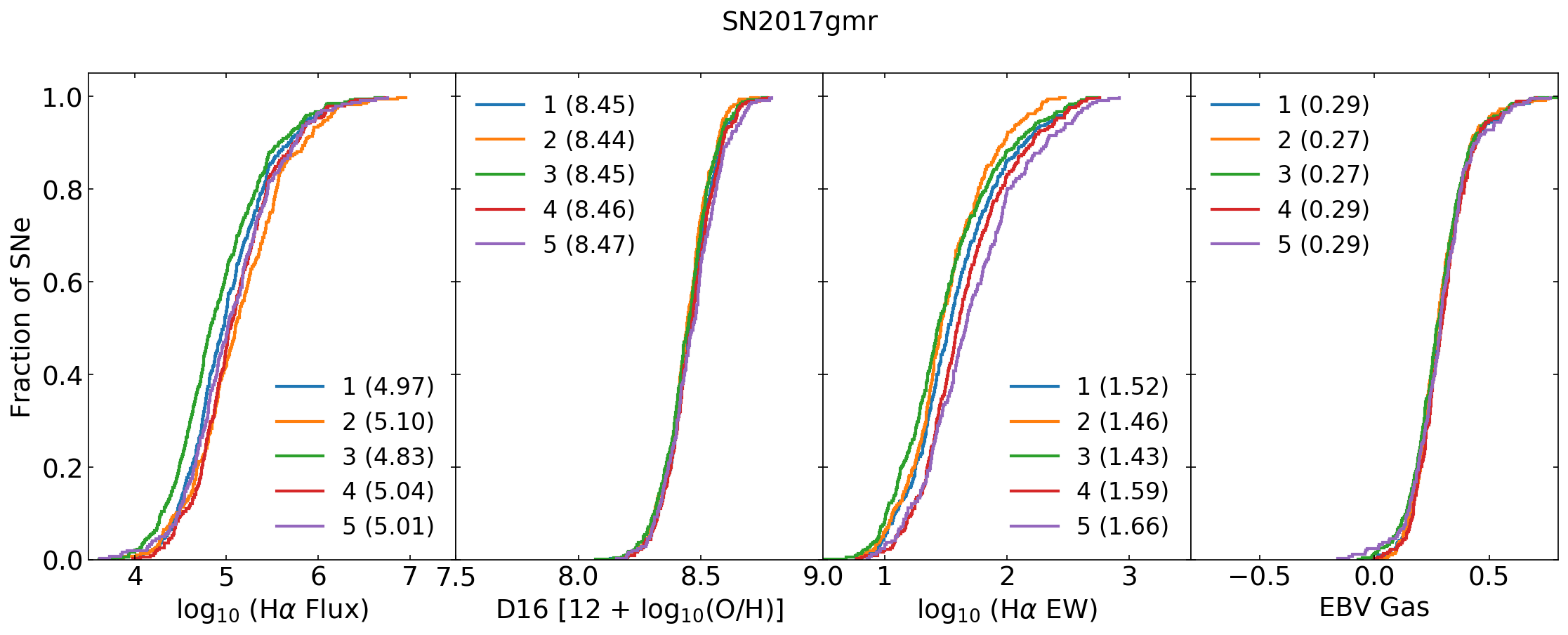}
\includegraphics[width=0.9\textwidth]{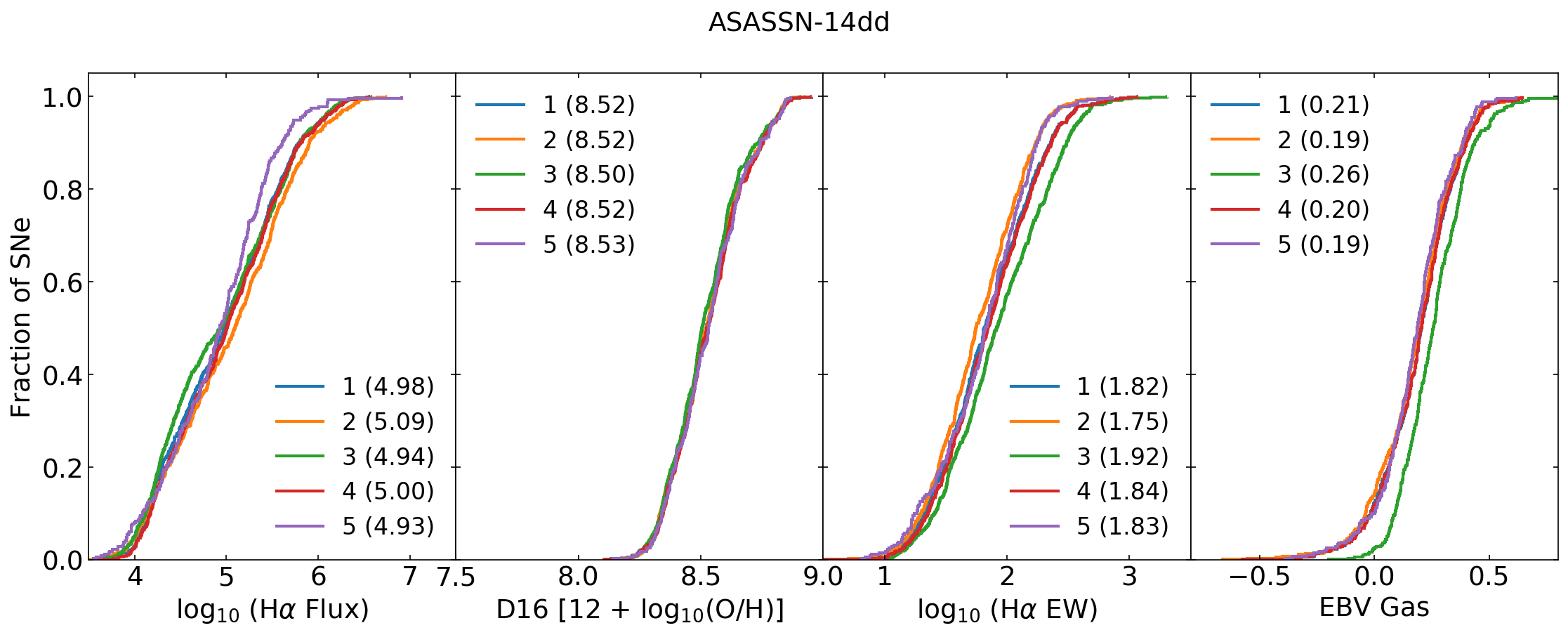}
\caption{Cumulative distributions for the different tests with the objects ESO 381-IG048/ASASSN-15bb (top), NGC 0988/SN 2017gmr (middle), and NGC 2466/ASASSN-14dd (bottom). The tests were run using the parameters described in \ref{tab:param_tests}. \label{fig:ASASSN-15bb_param_tests}}
\end{figure*}

\begin{table}
\caption{\textsc{IFUanal} parameters used in the tests. \label{tab:param_tests}}
\centering
\begin{tabular}{cccc}
\hline
Set & R$_{max}$ [px] & F$_{H\alpha, min}$ [$\times \sigma$] & F$_{H\alpha, max}$ [$\times \sigma$] \\
\hline
1 & 6 & 3 & 10 \\
2 & 12 & 3 & 10 \\
3 & 6 & 1.5 & 8 \\
4 & 6 & 4.5 & 12 \\
5 & Var. & Var. & Var. \\
\hline
\end{tabular}
\end{table}

\section{Metallicity gradients} \label{app:met_gradients}

In Figure \ref{fig:Z_grad_1} we show an example of the oxygen abundance gradient fit for ASASSN-16ba (hosted by MCG-03-25-015). 
The same method was applied for all the SNe where the nearest \ion{H}{ii} region was at a projected distance greater than $1 \times 10^3$~pc: ASASSN-15jp, SN2018yo, ASASSN-17oj, SN2017gbv, ASASSN-16ba, ASASSN-16al, LSQ15xp, SN2015W, and ASASSN-18ou.
A straight line is fit to all the oxygen abundances obtained from the \ion{H}{ii} regions. 
{The line fitting is weighted by the measured uncertainties from the oxygen abundance measurements. We note that systematic uncertainties from each index could input a larger error in the gradient fittings.}
We use pixel deprojected distances from the center of the galaxy following the method from \citet[][]{2009A&A...508.1259H}. 
Values of position angle (PA) and inclination angle ($i$) of each galaxy are taken from \textsc{HyperLeda} \citep[][]{2014A&A...570A..13M}. The resulting oxygen abundances in the D16, N2 and O3N2 indexes are reported in Table \ref{tab:HII_H_prop}.

\begin{figure*}[t!]
\centering
\includegraphics[width=0.92\textwidth]{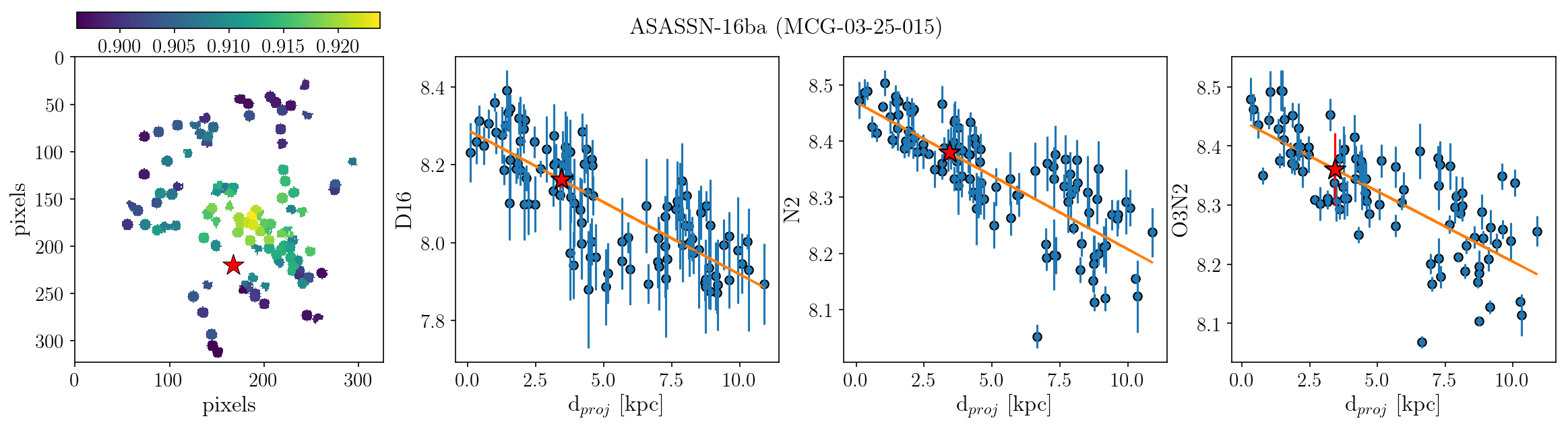}
\caption{Oxygen abundances fit for ASASSN-16ba, hosted by MCG-03-25-015.
The same method was applied for all the SNe where the extracted spectrum at their position has a SNR of $< 3$. The Left panel shows the spatial distribution in pixels of the \ion{H}{ii} regions extracted in the galaxy, where the color gradient indicates the oxygen abundance given by the D16 index. The spatial coordinates of the SN in relation to the \ion{H}{ii} regions is given by the red star. The other three panels show, respectively from left to right, the resulting fit to the oxygen abundance gradient in the D16, N2, and O3N2 indexes. The resulting value for the SN is given by the red star. The $x$-axis show the deprojected distance from the nucleus in kpc for each \ion{H}{ii}  region. \label{fig:Z_grad_1}} 
\end{figure*}

\section{Measured fluxes and physical parameters} \label{sec:fluxes_params}

In Table \ref{tab:HII_emission}, we report the measured fluxes of H$\alpha$, H$\beta$, [\ion{O}{III}] $\lambda 5007$, [\ion{N}{II}]$\lambda 6584$, and [\ion{S}{II}]$\lambda 6716$ for the \ion{H}{ii} region related to each SN. The individual line fluxes were corrected for Galactic, and host galaxy reddening. Table \ref{tab:HII_H_prop} reports the resulting H$\alpha$ EW, oxygen abundances, $\Sigma$SFR, and host galaxy extinction for the \ion{H}{ii} region related to each SN.

\begin{table*}
\caption{Measured fluxes of different emission lines at the neares \ion{H}{ii}  region to each SN. \label{tab:HII_emission}}
\centering
\begin{tabular}{lccccc}
\hline
SN Name & H$\alpha$ & H$\beta$ & [\ion{O}{III}]$\lambda 5007$ & [\ion{N}{II}]$\lambda 6584$ & [\ion{S}{II}]$\lambda 6716$ \\
\hline
SN2018eog & 19.92 (1.12) & 6.97 (2.02) & 3.80 (1.34) & 4.58 (1.05) & 5.73 (0.90) \\
SN2018ant & 61.91 (3.94) & 11.90 (24.92) & 12.46 (7.27) & 17.76 (3.58) & 17.35 (3.25) \\
ASASSN-18oa & 189.38 (3.61) & 66.22 (5.17) & 45.88 (4.45) & 57.99 (2.66) & 46.54 (2.31) \\
SN2018bl & 382.70 (5.33) & 133.82 (6.65) & 89.50 (5.41) & 98.01 (3.10) & 70.51 (3.34) \\
SN2018evy & 2398.58 (24.23) & 838.73 (35.76) & 157.37 (25.65) & 941.54 (16.49) & 348.78 (12.82) \\
SN2018cho & 140.11 (2.43) & 48.99 (6.81) & 19.77 (6.65) & 61.52 (2.10) & 27.41 (1.62) \\
SN2018ie & 629.83 (5.17) & 220.18 (6.65) & 150.08 (5.14) & 164.01 (2.83) & 111.36 (2.39) \\
SN2018yo & - & - & 35.18 (4.04) & 42.77 (2.60) & 33.53 (2.10) \\
... \\ 
\hline
\end{tabular}
\tablefoot{The entire table is available online. Flux units are given in 10$^{-17}$ erg s$^{-1}$ cm$^{-2}$ $\AA^{-1}$.}
\end{table*}

\begin{table*}
\caption{Physical properties of the \ion{H}{ii}  region related to each SN. \label{tab:HII_H_prop}}
\centering
\begin{tabular}{lcccccc}
\hline
SN Name & H$\alpha$ EW & D16 & N2 & O3N2 & $\Sigma$SFR &  $E(B-V)$ \\
 & (\AA) & ($12 \ +$ log(O/H)) & ($12 \ +$ log(O/H)) & ($12 \ +$ log(O/H)) & (log M$_\odot$ yr$^{-1}$ kpc$^{-2}$)  & (mag)  \\ 
\hline
SN2018eog & 3.45 (0.26) & 8.28 (0.12) & 8.45 (0.05) & - & -3.68 (-0.00) &  - \\
SN2018ant & 8.94 (0.75) & 8.37 (0.11) & 8.49 (0.04) & - & -2.90 (-0.02) &  - \\
ASASSN-18oa & 27.22 (0.79) & 8.49 (0.03) & 8.51 (0.01) & 8.46 (0.01) & -2.34 (-0.05) & 0.16 (0.07) \\
SN2018bl & 171.54 (5.87) & 8.54 (0.02) & 8.47 (0.01) & 8.44 (0.01) & -1.47 (-0.52) & 0.24 (0.05) \\
SN2018evy & 108.29 (3.37) & 8.86 (0.01) & 8.56 (0.00) & 8.60 (0.02) & -1.22 (-1.09) & 0.44 (0.04) \\
SN2018cho & 17.32 (0.52) & 8.79 (0.02) & 8.58 (0.01) & - & -1.90 (-0.15) & 0.27 (0.13) \\
SN2018ie & 342.46 (19.46) & 8.56 (0.01) & 8.47 (0.00) & 8.44 (0.00) & -1.15 (-1.35) & 0.28 (0.03) \\
SN2018yo & - & 8.28 (0.03) & 8.36 (0.05) & 8.36 (0.05) & - & - \\
... \\ 
\hline
\end{tabular}
\tablefoot{The entire table is available online. {Some CCSNe could not have their oxygen abundance in the O3N2 index and extinction estimated due to the low SNR ($<3$) of H$\beta$ or [\ion{O}{III}]$\lambda5007$ in their spectra.}}
\end{table*}

\section{Statistics results, KS, and AD tests.}\label{app:stats_KS_AD}

In Table \ref{tab:Z_stats} we show the number of events, median, and average values for each distribution used in the analysis for oxygen abundance. In Tables \ref{tab:Z_KS} and \ref{tab:Z_AD} we report, respectively, the results for the KS and AD tests for oxygen abundance. 
The KS and AD tests were performed through \textsc{SciPy} \citep[version 1.0][]{2020SciPy-NMeth}. 
In Table \ref{tab:EW_stats}
The same is reported for H$\alpha$ EW in Tables \ref{tab:EW_stats}, \ref{tab:EW_KS}, and \ref{tab:EW_AD}, $\Sigma$SFR in Tables \ref{tab:SFR_stats}, \ref{tab:SFR_KS}, and \ref{tab:SFR_AD}, and for $E(B-V)_{host}$ in Tables \ref{tab:ebv_stats}, \ref{tab:ebv_KS}, and \ref{tab:ebv_AD}.

\begin{table*}
\caption{Oxygen abundance statistics.\label{tab:Z_stats}}
\centering
\begin{tabular}{lccccccccc}
\hline
SN Type & D16 N & D16 Median & D16 Average
& O3N2 N & O3N2 Median & O3N2 Average
& N2 N & N2 Median & N2 Average\\
\hline
II & 78 & 8.28 (0.41) & 8.31 (0.33) & 78 & 8.34 (0.45) & 8.34 (0.36) & 78 & 8.41 (0.41) & 8.38 (0.33) \\
IIn & 9 & 8.35 (1.19) & 8.30 (0.95) & 9 & 8.39 (1.29) & 8.39 (1.03) & 9 & 8.38 (1.21) & 8.41 (0.96) \\
Ic & 6 & 8.55 (1.46) & 8.50 (1.16) & 6 & 8.44 (1.62) & 8.40 (1.29) & 6 & 8.49 (1.47) & 8.45 (1.18) \\
Ib & 7 & 8.43 (1.39) & 8.39 (1.11) & 7 & 8.22 (1.62) & 8.29 (1.29) & 7 & 8.46 (1.38) & 8.38 (1.10) \\
IIb & 7 & 8.68 (1.36) & 8.43 (1.09) & 7 & 8.40 (1.48) & 8.37 (1.18) & 7 & 8.50 (1.37) & 8.42 (1.09) \\
Ibn & 3 & 8.57 (2.10) & 8.43 (1.67) & 3 & 8.43 (2.10) & 8.43 (1.68) & 3 & 8.47 (2.10) & 8.48 (1.68) \\
II & 78 & 8.28 (0.41) & 8.31 (0.33) & 78 & 8.34 (0.45) & 8.34 (0.36) & 78 & 8.41 (0.41) & 8.38 (0.33) \\
IIn/Ibn & 9 & 8.32 (1.04) & 8.30 (0.83) & 9 & 8.39 (1.10) & 8.39 (0.87) & 9 & 8.44 (1.05) & 8.43 (0.84) \\
SESNe & 6 & 8.55 (0.80) & 8.50 (0.64) & 6 & 8.44 (0.88) & 8.40 (0.70) & 6 & 8.49 (0.80) & 8.42 (0.64) \\
\hline
\end{tabular}
\end{table*}

\begin{table*}
\caption{Oxygen abundance KS values.\label{tab:Z_KS}}
\centering
\begin{tabular}{lcccccc}
\hline
KS test & D16 $D$-value & D16 $p$-value & O3N2 $D$-value & O3N2 $p$-value & N2 $D$-value & N2 $p$-value \\
\hline
II x IIn & 0.34 & 0.32 & 0.21 & 0.80 & 0.34 & 0.32 \\
II x Ic & 0.46 & 0.21 & 0.38 & 0.31 & 0.46 & 0.21 \\
II x Ib & 0.40 & 0.35 & 0.18 & 0.95 & 0.40 & 0.35 \\
II x IIb & 0.33 & 0.50 & 0.29 & 0.56 & 0.33 & 0.50 \\
II x Ibn & 0.61 & 0.15 & 0.55 & 0.25 & 0.61 & 0.15 \\
IIn x Ic & 0.42 & 0.50 & 0.39 & 0.56 & 0.42 & 0.50 \\
IIn x Ib & 0.47 & 0.40 & 0.32 & 0.71 & 0.47 & 0.40 \\
IIn x IIb & 0.38 & 0.64 & 0.24 & 0.92 & 0.38 & 0.64 \\
IIn x Ibn & 0.50 & 0.56 & 0.56 & 0.45 & 0.50 & 0.56 \\
Ic x Ib & 0.40 & 0.87 & 0.26 & 0.93 & 0.40 & 0.87 \\
Ic x IIb & 0.33 & 0.82 & 0.26 & 0.93 & 0.33 & 0.82 \\
Ic x Ibn & 0.27 & 1.00 & 0.17 & 1.00 & 0.27 & 1.00 \\
Ib x IIb & 0.50 & 0.36 & 0.29 & 0.96 & 0.50 & 0.36 \\
Ib x Ibn & 0.60 & 0.46 & 0.43 & 0.70 & 0.60 & 0.46 \\
IIb x Ibn & 0.50 & 0.68 & 0.43 & 0.70 & 0.50 & 0.68 \\
II x IIn/Ibn & 0.38 & 0.10 & 0.26 & 0.44 & 0.38 & 0.10 \\
II x SESNe & 0.24 & 0.33 & 0.22 & 0.36 & 0.24 & 0.33 \\
IIn/Ibn x SESNe & 0.32 & 0.33 & 0.20 & 0.85 & 0.32 & 0.39 \\
\hline
\end{tabular}
\end{table*}

\begin{table*}
\caption{Oxygen abundance AD values.\label{tab:Z_AD}}
\centering
\begin{tabular}{lcccccc}
\hline
AD test & D16 Statistic & D16 Significance & O3N2 Statistic & O3N2 Significance & N2 Statistic & N2 Significance \\
\hline
II x IIn & -1.10 & 0.25 & -0.91 & 0.25 & -0.26 & 0.25 \\
II x Ic & 1.24 & 0.10 & -0.27 & 0.25 & -0.04 & 0.25 \\
II x Ib & -0.56 & 0.25 & -1.07 & 0.25 & -0.15 & 0.25 \\
II x IIb & -0.28 & 0.25 & -0.47 & 0.25 & -0.67 & 0.25 \\
II x Ibn & -0.43 & 0.25 & -0.15 & 0.25 & 0.24 & 0.25 \\
IIn x Ic & 0.71 & 0.17 & -0.59 & 0.25 & -0.65 & 0.25 \\
IIn x Ib & -0.64 & 0.25 & -0.81 & 0.25 & -0.05 & 0.25 \\
IIn x IIb & -0.59 & 0.25 & -0.92 & 0.25 & -0.55 & 0.25 \\
IIn x Ibn & -0.67 & 0.25 & -0.40 & 0.25 & -0.54 & 0.25 \\
Ic x Ib & -0.01 & 0.25 & -1.06 & 0.25 & 0.11 & 0.25 \\
Ic x IIb & -0.89 & 0.25 & -1.06 & 0.25 & -0.39 & 0.25 \\
Ic x Ibn & -1.04 & 0.25 & -1.32 & 0.25 & -1.39 & 0.25 \\
Ib x IIb & -0.57 & 0.25 & -1.01 & 0.25 & 0.25 & 0.25 \\
Ib x Ibn & -1.02 & 0.25 & -0.90 & 0.25 & -0.00 & 0.25 \\
IIb x Ibn & -0.32 & 0.25 & -0.77 & 0.25 & -0.35 & 0.25 \\
II x IIn/Ibn & -1.04 & 0.25 & -0.37 & 0.25 & 0.54 & 0.20 \\
II x SESNe & 1.39 & 0.09 & 0.13 & 0.25 & -0.17 & 0.25 \\
IIn/Ibn x SESNe & -0.18 & 0.25 & -0.67 & 0.25 & -0.40 & 0.25 \\
\hline
\end{tabular}
\end{table*}

\begin{table}
\caption{H$\alpha$ EW statistics.\label{tab:EW_stats}}
\centering
\begin{tabular}{lccc}
\hline
SN Type & N & Median (\AA) & Average (\AA) \\
\hline
II & 61 & 62.32 (1.75) & 80.55 (1.40) \\
IIn & 8 & 63.76 (8.63) & 155.24 (6.88) \\
Ic & 6 & 220.80 (8.93) & 199.77 (7.12) \\
Ib & 6 & 87.27 (11.82) & 280.30 (9.43) \\
IIb & 6 & 139.80 (8.56) & 176.73 (6.83) \\
Ibn & 3 & 44.94 (11.43) & 103.07 (9.12) \\
II & 61 & 62.32 (1.75) & 80.55 (1.40) \\
IIn/Ibn & 11 & 50.28 (6.77) & 141.01 (5.40) \\
SESNe & 19 & 109.04 (5.67) & 211.86 (4.53) \\
\hline
\end{tabular}
\end{table}

\begin{table}
\caption{H$\alpha$ EW KS values.\label{tab:EW_KS}}
\centering
\begin{tabular}{lcc}
\hline
KS test & $D$-value & $p$-value  \\
\hline
II x IIn & 0.36 & 0.25 \\
II x Ic & 0.48 & 0.11 \\
II x Ib & 0.32 & 0.54 \\
II x IIb & 0.43 & 0.19 \\
II x Ibn & 0.32 & 0.84 \\
IIn x Ic & 0.38 & 0.64 \\
IIn x Ib & 0.38 & 0.64 \\
IIn x IIb & 0.33 & 0.78 \\
IIn x Ibn & 0.38 & 0.84 \\
Ic x Ib & 0.33 & 0.93 \\
Ic x IIb & 0.33 & 0.93 \\
Ic x Ibn & 0.50 & 0.68 \\
Ib x IIb & 0.33 & 0.93 \\
Ib x Ibn & 0.50 & 0.68 \\
IIb x Ibn & 0.50 & 0.68 \\
II x IIn/Ibn & 0.33 & 0.21 \\
II x SESNe & 0.36 & 0.04 \\
IIn/Ibn x SESNe & 0.33 & 0.33 \\
\hline
\end{tabular}
\end{table}

\begin{table}
\caption{H$\alpha$ EW AD values.\label{tab:EW_AD}}
\centering
\begin{tabular}{lcc}
\hline
AD test & Statistic & Significance   \\
\hline
II x IIn & 0.74 & 0.16 \\
II x Ic & 2.45 & 0.03 \\
II x Ib & 0.95 & 0.13 \\
II x IIb & 2.05 & 0.05 \\
II x Ibn & -0.38 & 0.25 \\
IIn x Ic & -0.44 & 0.25 \\
IIn x Ib & -0.19 & 0.25 \\
IIn x IIb & -0.62 & 0.25 \\
IIn x Ibn & -0.95 & 0.25 \\
Ic x Ib & -0.23 & 0.25 \\
Ic x IIb & -0.92 & 0.25 \\
Ic x Ibn & -0.17 & 0.25 \\
Ib x IIb & -0.94 & 0.25 \\
Ib x Ibn & -0.19 & 0.25 \\
IIb x Ibn & -0.40 & 0.25 \\
II x IIn/Ibn & 0.43 & 0.22 \\
II x SESNe & 4.48 & 0.0054 \\
IIn/Ibn x SESNe & 0.67 & 0.17 \\
\hline
\end{tabular}
\end{table}

\begin{table}
\caption{$\Sigma$SFR statistics.\label{tab:SFR_stats}}
\centering
\begin{tabular}{lccc}
\hline
SN Type & N & Median & Average \\
 &  & (M$_\odot$ yr$^{-1}$ kpc$^{-2}$) & (M$_\odot$ yr$^{-1}$ kpc$^{-2}$) \\
\hline
II & 61 & -1.87 (0.11) & -1.87 (0.09) \\
IIn & 8 & -2.13 (0.17) & -1.83 (0.14) \\
Ic & 6 & -1.11 (0.50) & -1.51 (0.40) \\
Ib & 6 & -1.50 (0.30) & -1.31 (0.24) \\
IIb & 6 & -1.44 (0.57) & -1.56 (0.45) \\
Ibn & 3 & -1.97 (1.02) & -1.99 (0.81) \\
II & 61 & -1.87 (0.11) & -1.87 (0.09) \\
IIn/Ibn & 11 & -2.03 (0.15) & -1.87 (0.12) \\
SESNe & 19 & -1.27 (0.23) & -1.48 (0.18) \\
\hline
\end{tabular}
\end{table}

\begin{table}
\caption{$\Sigma$SFR KS values.\label{tab:SFR_KS}}
\centering
\begin{tabular}{lcc}
\hline
KS test & $D$-value & $p$-value \\
\hline
II x IIn & 0.22 & 0.83 \\
II x Ic & 0.49 & 0.11 \\
II x Ib & 0.29 & 0.64 \\
II x IIb & 0.27 & 0.72 \\
II x Ibn & 0.57 & 0.21 \\
IIn x Ic & 0.54 & 0.19 \\
IIn x Ib & 0.46 & 0.38 \\
IIn x IIb & 0.33 & 0.78 \\
IIn x Ibn & 0.50 & 0.56 \\
Ic x Ib & 0.33 & 0.93 \\
Ic x IIb & 0.33 & 0.93 \\
Ic x Ibn & 0.67 & 0.33 \\
Ib x IIb & 0.33 & 0.93 \\
Ib x Ibn & 0.83 & 0.10 \\
IIb x Ibn & 0.67 & 0.33 \\
II x IIn/Ibn & 0.30 & 0.30 \\
II x SESNe & 0.26 & 0.22 \\
IIn/Ibn x SESNe & 0.46 & 0.07 \\
\hline
\end{tabular}
\end{table}

\begin{table}
\caption{$\Sigma$SFR AD values.\label{tab:SFR_AD}}
\centering
\begin{tabular}{lcc}
\hline
AD test & Statistic & Significance   \\
\hline
II x IIn & -0.73 & 0.25 \\
II x Ic & 0.58 & 0.19 \\
II x Ib & 0.23 & 0.25 \\
II x IIb & -0.45 & 0.25 \\
II x Ibn & 0.21 & 0.25 \\
IIn x Ic & -0.03 & 0.25 \\
IIn x Ib & -0.03 & 0.25 \\
IIn x IIb & -0.65 & 0.25 \\
IIn x Ibn & -0.32 & 0.25 \\
Ic x Ib & -0.59 & 0.25 \\
Ic x IIb & -0.23 & 0.25 \\
Ic x Ibn & 0.21 & 0.25 \\
Ib x IIb & -0.91 & 0.25 \\
Ib x Ibn & 1.26 & 0.10 \\
IIb x Ibn & 0.21 & 0.25 \\
II x IIn/Ibn & -0.39 & 0.25 \\
II x SESNe & 1.11 & 0.11 \\
IIn/Ibn x SESNe & 1.00 & 0.13 \\
\hline
\end{tabular}
\end{table}

\begin{table}
\caption{Extinction statistics.\label{tab:ebv_stats}}
\centering
\begin{tabular}{lccc}
\hline
SN Type & N & Median & Average \\
& & (mag) & (mag) \\
\hline
II & 47 & 0.23 (0.10) & 0.23 (0.08) \\
IIn & 7 & 0.25 (0.24) & 0.23 (0.19) \\
Ic & 6 & 0.20 (0.39) & 0.31 (0.31) \\
Ib & 6 & 0.20 (0.35) & 0.26 (0.28) \\
IIb & 5 & 0.32 (0.32) & 0.32 (0.26) \\
Ibn & 1 & 0.27 (0.65) & 0.27 (0.52) \\
II & 47 & 0.23 (0.10) & 0.23 (0.08) \\
IIn/Ibn & 8 & 0.26 (0.22) & 0.23 (0.18) \\
SESNe & 18 & 0.22 (0.21) & 0.30 (0.16) \\
\hline
\end{tabular}
\end{table}

\begin{table}
\caption{Extinction KS values.\label{tab:ebv_KS}}
\centering
\begin{tabular}{lcc}
\hline
KS test & $D$-value & $p$-value \\
\hline
II x IIn & 0.28 & 0.64 \\
II x Ic & 0.17 & 0.99 \\
II x Ib & 0.22 & 0.92 \\
II x IIb & 0.34 & 0.54 \\
II x Ibn & 0.64 & 0.29 \\
IIn x Ic & 0.33 & 0.78 \\
IIn x Ib & 0.26 & 0.93 \\
IIn x IIb & 0.40 & 0.64 \\
IIn x Ibn & 0.57 & 0.56 \\
Ic x Ib & 0.17 & 1.00 \\
Ic x IIb & 0.43 & 0.59 \\
Ic x Ibn & 0.67 & 0.43 \\
Ib x IIb & 0.43 & 0.59 \\
Ib x Ibn & 0.83 & 0.21 \\
IIb x Ibn & 0.60 & 0.57 \\
II x IIn/Ibn & 0.28 & 0.58 \\
II x SESNe & 0.14 & 0.93 \\
IIn/Ibn x SESNe & 0.28 & 0.69 \\
\hline
\end{tabular}
\end{table}

\begin{table}
\caption{Extinction AD values.\label{tab:ebv_AD}}
\centering
\begin{tabular}{lcc}
\hline
AD test & Statistic & Significance\\
\hline
II x IIn & -0.61 & 0.25 \\
II x Ic & -0.79 & 0.25 \\
II x Ib & -0.81 & 0.25 \\
II x IIb & 0.07 & 0.25 \\
II x Ibn & nan & nan \\
IIn x Ic & -0.60 & 0.25 \\
IIn x Ib & -0.93 & 0.25 \\
IIn x IIb & -0.32 & 0.25 \\
IIn x Ibn & nan & nan \\
Ic x Ib & -1.11 & 0.25 \\
Ic x IIb & -0.45 & 0.25 \\
Ic x Ibn & nan & nan \\
Ib x IIb & -0.71 & 0.25 \\
Ib x Ibn & nan & nan \\
IIb x Ibn & nan & nan \\
II x IIn/Ibn & -0.41 & 0.25 \\
II x SESNe & -0.58 & 0.25 \\
IIn/Ibn x SESNe & -0.66 & 0.25 \\
\hline
\end{tabular}
\end{table}

\section{The SLSN 2015bn} \label{ref:app_15bn}

The SN 2015bn is a well studied Type I superluminous supernova (SLSN), located at a redshift $\sim 0.1136$ and is one of the closest events of this type \citep[][]{2016ApJ...826...39N, 2018ApJ...866L..24N}. Since SN 2015bn is part of the ASAS-SN catalog \citep[][]{2017MNRAS.467.1098H} and is the only SLSN event in the sample, we make a brief analysis of its environment in this section for a comparison with the other SN types described in this work. The transient was observed with the MUSE instrument on 2016 January 17, $\sim 390$ days after discovery and $\sim 296$ days after reaching peak brightness (see left panel of Figure \ref{fig:SN2015bn}).

We extracted the spectrum of SN 2015bn in a circular region with a 6 pixel radius centered at the SN coordinates, and show it in the right panel of Figure \ref{fig:SN2015bn}. The spectrum is dominated by emission from the SN, but a few emission lines originating from the underlying \ion{H}{ii}  region can be identified, such as H$\alpha$, H$\beta$, O[III]$\lambda5007$, and N[II]$\lambda6584$. We fit these lines using a Gaussian model with a local continuum and estimate the physical parameters using the methods described in Section \ref{sec:ana}. 

We measured an integrated flux for H$\alpha$ of {$5.96 \pm 0.05 \times 10^{-18} \ \textrm{erg s}^{-1} \textrm{cm}^{-2} \textrm{\AA}^{-1}$}, a luminosity of $2.42 \pm 0.12 \times 10^{38} \ \textrm{erg s}^{-1}$, and an EW $\sim 110$~\AA. 
We note that the EW value in this case is a lower limit, as the SN elevates the continuum level of the spectrum.
The value for the H$\alpha$ EW is similar to the median for the SESNe distribution, and slightly higher than the median for most of the CCSN distributions. 
This value is also similar to the distribution of H$\alpha$ EW of SLSNe-I found by \citet[][]{2015MNRAS.449..917L}.

We use the measured fluxes of  H$\alpha$, H$\beta$, [\ion{O}{III}] $\lambda 5007$, and [\ion{N}{II}]$\lambda 6584$ to estimate the oxygen abundance through the N2 and O3N2 indicators. We find $12 \ +$ log(O/H) $= 8.25 \pm 0.12$~dex for the N2 index, and $12 \ +$ log(O/H) $= 8.20 \pm 0.19$~dex for the O3N2 index. The values found for SN 2015bn are similar to the oxygen abundance of the H-poor SLSN reported by \citet[][]{2015MNRAS.449..917L}, and to the values in these two indicators of the environment around PTF12dam \citep[$12 \ +$ log(O/H) $\sim 8.00$~dex, ][]{2015MNRAS.451L..65T}. Our results agree with the occurrence of these types of events in regions with a lower abundance of oxygen, a strong indication of the preference of H-poor SLSN for metal-poorer environments.

\begin{figure*}[h]
\centering
\includegraphics[width=0.3\textwidth]{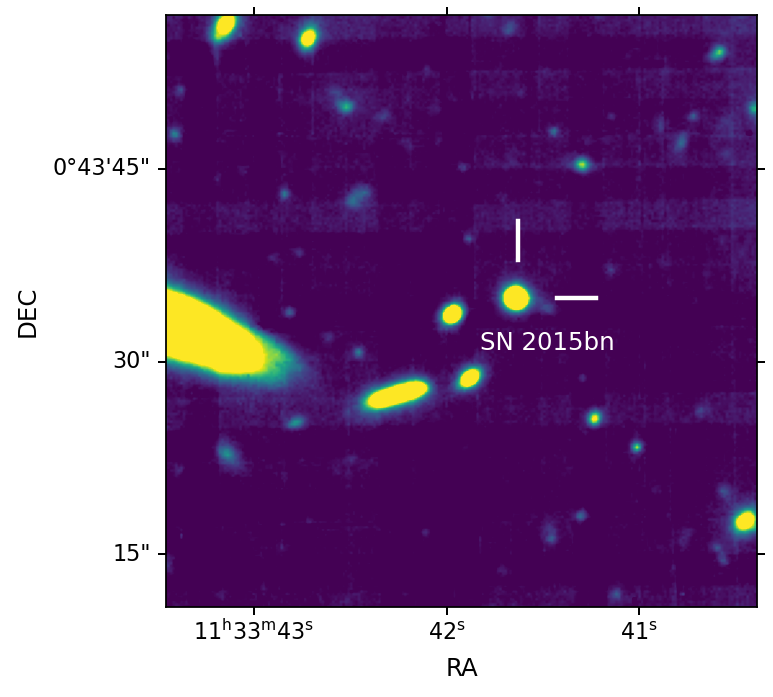}
\includegraphics[width=0.6\textwidth]{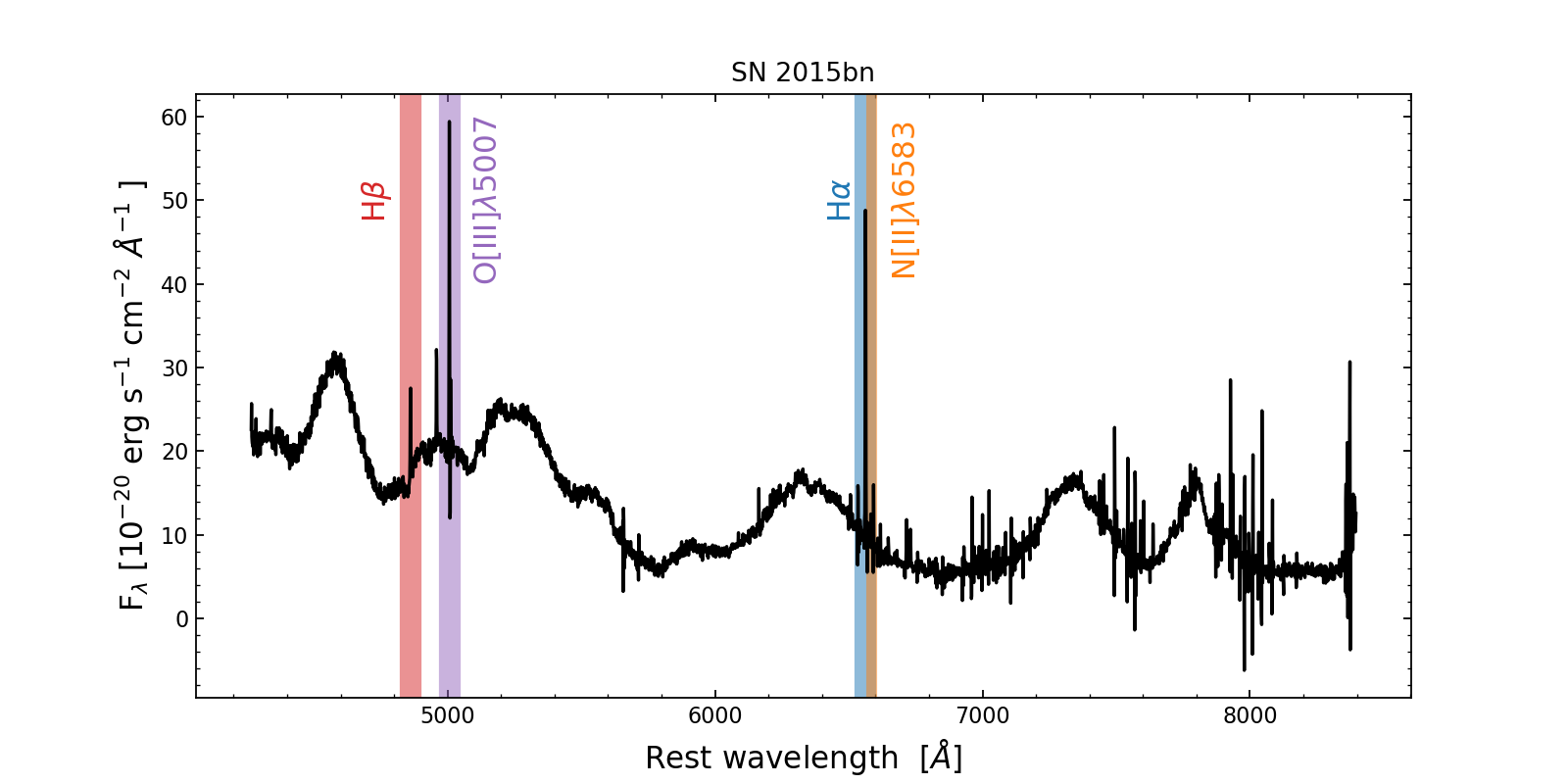}
\caption{Spectrum fitting for the SLSN 2015bn. Left panel: Position of SN 2015bn in the MUSE image of the observed field. Right panel: Extracted spectrum for SN 2015bn, where the broad emission lines of the SN are still clearly present. Emission lines of H$\alpha$, H$\beta$, O[III]$\lambda5007$, and N[II]$\lambda6583$ belonging to the underlying \ion{H}{ii} region can be identified.\label{fig:SN2015bn}} 
\end{figure*}

\section{Photometry and light curve parameters}\label{app:LC_params}

In Tables \ref{tab:phot_II}, \ref{tab:phot_SESNe}, and \ref{tab:phot_int} we present the LC derived parameters for the SNe II, SESNe, and SNe IIn, respectively. 
In most events, the $BVr$ photometry was retrieved from the ASAS-SN archive \citep[for a in depth discussion on ASASSN photometry, see][]{2017PASP..129j4502K}, but in some cases data was obtained from the literature, such as for ASASSN-14jb \citep{2019yCat..36290057M}, SN 2014cy, SN 2014dw, SN 2015W \citep[][]{2016MNRAS.459.3939V}, and SN 2017gmr \citep[][]{2021yCat..18850043A}.
A second order polynomial was used for fit the peak time and the magnitude at peak luminosity, as it is shown in Figure \ref{fig:example_LC_fit} for the SN II ASASSN-14jb. 
{The absolute magnitude is corrected by Galactic extinction.}
For SNe II, a linear fitting between peak and 30 days after was used to obtain the postmaximum decline rate. For SESNe the second order fit was used to obtain the $\Delta m_{15}$ values.

The relation between the LC parameters and the physical parameters derived at the SN position are shown in Figures \ref{fig:II_LCs}, \ref{fig:SESNe_LCs}, and \ref{fig:int_LCs}. The straight dashed lines represent linear fits obtained from the scatter of the values in each photometric band. Pearson correlation tests are performed through {\tt\string SciPy} to obtain the significance from each possible correlation. 

\begin{figure}[t!]
\centering
\includegraphics[scale=0.55]{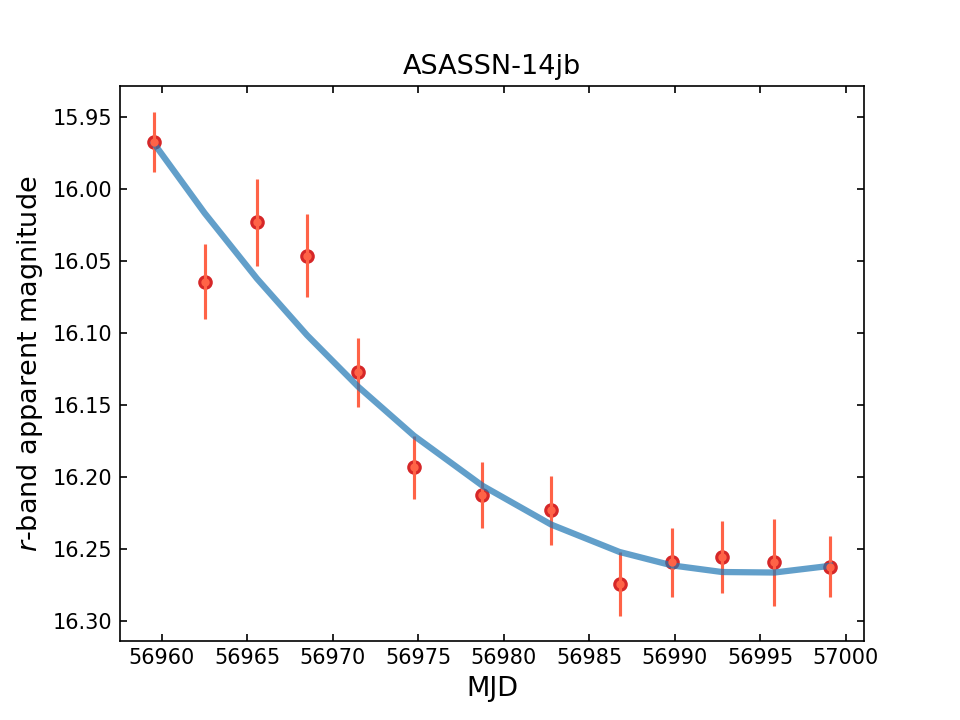}
\caption{Example of a second order polynomial fitting in $r$-band for ASASSN-14jb. The $y$-axis shows the apparent magnitude and $x$-axis gives the phase as modified Julian date (MJD).  \label{fig:example_LC_fit}} 
\end{figure}

\begin{table*}
\caption{LC parameters of the SNe II. \label{tab:phot_II}}
\centering
\begin{tabular}{lcccccc}
\hline
SN & \multicolumn2c{V} & \multicolumn2c{B} & \multicolumn2c{r}\\
 & M$_{peak}$ (mag) & $s$ (mag/days) & M$_{peak}$ (mag) & $s$ (mag/days) & M$_{peak}$ (mag) & $s$ (mag/days)\\
\hline
SN2018eog & -16.88 (0.06) & - & - & - & - & - \\
SN2018cho & -17.56 (0.14) & - & - & - & - & - \\
SN2018yo & -17.06 (0.09) & - & - & - & - & - \\
AT2018cuf & -16.52 (0.12) & - & - & - & - & - \\
ASASSN-18eo & -18.83 (0.08) & 0.0201 (0.0053) & - & - & - & - \\
ASASSN-18cb & -16.72 (0.07) & 0.0075 (0.0032) & - & - & - & - \\
SN2017jbj & -16.55 (0.45) & - & - & - & - & - \\
SN2017grn & -17.57 (0.09) & - & - & - & - & - \\
Gaia17chn & -18.28 (0.10) & - & - & - & - & - \\
SN2017gmr & -17.47 (0.03) & 0.0042 (0.0007) & -17.30 (0.04) & 0.0239 (0.0016) & -17.70 (0.03) & 0.0025 (0.0007) \\
SN2017fem & -17.12 (0.27) & - & - & - & - & - \\
SN2017faf & -19.01 (0.27) & - & - & - & - & - \\
ATLAS17hpc & -18.84 (0.08) & 0.0205 (0.0041) & - & - & - & - \\
... \\
\hline
\end{tabular}
\tablefoot{The entire table is available online.}
\end{table*}

\begin{table*}
\caption{LC parameters of the SESNe. \label{tab:phot_SESNe}}
\centering
\begin{tabular}{lcccccc}
\hline
SN & \multicolumn2c{V} & \multicolumn2c{B} & \multicolumn2c{r}\\
 & M$_{peak}$ (mag) & $\Delta m_{15}$ (mag) & M$_{peak}$ (mag) & $\Delta m_{15}$ (mag) & M$_{peak}$ (mag) & $\Delta m_{15}$ (mag)\\
\hline
SN2018ie & -17.72 (0.07) & 1.58 (0.34) & -17.02 (0.13) & 2.09 (0.23) & -17.92 (0.07) & 1.26 (0.32) \\
SN2018dfg & -16.81 (0.10) & - & - & - & - & - \\
SN2017gax & -18.24 (0.09) & 0.41 (0.15) & -17.69 (0.08) & 1.08 (0.22) & -17.94 (0.06) & 1.02 (0.16) \\
SN2017gat & -18.64 (0.07) & 0.72 (0.13) & -17.92 (0.09) & 0.64 (0.21) & -18.80 (0.05) & 0.54 (0.13) \\
ASASSN-14az & -17.89 (0.06) & 1.25 (0.12) & - & - & - & - \\
SN2016gkg & -16.86 (0.06) & 1.15 (0.20) & -14.93 (0.09) & 0.53 (0.25) & -16.72 (0.08) & 0.92 (0.14) \\
PSNJ1828582 & -16.90 (0.44) & - & - & - & - & - \\
ASASSN-15tu & -17.65 (0.19) & - & - & - & - & - \\
ASASSN-15ta & -18.62 (0.04) & - & - & - & - & - \\
ASASSN-15bd & -17.03 (0.10) & 0.79 (0.36) & - & - & - & - \\
\hline
\end{tabular}
\end{table*}

\begin{table*}
\caption{LC parameters of the SNe IIn. \label{tab:phot_int}}
\centering
\begin{tabular}{lccc}
\hline
SN & V & B & r\\
& M$_{peak}$ (mag)  & M$_{peak}$ (mag) & M$_{peak}$ (mag)\\
\hline
ASASSN-18oa & -18.33 (0.03) & - & - \\
ATLAS17lsn & -20.59 (0.03) & - & -20.69 (0.03) \\
ASASSN-16jt & -18.68 (0.04) & -18.60 (0.04) & -18.57 (0.04) \\
ASASSN-16in & -18.16 (0.03) & -18.08 (0.11) & -18.18 (0.04) \\
SN2016aiy & -17.47 (0.04) & -17.29 (0.04) & -17.60 (0.03) \\
ASASSN-15hs & -16.74 (0.03) & - & - \\
ASASSN-15ab & -19.35 (0.03) & - & - \\
ASASSN-15lx & -17.39 (0.12) & - & - \\
\hline
\end{tabular}
\end{table*}

%-------------------

\end{document}